\documentclass{Configuration_Files/PoliMi3i_thesis}

\usepackage[a4paper]{geometry}
\usepackage{parskip} 
\usepackage{setspace} 
\usepackage{emptypage} 
\usepackage{multicol} 
\usepackage{titlesec}
\titlespacing{\section}{0pt}{3.3ex}{2ex}
\titlespacing{\subsection}{0pt}{3.3ex}{1.65ex}
\titlespacing{\subsubsection}{0pt}{3.3ex}{1ex}
\usepackage{color}

\usepackage[english]{babel} 
\usepackage[utf8]{inputenc} 
\usepackage[T1]{fontenc} 
\usepackage[11pt]{moresize} 

\usepackage{graphicx}
\usepackage{transparent} 
\usepackage{eso-pic} 
\usepackage{tikz} 
\usetikzlibrary{}
\graphicspath{{./Images/}} 
\usepackage{caption} 
\usepackage{subcaption}
\usepackage{xcolor} 
\usepackage{amsthm,thmtools,xcolor} 
\usepackage{float}
\usepackage{epigraph}

\usepackage{amsmath}
\usepackage{amsthm}
\usepackage{amssymb}
\usepackage{amsfonts}
\usepackage{bm}
\usepackage[overload]{empheq} 
\usepackage{fix-cm} 

\usepackage{tabularx}
\usepackage{longtable} 
\usepackage{colortbl}
\usepackage{multirow}
\usepackage[table]{xcolor}
\usepackage{booktabs}
\definecolor{bluePoli}{cmyk}{0.4,0.1,0,0.4}

\usepackage{algorithm}
\usepackage{algorithmic}

\usepackage[colorlinks=true,linkcolor=black,anchorcolor=black,citecolor=black,filecolor=black,menucolor=black,runcolor=black,urlcolor=black]{hyperref} 
\usepackage{cleveref}
\usepackage[square, numbers, sort&compress]{natbib} 
\usepackage{pdfpages} 
\usepackage{afterpage}
\usepackage{lipsum} 
\usepackage{fancyhdr} 
\usepackage{comment}
\usepackage{epstopdf}

\definecolor{bluepoli}{cmyk}{0.4,0.1,0,0.4}

\declaretheoremstyle[
  headfont=\color{bluepoli}\normalfont\bfseries,
  bodyfont=\color{black}\normalfont\itshape,
]{colored}

\theoremstyle{colored}

\newcommand\T{\rule{0pt}{2.6ex}}
\newcommand\B{\rule[-1.2ex]{0pt}{0pt}}

\newcounter{algsubstate}

\newcommand\numfontsize{\@setfontsize\Huge{200}{60}}

\titleformat{\chapter}[hang]{
\fontsize{50}{20}\selectfont\bfseries\filright}{\textcolor{bluepoli} \thechapter\hsp\hspace{2mm}\textcolor{bluepoli}{|   }\hsp}{0pt}{\huge\bfseries \textcolor{bluepoli}
}

\titleformat{\section}
{\color{bluepoli}\normalfont\Large\bfseries}
{\color{bluepoli}\thesection.}{1em}{}

\titleformat{\subsection}
{\color{bluepoli}\normalfont\large\bfseries}
{\color{bluepoli}\thesubsection.}{1em}{}

\titleformat{\subsubsection}
{\color{bluepoli}\normalfont\large\bfseries}
{\color{bluepoli}\thesubsubsection.}{1em}{}

\newcommand{\hsp}{\hspace{0pt}}

\makeatletter
\renewcommand*\cleardoublepage{%
  \clearpage\if@twoside\ifodd\c@page\else
  \null
  \AddToShipoutPicture*{\BackgroundPic}
  \thispagestyle{empty}%
  \newpage
  \if@twocolumn\hbox{}\newpage\fi\fi\fi}
\makeatother

\numberwithin{algorithm}{chapter}

\newcommand{\bea}{\begin{eqnarray}} 
\newcommand{\eea}{\end{eqnarray}}

\newtheorem{definition}{Definition}

\begin{document}

\fancypagestyle{plain}{%
\fancyhf{} 
\fancyhead[RO,RE]{\thepage} 
\renewcommand{\headrulewidth}{0pt}
\renewcommand{\footrulewidth}{0pt}}


\pagestyle{empty} 
\frontmatter 

\puttitle{
	title=GPU Acceleration and Portability of the TRIMEG Code for Gyrokinetic Plasma Simulations using OpenMP, 
	name=Giorgio Daneri, 
	course=High Performance Computing Engineering - Ingegneria del Calcolo ad Alte Prestazioni, 
	ID  = 251529,  
	advisor= Prof. Gianluca Palermo, 
	coadvisor={Dr. Zhixin Lu}, 
	academicyear={2024-25},  
} 

\startpreamble
\setcounter{page}{1} 


\chapter{Abstract}

The field of plasma physics heavily relies on simulations to model various phenomena, such as instabilities, turbulence, and nonlinear behaviors that would otherwise be difficult to study from a purely theoretical approach. Simulations are fundamental in accurately setting up experiments, which can be extremely costly and complex. As high-fidelity tools, gyrokinetic simulations play a crucial role in discovering new physics, interpreting experimental results, and improving the design of next-generation devices. However, their high computational costs necessitate the use of acceleration platforms to reduce execution time. This work revolves around the TRIangular MEsh based Gyrokinetic (TRIMEG) code, which performs high-accuracy particle-in-cell plasma simulations in tokamak geometries, leveraging a novel finite element approach. The rise of graphical processing units (GPUs) constitutes an occasion to satisfy such computational needs, by offloading the most expensive portion of the code to the accelerators. The chosen approach features GPU offloading with the OpenMP API, which grants portability of the code to different architectures, namely AMD and NVIDIA. The particle pushing as well as the grid-to-particle operations have been ported to GPU platforms. Compiler limitations had to be overcome, and portions of the code were restructured to be suitable for GPU acceleration. 
The kernels were analyzed with the available profiling tools to gather metrics about resource occupancy and memory usage. Their performance was evaluated by carrying out GPU grid size exploration, as well as scalability studies. In addition, the efficiency of hybrid MPI-OpenMP offloading parallelization was assessed. The speedup of the GPU implementation was calculated by comparing it with the pure CPU version using different rationales. The Ion Temperature Gradient (ITG) mode was simulated using the GPU-accelerated version, and its correctness was verified by comparing the physics results in terms of the energy growth rate and the two-dimensional mode structures. \\
\\
\textbf{Keywords:} openmp offloading; particle-in-cell; gpu portability.

\chapter*{Abstract in lingua italiana}
Il campo della fisica del plasma necessita di simulazioni numeriche per modellare fenomeni rilevanti come instabilità e turbolenze, difficili da studiare con un approccio puramente teorico. Le simulazioni hanno un ruolo cruciale nell'impostare gli esperimenti, spesso complessi e costosi. In virtù della loro precisione, le simulazioni girocinetiche sono fondamentali per scoprire nuovi fenomeni fisici, interpretare i risultati sperimentali e migliorare il design dei reattori. Ciononostante, il loro elevato costo computazionale richiede l'utilizzo di piattaforme per accelerare il codice e ridurre il tempo di esecuzione. Questo lavoro è basato sul codice TRIangular MEsh based Gyrokinetic (TRIMEG), che effettua simulazioni del plasma ad alta accuratezza tramite il metodo particle-in-cell in geometrie tokamak. L'avvento delle graphical processing units (GPUs) costituisce un'occasione fondamentale per soddisfare i requisiti computazionali, effettuando offloading della parte più onerosa del codice sugli acceleratori. Questo lavoro implementa una soluzione accelerata su GPU con OpenMP e portabile su architetture NVIDIA a AMD. Le operazioni di evoluzione delle particelle e particella-griglia sono state adattate per l'esecuzione su GPU. Durante il processo è stato necessario gestire le limitazioni dei compilatori, che hanno richiesto una ristrutturazione di porzioni del codice. I kernel sono stati analizzati con gli strumenti di profiling disponibili, al fine di raccogliere metriche di valutazione delle prestazioni. Inoltre, le prestazioni sono state vagliate effettuando esplorazione della griglia su GPU, studi di scalabilità e valutazione dell'efficienza di un approccio ibrido MPI-OpenMP offloading. L'accelerazione (speedup) dell'implementazione GPU è stato calcolato comparandola con la versione originale utilizzando diverse strategie. L'instabilità Ion Temperature Gradient (ITG) è stata simulata utilizzando la versione accelerata su GPU e la sua correttezza è stata verificata comparando i risultati fisici con la versione originale in termini di rateo di crescita dell'energia e struttura bidimensionale dei modi.  
\\
\\
\textbf{Parole chiave:} offloading con openmp; simulazioni particle-in-cell; portabilità su gpu.


\thispagestyle{empty}
\begingroup
  \titlespacing*{\chapter}{0pt}{-1.2cm}{20pt} 
  \tableofcontents
\endgroup
\thispagestyle{empty}
\cleardoublepage

%
%
%

\mainmatter 


\chapter{Introduction}
\label{chap:introduction}
\ifpdf
    \graphicspath{{Introduction/Figures/PNG/}{Introduction/Figures/PDF/}{Introduction/Figures/}}
\else
    \graphicspath{{Introduction/Figures/EPS/}{Introduction/Figures/}}
\fi




\section{Motivations}
\label{motivations}
High Performance Computing (HPC) platforms are fundamental in order to satisfy the rising computational needs of large-scale applications and simulations. These are especially required in the context of scientific computing, which is often based on high-accuracy simulations of physical phenomena. Traditional computational paradigms, leveraging multi-process and multi-node execution, have been flanked by newer acceleration platforms, such as Graphical Processing Units (GPUs) and Accelerated Processing Units (APUs), for the purpose of reducing execution time by harnessing their massively parallel capabilities, hence enabling larger and more realistic simulations. \\
Plasma physics is a field that perfectly exemplifies this concept, being the foundation for a deeper understanding of magnetic fusion. The future of green energy lies in fusion power, but the path ahead is still long and technology aid is necessary. 
TRIMEG is a gyrokinetic code written in Fortran with the purpose of modeling the plasma at a particle level, and has been developed at the Max Planck Institute for Plasma Physics (IPP). It solves the gyrokinetic equation by tracking the motion of particle gyro centers. MPI parallelism was already implemented, allowing the code to be simulated on multiple nodes. The code employs a massive number of particles, which can be treated independently during the pushing and grid-to-particle operations, thus it is well suited for GPU offloading.

\section{Context of the Study}
\label{context}
The application domain is plasma physics applied to fusion energy, a field that studies the state of an extremely hot, ionized gas inside a fusion device. Several simulation codes are developed to model the behavior of plasma on different spatial and time scales, and are instrumental to understanding how to mitigate turbulence, non-linear behaviors, and instabilities inside the fusion device, with the final goal of achieving net energy gain. The plasma can be modeled in different ways, with a varying degree of approximation. Magnetohydrodynamics is a coarse-grained model that treats the plasma as a fluid and is useful to simulate large-scale phenomena on longer time scales, due to its reduced computational costs. 
Gyrokinetic is a framework that models the plasma as particles, and reduces the velocity space dimensions from three to two, averaging over the fast gyration of particles around their magnetic axis. This approach demands higher computational needs and is more suitable for simulating smaller-scale phenomena on a shorter time scale. 
Finally, the plasma can also be modeled with fully kinetic particles, which grant more precision but incur even higher computational costs. 
Each approach has their own strength and weaknesses, thus requiring specific numerical methods. To achieve a realistic and overall modeling of the fusion plasma, integration between the models is necessary.

\section{Objectives and Contributions}
\label{objectives}
The contribution of this work is to solve the challenge of achieving a GPU version of TRIMEG that is portable to AMD and NVIDIA GPUs. OpenMP has been chosen as the GPU offloading paradigm to obtain this. The main bottleneck is the compilers that support OpenMP offloading for the target architectures, which required extensive compiler exploration. Advanced programming features like polymorphism are hardly supported inside GPU kernels, hence why a compromise between complexity and code capabilities was necessary. The GPU implementation targets the two most common GPU architectures on the market, NVIDIA and AMD, which implement different memory models, require different compilers, and thus pose distinct implementation problems. The objective of achieving portability across different platforms strongly depends on such differences, and code divergence cannot be avoided altogether. However, the concept of portability is fundamental to address the transition between vendors when selecting GPU hardware for new HPC clusters, dictated by financial policies and strategic decisions. Acceleration is a key aspect of GPU offloading, and is achieved by the kernels we implemented, as shown in Section \ref{chap:performance-evaluation}.

\section{Overview of the Thesis}
\label{overview}
First, a background on plasma physics, magnetic confinement fusion, tokamak geometry, and gyrokinetic theory is provided. The purpose is to create a framework that enables the reader to have a basic understanding of the problem at hand. Then, an outline of the TRIMEG code structure, the physics it tackles, and its foundational algorithms is given. Gaining a deep understanding of the code is a preliminary step in adapting its structure for GPU offloading.
An overview of GPU architectures and their massively parallel capabilities is given, as well as the required programming model to efficiently exploit their hardware capabilities. The choice of the OpenMP Application Programming Interface (API) over potential competitors is motivated, and its foundations in terms of parallel constructs, GPU offloading, and data transfers are described. Note that the choice was not made by the author of this work, but was the result of a careful evaluation by the main contributor to the TRIMEG code and their collaborators.
Afterwards, a description of the code sections that have been ported to GPU architectures is given, including compiler limitations, needed workarounds, and code restructuring for improved GPU suitability. 
Several metrics to assess the performance of the GPU kernels were used to compute the speedup of the GPU-accelerated implementation compared to the original CPU model, and to evaluate the workload scalability. Finally, an assessment of the GPU implementation correctness is conducted by performing benchmark simulations to evaluate the physical results it produces.
The appendix provides further details about the theoretical background, namely the mixed variable model to mitigate the cancellation problem, an in-depth description of magnetic flux coordinates, the transformation from cylindrical to field-aligned coordinates, and the gyro centers' equations of motion that are solved inside the main GPU kernel.
\chapter{Foundations of Magnetic Confinement Fusion}
\label{chap:first}%
\graphicspath{Images/}
    
\section{Magnetic Confinement Fusion}
\label{sec:sec21}
Nuclear fusion might be the key to completing the transition to zero-emission, renewable energy sources.
Magnetic Confinement Fusion (MCF) is a method for generating thermonuclear energy that leverages electromagnetic fields to confine plasma inside a specially designed structure. Plasma is an extremely hot, electrically charged ionized gas. The most reactor-relevant fusion reaction is based on the combination of the light atomic nuclei deuterium and tritium, which are hydrogen isotopes, to alpha particles and neutrons, which release energy kinetically. The neutrons bombard the internal blanket of the confining vessel, transferring a huge amount of heat, which is then transferred to a coolant to eventually harvest energy. \\
A legitimate question would be whether fusion is indeed a renewable source of energy. As previously stated, the main fusion fuels are deuterium and tritium. The former is present in enormous amounts in sea water, with a density of about 30 mg per liter \cite{hagemann1970absolute}, making it virtually unlimited and completely safe to use. The latter is radioactive with a half-life of 12.3 years, and must be produced within the reactor, also due to its unavailability as a natural resource. The internal part of the reactor is coated with a layer of lithium, called the lithium blanket. Several neutrons are emitted from the plasma, which react with lithium to produce a helium nucleus and a tritium nucleus, which can then be extracted and used as fusion fuel. \cite{romanelli2022fusion} \\
Current magnetic fusion devices are experimental, and, as such, do not have the facilities to generate electricity as a commercial reactor design would. A fusion reaction requires overcoming the electrostatic repulsion between the nuclei, thus requiring temperatures on the order of tens of millions of degrees, exceeding those in the core of the Sun. In these conditions, any gas is ionized, forming a plasma that is too hot to be confined by any material container. For this reason, it is necessary to use magnetic confinement to confine the plasma well enough for fusion reaction to occur. This is possible since plasma is electrically conductive, unlike electrically neutral atoms. The most common geometry of magnetic confinement devices is that of a \textbf{tokamak}. However, several challenges need to be overcome to achieve a net energy gain with nuclear fusion. Among these are turbulence, disruptions, runwaway electrons, and edge localized modes, which hinder plasma stability and pose a threat to internal materials and instrumentation. The TRIMEG code is designed to model gyrokinetic plasma behavior in a tokamak vessel, and thus can contribute particularly well to addressing questions related to turbulence and transport.

\subsection{Tokamak Geometry}
\begin{figure}
    \centering
    \includegraphics[width=0.8\linewidth]{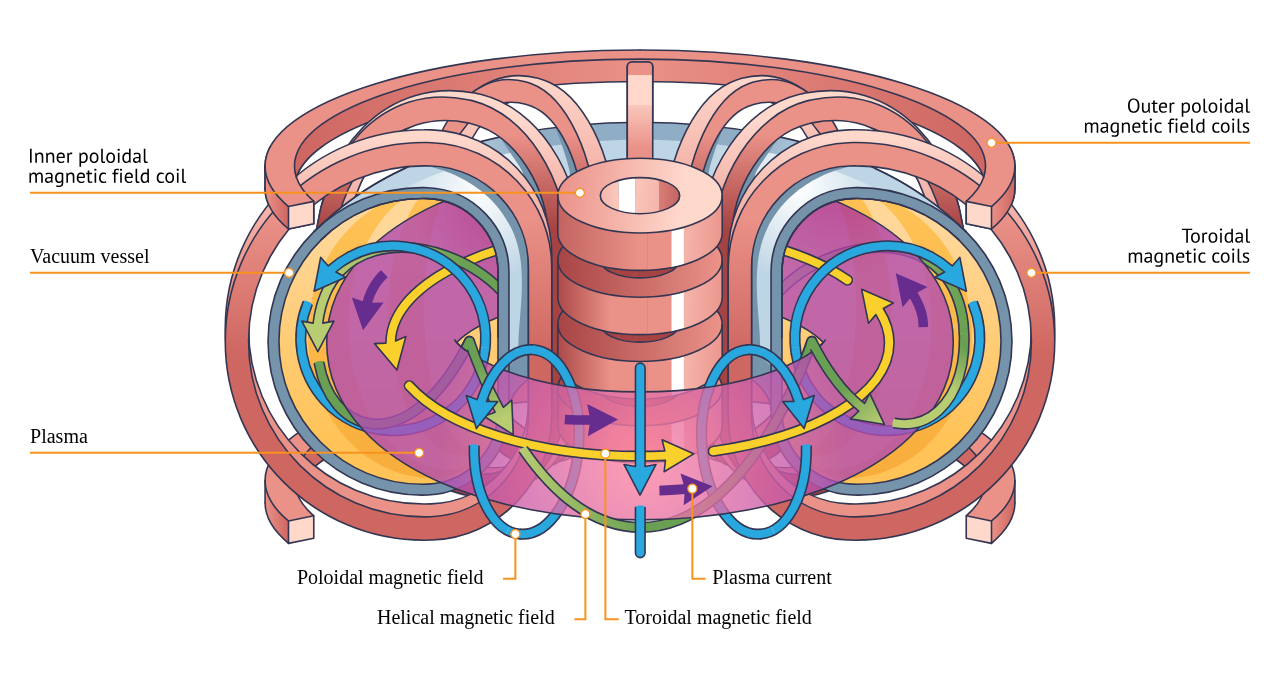}
    \caption{Magnetic field components in a tokamak \cite{energyencyclopedia2025tokamaks}}
    \label{fig:poloidal_toroidal_directions}
\end{figure}

Tokamaks are devices used for the magnetic confinement of plasma in the shape of an axially symmetric torus, achieved by means of external magnets. Given that the plasma is electrically conductive, it can be effectively manipulated by such magnetic fields. The charged particles of the plasma can be controlled by the magnetic coils placed around the vessel. This is fundamental to separate the plasma from the vessel walls, preventing material damage and ensuring the correct functioning of the reactor. To kickstart the process, air and impurities are first evacuated from the vacuum chamber to avoid any interference with the subsequent chemical reactions. Then, the magnet systems for plasma confinement are activated and charged up, as the fuel is injected inside the chamber in the form of gas. A strong electrical current runs through the vessel and causes the gas to break down electrically and become ionized, thus forming the plasma. 
When magnetically confined, the particles describe helical paths along the field lines. While following such orbits for long time scales, particles will eventually collide and fuse. \\
The plasma is confined with a magnetic field composed of both toroidal and poloidal components. The former is induced by the electric current that flows through the external coils, while the latter is mainly produced by a current driven inside the plasma itself. This is generated by the action of a transformer, a central solenoid that acts as the primary winding. The toroidal component $B_{\phi}$ dominates the total magnetic field, which loops around the torus. Fig. \ref{fig:poloidal_toroidal_directions} shows the poloidal and toroidal directions of the magnetic field in a tokamak.

\begin{figure}
    \centering
    \includegraphics[width=0.3\linewidth]{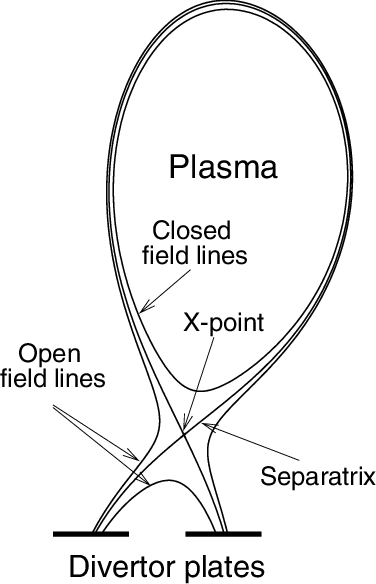}
    \caption{Poloidal section of a tokamak \cite{abdullaev2006mappings}}
    \label{fig:poloidal_section}
\end{figure}

\noindent The separatrix identifies the boundary between the region with closed field lines and that with open field lines; the former corresponds to the magnetically confined plasma core, where the field lines are continuous along the toroidal direction. The latter is characterized by field lines that connect with the surface material of the device. Outside of the separatrix, there is the scrape-off layer (SOL), where the field lines are directed towards a region called divertor, used to extract energy from the device. The value of the poloidal field is equal to zero at the \textbf{x-point}, the initial portion of the divertor, which can be visualized in Fig. \ref{fig:poloidal_section}. For this reason, modeling the x-point may cause numerical instabilities in simulations, primarily due to the Jacobian behavior, but it is crucial for optimizing power exhaust stability. \cite{zheng2025x}

\noindent A divertor geometry is fundamental to extract heat and ash from the fusion reactor while protecting the main chamber from thermal loads, and to reduce the plasma contamination level due to impurities. This is useful to achieve high-confinement modes. A problem fusion reactors will face in the near future is the inability to handle divertor heat loads that exceed the thermal capabilities of plasma-facing components. Because of a divertor, a particle can escape through the separatrix, allowing energy to be absorbed by the part of the divertor suitably placed outside the plasma. In this way, the plasma-facing material faces less stress compared to the rest of the internal surface. Reducing such material erosion is important to contain the amount of impurities that can enter the plasma, which lowers its temperature as well as the fusion rate and efficiency. \\
One of the main objectives and challenges in tokamak operation is to achieve and maintain equilibrium in the plasma configuration. The equilibrium arises when the pressure and magnetic forces balance each other, enabling a stable plasma. Problems arise from instabilities within the plasma, which cause the degradation or disruption of the equilibrium. One of the objectives of plasma simulations is to correctly reproduce and study these instabilities, so that ways to overcome them can be found. 

\section{Gyrokinetic Framework}
\label{sec:subsec21}
Gyrokinetics is a self-consistent kinetic model that studies plasma behavior on scales comparable to the gyroradius with which particles rotate around the magnetic field lines, where the typical frequencies of the dynamical system are lower than the particle cyclotron frequencies. 
\cite{brizard2007foundations,garbet2023gyrokinetics} 

\begin{definition}
    A \textbf{cyclotron} is a device that leverages an electromagnetic field to accelerate particles by increasing their energy.
\end{definition}

\begin{definition}
    The \textbf{cyclotron frequency} is the frequency at which a charged particle orbits in a uniform magnetic field, completing a full revolution around the magnetic field lines inside a cyclotron.
\end{definition}

\begin{definition}
    The \textbf{gyroradius} is the radius of circular motion of the particle in the presence of a magnetic field; in fact, the particles describe a helix when traveling through the plasma. 
\end{definition}

Gyrokinetics is complementary to magnetohydrodynamics, as it considers the individual motion of the particles, specifically their gyration around magnetic field lines, and the wave-particle interactions. The gyrokinetic framework employs an approximation that consists of averaging the fast gyration of particles, which occurs on a faster time scale than the phenomena of interest, namely, the turbulent transport and wave propagation. The dimensionality of the problem is therefore reduced, since it is not required to resolve the full motion of the particles along the magnetic field lines. It also allows larger time steps in the simulations while still capturing essential drift motions and wave-particle interactions. The advantage it retains over magnetohydrodynamics is the fact that it is often necessary to assess particle kinetics to correctly model plasma evolution and phenomena therein. This requires the use of gyrokinetic simulations for specific particle species, which can characterize small-scale as well as global phenomena, such as microturbulences \cite{mishchenko2017mitigation}. 
The gyrokinetic framework proves extremely appropriate in modeling plasma turbulence. As previously mentioned, particles in a uniform magnetic field follow helical trajectories, spiraling around the field lines. A field line is a graphical visualization for a vector field, an integral curve that is tangent to the field vector at each point along its length. The trajectory of a particle can be decomposed into a fast circular motion, the gyromotion, and a comparatively slower motion of the guiding center along the magnetic field lines. 

\begin{definition}
The \textbf{guiding center} is the point around which the particle revolves with its gyromotion.    
\end{definition}

\begin{definition}
The \textbf{gyrocenter} is a more refined position than the guiding center, as it accounts for electromagnetic perturbations and higher-order terms. 
\end{definition}

For most plasma behaviors, the gyromotion can be ignored; hence, averaging over it reduces the dimensionality of the velocity space, thus accounting for a total of five dimensions: three for space and two for velocity. This approximation is the foundation of gyrokinetics, which uses the gyrocenter to model the particles and treats their gyromotion in a simplified and computationally more efficient way. 
The spatial variables are then changed from the actual particle positions \textbf{r} to their guiding center position \textbf{R}. For velocity coordinates, we operate the change from $(v_x,v_y,v_z)$ to the velocity parallel to the magnetic field lines, or parallel velocity for short, defined as $v_{||} \equiv \textbf{v}\cdot \hat{\textbf{b}}$, the magnetic moment $\mu ={m_s v_\perp}/(2B)$, and the gyrophase angle $\alpha$, where $m_s$ is the mass of the particle species that is considered. Averaging over the gyrophase angle at a constant gyrocenter position yields the gyrokinetic equation. Gyrokinetic simulations are particularly useful in predicting the transport level due to neoclassical physics or turbulence \cite{lu2019development}\\

\begin{definition}
    \textbf{Transport level} refers to the rate at which particles, momentum, and energy move across magnetic field lines in a magnetized plasma, to bring a system towards thermodynamic equilibrium with its surroundings. The transport level determines the confinement quality in devices as tokamaks, as it is crucial to preserve the necessary conditions for an efficient fusion reaction.
\end{definition}

Classical transport models a magnetized plasma as a swarm of particles moving along the magnetic field lines. Given the extremely high density, the particles eventually collide and scatter, leading to random movements and runaway effects.
Neoclassical transport also arises from collisions in a toroidal magnetic field, as a result of the guiding center motion of particles. It constitutes an improved version of the classical transport, as it accounts for several additional factors. One is trapped particle orbits, which can happen in toroidal geometries, where some particles get trapped in magnetic wells. 
Also, it considers collisional effects, i.e. collisions that cause particles to diffuse radially, leading to a finite level of transport. Note that neoclassical transport does not account for turbulent effects and is used to predict the lowest transport level in optimized devices. Moreover, it incorporates the effects of the field geometry, which is fundamental for a tokamak device. In a toroidal geometry, since the magnetic field is divergence-free, the field is stronger in the region closer to the symmetry axis when compared to the outside. \\
However, turbulent transport occurs as a result of electromagnetic and electrostatic fluctuations generated by small instabilities within the plasma. In most experimental conditions, it dominates over neoclassical transport, because of its effectiveness in transporting energy and mass. The goal of plasma confinement is to reduce the "anomalous" turbulent transport phenomena as much as possible to improve plasma confinement \cite{weiland2019drift}.\\
It is crucial to be able to simulate plasma in high-confinement mode, where particle confinement is improved due to intrinsic turbulence suppression by sheared plasma rotation, theoretically allowing for sustained energy generation.

\begin{definition}
    \textbf{High-confinement mode} (H-mode) is an operating regime of enhanced confinement in toroidal plasma devices, e.g., tokamaks. It was first observed in the ASDEX tokamak and has been studied intensively in the past several decades \cite{wagner1982regime,wagner1984development,wagner2007quarter}. The energy confinement time is approximately doubled in magnitude. In this mode, the change in confinement is evident, especially at the plasma edge, where the pressure gradient increases substantially due to a rise in edge density.  
\end{definition}

This also yields the formation of pedestal structures in the plasma edge region with reduced transport levels and thus improved confinement. 

\begin{definition}
    \textbf{Edge-localized modes} (ELMs) are periodic disturbances of the plasma periphery that occur in tokamaks with an H-mode edge transport barrier. \cite{connor2008edge}
\end{definition}

This causes a portion of the plasma to transition to the OFL region and the divertor target, following the so-called ELM bursts. These instabilities can potentially damage walls, especially those in the divertor region, due to high energy transfer, as well as triggering other instabilities.
This phenomenon was first observed in the ASDEX tokamak by the same team that first discovered the H-mode. \cite{wagner1990recent_elms}. Theoretical and numerical work has been reported to explain the physics mechanism of ELMs, such as the peeling ballooning mode \cite{snyder2002edge}. \\

Gyrokinetic codes like TRIMEG aim to address the turbulence arising in the plasma and its partial suppression in the H-mode pedestals. However, present developments aim to also capture macroscopic instabilities like ELMs, that were previously only accessible in fluid models. The large-scale separations require highly optimized code running on large HPC systems, which is a strong argument for adapting the code to run on hardware accelerators like GPUs.

\section{Coordinate Systems}
The most used coordinate systems for gyrokinetic simulations are the following: \textbf{cylindrical coordinates} $(R,\varphi,Z)$ and \textbf{magnetic flux coordinates} $(\psi,\varphi,\theta)$. In the former coordinate system, we have that:
\begin{itemize}
    \item R is the major radius, i.e. the distance from the tokamak central axis
    \item $\varphi$ is the toroidal angle
    \item Z is the vertical height 
\end{itemize}

\noindent 
The transform between the cylindrical coordinates and the cartesian coordinates $(X,Y,Z)$ is readily obtained as $X=R\cos\phi$, $Y=-R\sin\phi$, $Z=Z$.
The three-dimensional global simulations are carried out in tokamak geometry using the cylindrical coordinates. 
For the latter coordinate system, we have that:
\begin{itemize}
    \item $\psi$ is the poloidal magnetic flux function
    \item $\varphi$ is again the toroidal angle
    \item $\theta$ is the poloidal angle
\end{itemize}
A more in-depth analysis of the coordinate systems and a description of the field-aligned coordinates are given in Appendix \ref{field-aligned-coordinates}.


\chapter{Structure of the TRIMEG Code}
\label{chap:second}%
\graphicspath{Images/}

\section{Introduction to the Code}
TRIMEG is a gyrokinetic, particle-in-cell code that models the plasma, including the open field lines (OFL) region. The most recent version of TRIMEG features a high-order $C^1$ finite element method for unstructured meshes in the poloidal cross section, but the cubic B-spline in the toroidal direction \cite{lu2024gyrokinetic}. 

\begin{definition}
    \textbf{An unstructured mesh} is a mesh with general connectivity, which must be explicitly defined since its structure is arbitrary. General connectivity element types are nonorthogonal, e.g. triangles and tetrahedra. This requires mapping more information to each element, e.g. adjacency lists. Grids of this type can be used in finite element analysis when the input is of irregular shape.
\end{definition}

TRIMEG leverages the particle-in-cell (PIC) method, a technique used to solve a class of partial differential equations featuring an efficient particle positioning scheme for charge and density deposition, called scattering, and field interpolation to the particle positions, called gathering. The core concept of a PIC method consists of tracking individual particles in a continuous space, while physical quantities such as densities and current are computed concurrently on stationary mesh points, also called Eulerian points. Applying this concept to plasma physics amounts to computing the particle trajectories inside an electromagnetic field solved on a stationary mesh.
The iterative structure of the PIC method is as follows.
\begin{itemize}
    \item \textbf{particle push}: the position and velocity of each particle are updated based on the force acting on it, which is usually computed by solving the gyro center's equation of motion for the gyrokinetic model or the Lorentz equation for the fully kinetic model.
    \item \textbf{charge deposition/scattering:} the charge of each particle is distributed on neighboring mesh points using a weighting function. When using a finite element method, the charge of each particle is projected onto the basis functions. 
    \item \textbf{field solve:} the field is computed on the stationary mesh, typically by solving the perturbed magnetic field with Amp\`ere's law and the electrostatic field with Poisson's equation.
    \item \textbf{field gathering/interpolation:} the field values computed in the previous step are interpolated to the particle positions and used to compute the forces acting on them in the following time step.
\end{itemize}

These operations are the fundamental steps of the so-called gather-scatter scheme. The \texttt{gather} phase consists of information flow from the grid to the particles, in order to compute the value of the field at the particle coordinates. In the \texttt{scatter} phase, the flow of information is inverse.
There exist two main formulations for the PIC method. The $p_{||}$ formulation is widely used, whereas the $v_{||}$ formulation suffers from numerical instabilities and requires a fully implicit time solver. Still, the former has the disadvantage of the cancellation problem,  which limits electromagnetic gyrokinetic simulations to low-beta cases, i.e. $\beta < \sqrt{m_e / m_i}$ \cite{mishchenko2017mitigation}. 

\begin{definition}
    The beta of a plasma is the ratio of the plasma pressure to the magnetic field pressure. In nuclear reactors, the temperature of the fuel scales with pressure, which is why reactors aim at achieving the highest possible pressure. Since the cost of large magnets scales approximately as $\beta^{1/2}$,  this factor can be interpreted as an economical indicator of reactor efficiency, to keep the magnetic field as small as possible.
\end{definition}

The cancellation problem arises when the cancellation between the electron skin depth term and the corresponding term in the $p_\|$ current term can hardly be achieved in solving the parallel Amp\`ere's law, potentially leading to inaccuracy and numerical instabilities. 
This can happen when the plasma beta is high, and thus the electron skin depth term is much larger than the physical term, leading to inexact cancellation. More details about it can be found in Appendix \ref{mixed-variable-and-pullback-scheme}.

In order to simulate the complex physical events that occur in H-mode and along the edge region, numerical schemes for unstructured meshes were developed, especially to deal with the OFL region. TRIMEG also features an efficient particle positioning method based on an intermediate structured grid as the search index for the mesh triangles where the particle is actually located. 

\section{Structure of the Code}
\label{subsec:structure_code}
There exist two branches of the TRIMEG code. The first is TRIMEG-C1, which uses triangular meshes with $C^1$ continuous finite elements for the whole volume, up to the OFL, as previously mentioned. 
The second branch is TRIMEG-GKX, which leverages structured meshes, focusing on the core plasma. It serves as a testing environment for new physics models and numerical methods that would eventually be applied to the C1 version. An explicit integration method and a field-aligned finite element approach have been implemented for multi-\textit{n} simulations in TRIMEG-GKX \cite{lu2025piecewise}.
TRIMEG also performs effective noise reduction on full \textit{f} and mixed $\delta f$-full \textit{f} simulations \cite{lu2023full_f_delta_f}. In the former case, the total distribution function \textit{f} evolves directly using kinetic equations. This includes both equilibrium and perturbations, and is necessary when studying phenomena that are driven by large perturbations and strongly deviate from equilibrium. This simulation approach is computationally expensive, since one needs to solve the whole velocity-space dynamics, including small fluctuations, with an acceptable signal-to-noise ratio.
The $\delta f$ method is a reduced kinetic approach in which the distribution function is split into an equilibrium component $f_0$ and a small perturbation $\delta f$: $f = f_0 + \delta f$. Instead of computing the evolution of the whole distribution, only the evolution of $\delta f$ is computed. This is useful for studying small perturbations, which applies to micro-turbulence and various large-scale Alfv\'enic modes. It is also capable of capturing the linear and non-linear dynamics. 

To solve the field equations, the code makes use of \texttt{PETSc}, a library designed for scientific applications modeled by partial differential equations. Specifically, it provides linear and nonlinear solvers, efficient preconditioners, time integrators, structured and unstructured grids, matrix and vector structures, as well as utilities to assemble them. 
In pushing the particles, the perturbed field is interpolated at each particle position using the field variables on the grids. This process is done in \texttt{G2P}-like subroutines (grid-to-particle). 
In solving the field equations, the charge and current on the grid are calculated using the projection operation from particles to the finite element basis functions.  This process is done in \texttt{P2G} (particle-to-grid) functions. The \texttt{G2P} interpolation and the \texttt{P2G} projection implement the gather-scatter method. The time evolution of the particle coordinates requires an iterative procedure, where particles are pushed, field equations are solved, then particles are pushed again, and so forth, as described earlier. Eventually, the field equations are discretized and result in a linear system, where the \texttt{solver} class sets up the values of the system matrix corresponding to the field equations, such as the quasi-neutrality equation and the parallel Amp\`ere's law. 

\begin{figure}
    \centering
    \includegraphics[width=1.1\linewidth]{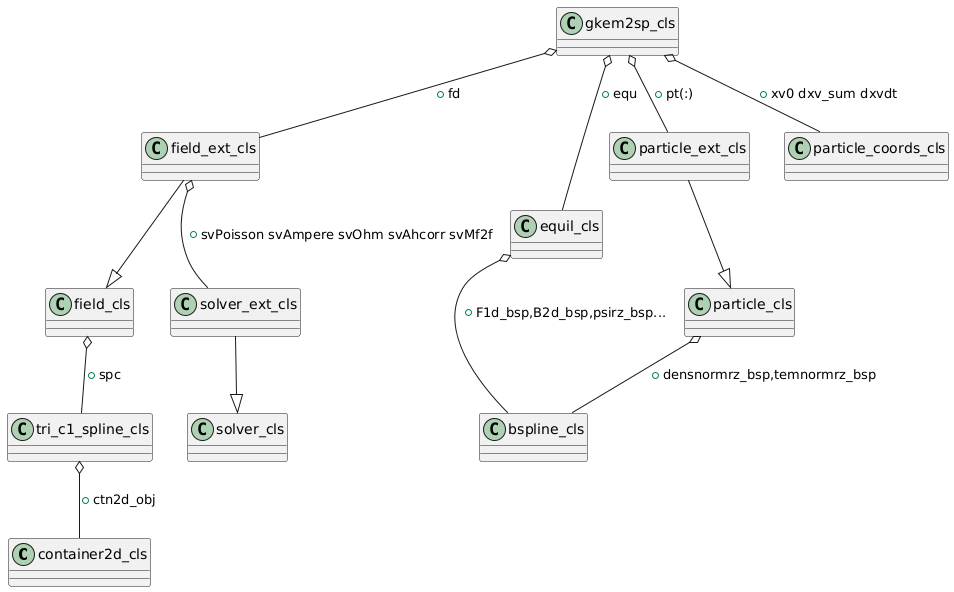}
    \caption{Class Hierarchy of TRIMEG}
    \label{fig:class-uml}
\end{figure}

Figure \ref{fig:class-uml} depicts the hierarchical structure of the classes implemented in the TRIMEG code.
A high-level description of the code is provided below, focusing on the main functionalities implemented by each class. 
The \textbf{container class} mainly generates the intermediate grid and computes and stores the mapping between the triangular mesh and the square elements of the intermediate grid. The \textbf{triangular C1 spline class} offers an ensemble of routines to implement $C^1$ continuous triangular elements in the $(R,Z)$ plane, and cubic spline in the $\varphi$ direction. The \textbf{solver class} has the purpose of solving the field equations using PETSc routines. It allocates the necessary data structures, assembles the system matrices in a distributed way among MPI ranks, then solves the linear system corresponding to the field equations. The \textbf{field class} implements g2p operations for field interpolation to particle positions, by solving Ampère's law and Ohm's law. The \textbf{equilibrium class} computes the equilibrium of the magnetic field in the poloidal section, then interpolates over the $\varphi$ coordinate to obtain its toroidal component. It also computes the contributions that appear in the gyrocenters' equations of motion. The \textbf{particle class} implements the p2g routines, and computes few terms of the quasi-neutrality equation. The \textbf{gkem2sp class} is the high-level wrapper that computes the time evolution of the system using the integration method of choice between forward Euler, Runge-Kutta 4th order, and a semi-implicit scheme. In general, for each time step the equation for the pullback treatment is first solved. If the chosen model is electromagnetic, then both the magnetic vector potential and electrostatic scalar potential are computed, together with the plasma and current density. Then, Ohm's law is solved with a subroutine of the field module, so that the time derivative of the simplectic part of the magnetic potential is calculated. Subsequently, the particle pushing procedure is performed, first calculating the derivative of the particle position and velocity at the current time step. Finally, the position, parallel velocity, and weight coefficient is computed for each particle at the next time step.

\section{Particle Positioning Method} \label{Particle positioning method}
Particle positioning is the process of mapping the particle markers to the corresponding elements of the mesh. Its efficiency affects the calculation of the toroidal component of the charge density perturbation $\delta n(R,Z)$ in TRIMEG-C0 \cite{lu2019development} or the 3D charge and current density perturbation in TRIMEG-C1 \cite{lu2024gyrokinetic}, as well as the interpolation of the field value at the particle position using the value previously computed for the whole grid.  
A naive approach is the following: for every particle marker, iterate over all triangles in the mesh and check whether the particle position is inside the current element. The brute-force scheme has a complexity of $NN_t$, where $N$ is the number of particle markers, and $N_t$ is the number of triangles in the unstructured mesh. In TRIMEG, a more efficient approach is implemented. An intermediate rectangular grid of square elements, referred to as bounding boxes, is superimposed on the unstructured mesh composed of triangular elements. This grid is generated in the $(R,Z)$ cross section, and mapping $\{\text{Box}_i\rightarrow\text{Triangle}_j\}$ is constructed when triangle $j$ overlaps with box $i$. First, the procedure requires finding the box containing the particle marker $p$, then identifying the corresponding triangle using the box-triangle mapping previously built.
The computational cost is reduced to $O(\alpha N)$, where $\alpha$ is a constant. The rectangular elements of the intermediate grid are generated according to the input parameters that correspond to the dimensions of the rectangular grid along the $R$ and $Z$ dimensions. If these parameters are set to 2 the positioning scheme turns into a brute-force scheme, since only one bounding box is generated. A reasonable approach consists of choosing $N_x \approx N_r$. 
A pseudo-algorithm of this method can be found below:

\begin{verbatim}
! two-dimensional vector to hold the mapping between boxes and triangles
! dimensions are x and y coordinates of the box
! every element is itself a VECTOR holding a number of triangles
VECTOR box_triangle_mapping(:,:)
! first construct the mapping between the boxes and the triangles
for box j in intermediate grid
    for triangle i in mesh
        if(i superimposes on j)
            box_triangle_mapping(j.x,j.y).add(i)
        endif
    endfor
endfor

! push the particle markers on the mesh
! discretize the position of the particles so that the x and y coordinates
! identify a box in the intermediate grid
int triangle_index(:)
VECTOR temp_box
for particle p in particles
    temp_box = box_triangle_mapping(p.x,p.y)
    for triangle t in temp_box
        if(contains(t,p))
            triangle_index = t.index
        endif
    endfor
endfor
\end{verbatim}
The \texttt{contains()} function calculates if the particle is located inside a triangle.
During simulations, it is necessary to find the triangular element where each particle is located at each time step, by calling the actual implementation of the \texttt{contains()} function. Therefore, this algorithm is fundamental to achieve good performance in the particle push procedure, which is the primary focus of GPU porting. The data structure used to store the mapping was restructured to be more easily mapped to the GPU device, as described in Section \ref{implementation-problems-code-restructuring}.

\section{Discretization of the Distribution Function}
\begin{definition}
    The distribution function $f$ of a plasma is the probability of finding a particle at a given position, with a given velocity at a given time step. It is a statistical representation of the plasma particles.
\end{definition} 

\begin{definition}
    In tokamak geometry, the phase space is a multi-dimensional space that describes the possible states of the particles inside the plasma, in terms of their positions and momenta. It is used to characterize the trajectories of particles over time.
\end{definition}

The $\delta f$ model describes the distribution of the $N$ particle markers as follows.
$$f(z,t) \approx \sum\limits_{p=1}^{N} \frac{\delta[z_p - z_p(t)]}{J_z}\;\;,$$
where $z$ is the phase space coordinate, $\delta$ is the Dirac delta function, $J_z$ is the corresponding Jacobian and $z=(\textbf{R},v_{||},\mu)$ is the phase space coordinate adopted in \cite{lu2023full_f_delta_f}, $\mu = v_\perp^2/2B$ is the magnetic moment, $v_{||}$ is the parallel velocity, and $\textbf{R}$ is the real space coordinate \cite{lu2024gyrokinetic}. 
\begin{definition}
    The magnetic moment is a property associated with atoms as a result of the presence of electrons. These have two types of motion, spin and orbital, where the latter is associated with magnetic moment. The orbital motion of an electron around the nucleus can be described as a current in a loop of wire that has no resistance. The magnetic moment is the combination of strength and orientation of a system that exerts a magnetic field.  
\end{definition}

The particle markers used in TRIMEG are randomly initialized, uniformly distributed in the toroidal direction and in the $(R,Z)$ plane.
For the $\delta f$ model, the total distribution function can be decomposed into the background and perturbed parts: $f(z,t) = f_0(z,t) + \delta f(z,t)$. The background contribution is chosen to be time independent, i.e., $f_0(z,t)= f_0(z)$, and a usual choice would be the Maxwellian distribution. 
In general, the gyrocenter's equation of motion can be decomposed into an equilibrium part and a perturbed part due to the perturbed field.
\begin{align}
    &\dot{\textbf{R}} = \dot{\textbf{R}}_0 + \delta \dot{\textbf{R}} \label{eq:position_equation_of_motion} \;, \\
    &\dot{v}_{||} = \dot{v}_{||,0} + \delta \dot{v}_{||} \label{eq:velocity_equation_of_motion_deltaf} \;.
\end{align}

The time derivative is equal to zero for the equilibrium part $f_0$, since $f_0$ is chosen as a steady state solution.

\section{C1 Finite Elements in Triangular Meshes} \label{C1 finite elements in triangula meshes}
\begin{figure}
    \centering
    \includegraphics[width=0.3\linewidth]{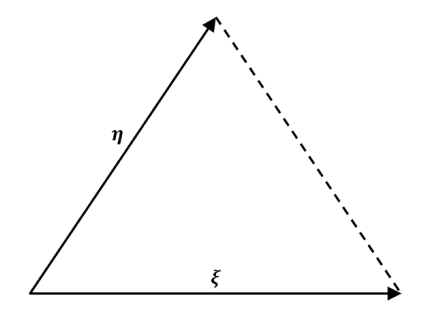}
    \caption{Local coordinates $(\xi,\eta)$ in a mesh triangle \cite{lu2024gyrokinetic}}
    \label{fig:triangle-local-coordinates}
\end{figure}

The code implements a finite element method in all three dimensions of the cylindrical coordinate system. An unstructured triangular mesh is used in the poloidal plane, while a uniform structured mesh is used in the toroidal direction $\phi$. A periodic boundary condition is applied in the toroidal direction, while a homogeneous Dirichlet boundary condition is imposed on the outer boundary of the simulation domain in the \((R, Z)\) cross section.
The total size of the grid is obtained by summing up the number of grid points in the $\phi$ direction and the vertices in the $(R,Z)$ poloidal plane. \\
Each triangle in the poloidal plane is characterized by local coordinates $(\xi,\eta)$. Fig. \ref{fig:triangle-local-coordinates} shows that the triangle vertices are located at $(\xi,\eta) = (0,0),(1,0),(0,1)$, respectively.

\noindent The basis functions of the C0 finite elements are the following. \cite{lu2024gyrokinetic}
\begin{align}
    &A_1^{lin} = 1 - \xi - \eta\;\;, \\
    &A_2^{lin} = \xi\;\;,\\
    &A_3^{lin} = \eta\;\;.
\end{align}

For the finite element method implemented in TRIMEG C1, 18 basis functions are used for each triangle in the form of quintic polynomials. 
\begin{definition}
    For a one-dimensional problem, a quintic polynomial is a function of the form:
$$g(x)=ax^5+bx^4+cx^3+dx^2+ex+f$$
where the coefficients are chosen for certain purposes such as the partition of unity. 
For a two-dimensional problem, the quintic polynomial refers to the following form.
$$g(x,y)=\sum_{0\le i+j\le5} A_{ij}x^iy^j \;\;.$$ 
\end{definition}
The one-dimensional quintic polynomials appear similar to cubic functions, but can exhibit two additional local maximum/minimum values. 
Note that in general there is no algebraic expression for the solutions of a quintic function, while it always exists for lower-degree polynomials.  
Clearly, this finite element method requires $C^1$ continuity along the edges of the triangle. Each triangle vertex, or node, has six degrees of freedom: $$f\;\;,\;\; \frac{\partial f}{\partial R}\;\;,\;\; \frac{\partial f}{\partial Z}\;\;,\;\; \frac{\partial^2 f}{\partial^2 R}\;\;,\;\; \frac{\partial^2 f}{\partial R\partial Z}\;\;,\;\; \frac{\partial^2 f}{\partial^2 Z}\;\;.$$ \\

By means of the aforementioned basis functions, any variable $y(R,Z)$ can be written as a linear combination of the basis functions with suitable coefficients: \cite{lu2024gyrokinetic}
$$y(R,Z) = \sum_{k=1}^{18}\sum_{j=1}^{N_{tri}}\Lambda_{j,k}(\xi,\eta)Y_{j,k} = \sum_{i=1}^{18N_{tri}}\Lambda_{j,k}(\xi,\eta)Y_i \;\;,$$
where $N_{tri}$ is the number of triangles, and each $\Lambda_{j,k}$ is the basis function of triangle $j$ for the degree of freedom $k$ at local coordinates $(\xi,\eta)$. The coefficients $Y_i$ are solved with the constraints that the variables 
$$y(R,Z), \frac{\partial y(R,Z)}{\partial R}, \frac{\partial y(R,Z)}{\partial Z}, \frac{\partial^2 y(R,Z)}{\partial^2 R}, \frac{\partial^2 y(R,Z)}{\partial R\partial Z}, \frac{\partial^2 y(R,Z)}{\partial^2 Z}\;\; $$are $C^1$ continuous at any vertex.\\
For the three-dimensional variable $Y(R,\Phi,Z)$ the $C^1$ continuous triangular element is used in the $(R,Z)$ directions, whereas the cubic spline basis functions are adopted in the $\Phi$ direction, resulting in the following formulation. \cite{lu2024gyrokinetic}
$$Y(R,\Phi,Z) = \sum_{i=1}^{6N_{vert}}\sum_{j=1}^{N_\Phi}\hat{\Lambda}_i(\xi,\eta)N_j(\Phi)\hat{Y}_{i,j}\;\;.$$
The weak form of the field equations can be readily obtained, and the details can be found in the recent work \cite{lu2024gyrokinetic}. 

\chapter{GPU Porting with OpenMP}
\label{chap:third}%
\graphicspath{Images/}

\section{GPU Architectures} 
\label{sec:gv1}
\subsection{General Features of GPUs}
The rise in computational demands requires new programming paradigms as well as the exploitation of massively parallel architectures. GPUs particularly stand out in this regard, given their capabilities of multi-layered parallelism, massive number of workers, high throughput and low-latency memory access. The concept of heterogeneous computing has become more and more established, and consists in executing different portions of the code on the best suited architecture, depending on the available parallelism, memory access patterns and computational demands. In the case of TRIMEG, we limited our architectures of choice to traditional CPUs and we accelerated the hotspots on GPUs. The concept of GPU offloading implies the transfer of a resource-demanding workload to the available accelerator device, which is capable of performing a type of computation in a more efficient way, both in terms of time and energy consumption. First of all, it is important to assess whether the code portion to be offloaded satisfies a set of criteria, thus justifying this implementation choice. An important requirement is a high degree of parallelism, i.e., few or no data dependencies are present within the code section, hence the amount of computations that can be performed in parallel allows to exploit the available workers as best as possible. The control divergence due to the presence of conditional statements needs to be as low as possible. If control divergence is present within the same block, the execution of its threads will be serialized. Specifically, the threads for which the condition evaluates to true will first execute, than all those for which the condition evaluates to false, or vice versa. Also, the code needs to have a high ratio of arithmetic operations per memory access; a memory intensive application might not be suited for offloading if the data transferred to the device is not frequently reused. The memory allocation and transfer cost, as well as under-utilization of the resources could nullify the performance gain, since most of the time is spent in waiting for the necessary data to be transferred. Memory accesses can incur high latency, especially if the data is not located in the cache, but can be partially hidden by computations that do not depend on them. Even in the case of a memory-bound application, it is still possible to benefit from GPU acceleration by leveraging shared memory and coalesced accesses; however, these usually require a restructuring of the code that was originally conceived to run on the CPU. A coalesced memory implies that consecutive memory accesses will request data that is stored contiguously, allowing for the aggregation of multiple accesses. It is the key to achieve the best possible memory access pattern, by maximizing the utilization of the bus bandwidth. The number of memory accesses is reduced, given that a thread accesses a whole memory sector, not a single element, thus reading or writing data that are also needed by other threads. Note that memory accesses involve a non-negligible latency, hence minimizing their amount results in a potentially significant speedup of the program execution time. The suitability of TRIMEG for GPU architectures according to these parameters will be discussed in Section \ref{gpu-offloading-of-trimeg}.\\
Although GPUs were originally designed to handle computer graphics tasks, their usage has evolved to perform computations traditionally handled by CPUs. This gave rise to the concept of general purpose GPU (GPGPU) programming, which is a processing paradigm applied to massively parallel programs. GPUs generally have a much higher number of threads than CPUs. A GPU consists of several streaming multiprocessors (SMs), which are independent working units with dedicated memory, shared among all their threads. SMs have a function analogous to that of the CPU cores, since they both execute computations and store the current state in the available registers and caches. The instructions are logically partitioned into blocks, where each block is a collection of threads mapped to a single SM.
The typical unit of execution on the GPU is called warp, which follows the single-instruction-multiple-thread (SIMT) execution model, since all the threads in a warp share the instruction fetch phase and execute each one on a different set of inputs. A single SM is capable of executing multiple warps, which can be interleaved by the warp scheduler for latency hiding.  When the instruction executed by a warp performs a memory access, another warp is scheduled in place of the current one, while the required data are being fetched.  
While having a high number of threads, SMs are not designed to support speculative execution nor instruction pointer predictions, optimizations that are limited to traditional CPU hardware. Within a single warp, different threads may evaluate a branch differently based on the data unit that they are processing. This causes warp divergence, that is, the warp executes both paths sequentially, a phenomenon to be avoided as it can seriously affect performance. It corresponds to the concept of control divergence applied to GPU architectures.  \\
While each SM has a dedicated shared memory for all its warps, all threads can use the global memory, which is separated from that of the CPU. For this reason, on most GPUs data transfers play a crucial role in optimizing the code and constitute one of the primary bottlenecks. 

\subsection{APUs}
An innovative concept is that of accelerated processing units (APUs), which leverage a unified shared memory (USM) between the CPU and the GPU. USM refers to a single, coherent memory space that is accessible to both CPU and GPU components. The primary advantage is the simplification of the programming model, as it does not require to explicitly map the data between host and device, thus avoiding synchronization barriers. The design reduces latency thanks to the interconnect that preserves cache coherence across the different chips, improves data locality, and reduces data redundancy. In the case of TRIMEG, directives to map data from the host to the device and vice versa are still necessary, since the code must also be executable on architectures that do not feature USM.
In general, AMD adopts the terminology of the OpenCL programming model for GPUs, which differs from that conceived by NVIDIA, so it is beneficial to briefly introduce it.
A compute unit (CU) is a parallel vector processor that contains multiple APUs, a concept analogous to that of streaming multiprocessor. The command processor sends workgroups of work-items to each CU, where these are executed in wavefronts. A wavefront, also known as wave, is a group of work-items that is executed on a SIMD processor. All wavefronts in a block are bounded to the same CU. It corresponds to a warp of an NVIDIA GPU, in that all hardware threads that are part of it execute the same instructions on different data and follow the same control flow path. The difference is that warps are constituted by 32 threads, while a wavefront is usually formed by 64 threads. A collection of wavefronts forms a workgroup, analogous to a thread block. Multiple workgroups are executed on the GPU at the same time, can synchronize and communicate through local memory. In this model of computation, a thread is an individual lane in a wavefront, which must run in a lockstep with the other threads of the same workgroup, i.e., must be synchronized with them at each barrier.

\section{Introduction to OpenMP}
There exist several ways to achieve GPU offloading, which vary in the level of abstraction, performance, and flexibility in terms of supported platforms. Due to the heterogeneity in the hardware of the different HPC systems available, the decision was to opt for the OpenMP library to carry out the task of porting TRIMEG to GPU architectures. Specifically, OpenMP supports offloading for both NVIDIA and AMD GPUs, which are part of the Max Planck Computing and Data Facility (MPCDF) Raven and Viper cluster hardware, respectively. The institute also has access to other third-party clusters, such as Pitagora, an NVIDIA-based cluster managed by Cineca.
OpenMP is an application programming interface (API) that allows to exploit shared-memory multiprocessing, as well as GPU offloading for the C/C++ and Fortran programming languages, from OpenMP 4.0 onward. Similarly to OpenACC, it is a high-level standard mainly composed of compiler directives, which allow the user to specify the portions of the code to be parallelized, and the strategy to be used. It also features environment variables and library functions for various purposes, which will be described afterwards. The main difference between OpenMP and OpenACC is that the former relies more on the ability of the programmer to choose the right parallelization strategy, while the latter is more focused on the capability of the compiler to exploit the available parallelism. However, although OpenACC support for NVIDIA GPUs is advanced, that for AMD GPUs leaves much to be desired, as this vendor is investing far more resources in broadening the OpenMP counterpart.
Other options for GPU programming in Fortran would include CUDA Fortran and Kokkos. The former is obviously not compatible with AMD GPUs, thus requiring the translation of the CUDA kernels to a compatible language via automated tools, e.g. GPUFORT. In the case of complex kernels, deeply integrated with Fortran code, translation is not fully automated and requires the programmer to manually tune and assess the semantic correctness of the result. This approach was avoided because the tools were not mature enough and because of the widespread duplication of the code that this approach would require.

Kokkos has a C++ core, but a Fortran compatibility layer is available, despite still being in early-stage development. It requires writing Fortran wrappers that call the C++ Kokkos API routines, which increases the complexity of the approach. Moreover, compiler support may be poor, especially for AMD architectures. 
All of these reasons contribute to making a strong argument in favor of choosing OpenMP as GPU programming model. Although it is theoretically the easiest approach to GPU offloading, it bears several limitations in terms of performance, debugging, and low-level instruction control. The abstraction introduced by the OpenMP API does not give complete control over memory transfers, as some data structures are mapped automatically in an implicit way, potentially introducing unnecessary memory movements and runtime overhead. There is no way to explicitly harness the reduced latency of shared memory (local data cache in OpenCL terminology), as this is done under the hood once the \texttt{target} regions are translated into CUDA/HIP kernels, which might not reflect compile-time optimizations that would be applicable otherwise. Finally, portability across different platforms relies on the actual compiler support for OpenMP offloading, making it a potentially fragile solution. Simultaneously developing and testing on both NVIDIA and AMD architectures helps in minimizing code divergence due to different compiler limitations. \\
Other gyrokinetic codes have been ported to GPU architectures, but prioritized a working solution on NVIDIA platforms and used different programming strategies \cite{ohana2021gyrokinetic, germaschewski2021toward}, making this work a pioneering effort in the field of plasma physics. 

\subsection{Programming Paradigm}
First, let us explain the basics of multithreaded execution on the CPU, which forms the basis of the parallel programming framework to be applied for GPU offloading. The master thread executes the program sequentially until it reaches a parallel construct, which is implemented as follows.
\begin{verbatim}
    !$omp parallel
    ...
    !$omp end parallel
\end{verbatim}
When the master thread reaches the parallel region, it spawns as many threads as those specified in the configuration. The user can modify this quantity in three ways:
\begin{enumerate}
    \item set the environment variable \texttt{OMP\_NUM\_THREADS} either from the command line or inside the SLURM batch script that is used to configure and launch the job, e.g. \texttt{export OMP\_NUM\_THREADS=\${SLURM\_CPUS\_PER\_TASK}}. This is applied globally to all parallel regions.
    \item use the library function \texttt{omp\_set\_num\_threads} directly inside the code at compile time. This method overwrites the previous setting.
    \item use the \texttt{num\_threads(N)} for each parallel region, which enables fine-grain control by manually setting the number of threads based on the estimated workload size of each parallel region.
\end{enumerate}

The content of the region is then executed by all the threads, which stop their execution as soon as they reach the implicit barrier at the end of the parallel construct. The barrier serves as a synchronization point for all threads, which need to be joined, as the execution after that point will be sequential again.  
The main target of parallelization are loop structures, since they usually enclose most of the computations. In the case of Fortran, the \texttt{do loop} construct is of interest, and it can be made parallel by nesting the following directive inside a parallel region as follows.
\begin{verbatim}
    !$omp parallel
    !$omp do
    ...
    !$omp end do
    !$omp end parallel
\end{verbatim}
Usually these directives are merged into a single one:
\begin{verbatim}
    !$omp parallel do
    ...
    !$omp end parallel do
\end{verbatim}

Vectorized instructions can be used with the \texttt{simd} directive, which enables the execution of multiple iterations of the associated loops concurrently by means of single-instruction multiple-data (SIMD) instructions. This can be used in combination with the \texttt{collapse} clause, which combines the iteration space of two or more loops into a single one, exposing more parallelism. This allows one to parallelize multiple loops without inserting nested parallelism. In order to be used, the loops must form a rectangular iteration space and their bounds and strides must be invariant. If the loop indices have different sizes, the largest index will be the one used for the collapsed loop. Furthermore, the loops must be perfectly nested, i.e. there is no code nor OpenMP directives between the loops that are collapsed. A toy example is the following. 
\begin{verbatim}
!$omp parallel do collapse(2) private(i,j,k) 
   do k = 1, Ksize
     do j = 1, Jsize
       do i = 1, Isize 
         call dummy_computation(a,i,j,k) 
       enddo 
     enddo 
   enddo 
!$omp end parallel do 
\end{verbatim}

The example makes use of a clause to specify the context of the variables that are accessed inside the parallel construct. Variables in a \texttt{shared} context can be accessed by all threads that execute the parallel region. On the other hand, variables in a \texttt{private} context require to create a copy for each thread at the beginning of the parallel region. Modifications made by a thread to its own copy of the variable are not seen by the other threads. Finally, \texttt{firstprivate} specifies that each thread should have its own instance of a variable, and each instance should be initialized with the value that the variable had before the parallel construct, since it had already been declared. There exists a default scope depending on the location of the variable declaration. If a variable has been declared before the work-sharing construct, then its default scope is shared among threads. If a variable is declared inside the work-sharing region, each thread creates its own copy; thus, its scope is private. 
A best practice is to use the \texttt{default(none)} clause, which requires that each variable accessed inside the OpenMP construct, whose scope was not specified previously, must have its data-sharing attribute explicitly determined by the programmer. Since this forces to list the scope of all the variables, it avoids mistakes like not specifying that a pre-existing variable should be private instead of shared.

\subsection{Data Dependencies}
The programmer should be mindful when wrapping loops with a work-sharing construct, especially if there are so-called loop-carried dependencies. These are present when an iteration reads a value that is written by an earlier iteration. When the loop is executed sequentially, there is no problem. In contrast, if the loop iterations are executed in parallel by multiple threads, there is no certainty that the given value will be ready before the iteration that depends on it is executed. However, in some cases it is still possible to exploit a degree of parallelism, at least if the dependency distance is greater than one. This is defined as the distance between the iteration that produces a value and the one that uses it. An example is the following.
\begin{verbatim}
do i = 1,n
    do j = 3,m
        b(i,j) = b(i,j-2) * a(i,j)
    enddo
enddo
\end{verbatim}
In the above case, there is a loop-carried dependency with distance = 2 for the index \texttt{j}. This still allows to execute two consecutive instructions of the inner loop at a time. Nonetheless, OpenMP cannot be made aware of the dependency distance, so it cannot be used to parallelize the loop. To achieve a degree of parallelism, we must rely on the ability of the compiler to perform low-level optimizations. 

\subsection{GPU Offloading} \label{GPU_offloading_subsubsection}
OpenMP has introduced GPU offloading since the 4.0 version and has continued to improve support in its successive 5.0 and 6.0 versions. However, real-life limitations arise due to incomplete compiler support for the latest standards. The programming paradigm is analogous to that of multi-threaded execution on the CPU, based on compiler directives, thus the user can declare a region of code that will be run on the available accelerator with the \texttt{target} directive:
\begin{verbatim}
!$omp target
...
!$omp end target
\end{verbatim}
Under the hood, all the instructions enclosed in this region will be translated into a CUDA kernel on NVIDIA platforms, or into a HIP kernel on AMD platforms. During program execution, the OpenMP runtime handles calls to the native CUDA/HIP runtime.  
The directive above does not introduce any level of parallelism per se and needs additional work-sharing constructs, some of which are specifically designed for GPU offloading. In this context, there are two main levels of parallelism: block-level parallelism and thread-level parallelism. The first is achieved with the \texttt{teams} construct, which needs to be strictly nested inside a \texttt{target} region, i.e., there should be no code between the two directives. This directive is the foundation of hierarchical parallelism, since each team works independently from the others. Inside each team, further parallelism can be exploited, which will be described later. When the master thread encounters this directive, it creates a league of teams. If the number of requested teams exceeds either of the following values, the number of teams that will be generated is the minimum between the two:
\begin{enumerate}
    \item the value specified in the \texttt{num\_teams} clause, which follows the aforementioned directive
    \item 65536, which is the theoretical implementation limit
\end{enumerate}
The number of threads per team can be specified by the user in different ways, but analogous restraints apply. If the requested number exceeds one of the following values, the number of threads per team will be the minimum between the two:
\begin{enumerate}
    \item the value specified in the \texttt{thread\_limit} clause, which follows the aforementioned directive
    \item 992, which is the theoretical implementation limit
\end{enumerate}
Note that the values above may vary depending on the compiler and the corresponding architecture. In order to set these quantities, the most effective way is to use environment variables as follows.
\begin{verbatim}
export OMP_NUM_TEAMS=256
export OMP_TEAMS_THREAD_LIMIT=256
\end{verbatim}
Intuitively, the first variable is used to set the number of teams, which are mapped to CUDA blocks, or workgroups in HIP/OpenCL terminology. The second variable is used to specify the number of threads per team, determining the size of each block or workgroup. Using environment variables proves to be more effective for parameter exploration, where the number of GPU workers in terms of block and grid size needs to be tweaked to find the best configuration for a given workload. These should be modified inside the SLURM script that is used to launch the simulations. Fortran routines are also available to achieve this purpose, e.g. \texttt{
omp\_set\_num\_teams(int num\_teams)}, but they require recompiling the code if the programmer needs to modify either the grid or block size, which implies potentially high overhead, especially in the case of relatively large codes like TRIMEG. Nonetheless, the actual support for the environment variables depends on the compiler, while the second approach is universally supported. Finally, as mentioned above, a fine-grained approach consists of utilizing the ad hoc clauses for each \texttt{teams} region, specifically \texttt{num\_teams} and \texttt{threads\_limit}, if an estimate of the number of loop iterations is available a priori. Note that the last approach overwrites any value that is set using Fortran utilities, which in turn have greater priority than the environment variables.\\
A directive that is often used in conjunction with the previous one is the \texttt{distribute} directive, which specifies that the iteration space of one or more loops will be distributed among teams. The portion of the iteration space assigned to each team is traversed in parallel by its threads. This directive is useful to create independent work chunks assigned to different teams, provided that there is no loop-carried dependency. Note that no implicit barrier is present at the end of such a region, that is, teams are not implicitly synchronized. On the other hand, threads within a team are necessarily synchronized, unless the \texttt{nowait} clause is specified. This directive is usually combined with the \texttt{parallel do} directive that has been described before, so that an additional level of parallelism is achieved. In principle, each team executes the code sequentially, so this is necessary to achieve intra-team parallelism. To summarize, \texttt{distribute} assigns work to each team, while \texttt{parallel do} assigns work across threads within a team.\\

It is also necessary to be able to call a function from within a target region. Since the portion of code to be GPU-accelerated is vast and spans multiple modules, the compiler must generate a device-callable version of each function called inside a kernel. The \texttt{omp declare target} directive has this purpose, and should be put right at the beginning of the desired subroutine, as follows.
\begin{verbatim}
subroutine compute(this)
!$omp declare target
implicit none
...
end subroutine compute
\end{verbatim}
Clearly, if the user calls the function from outside a target region, the CPU version will be called at runtime. Among the different compilers that have been used for the corresponding architectures, the \texttt{nvfortran} compiler showed a limitation in this regard. A function that has been declared as target cannot contain any other target directive. This does not allow for performing data transfers while the native runtime has execution control, implying that all the necessary data needs to be present on the device before the beginning of the kernel, if managed memory support is not present. In contrast, the \texttt{amdflang} compiler allows inserting a target data directive inside a target function, offering a higher degree of flexibility.

\subsection{Data Transfers}
A fundamental part of the GPU offloading paradigm is the handling of data transfers between host and device, in both directions. This is one of the main limitations to porting complex simulation codes to the GPU, especially if the available compilers do not fully support OpenMP directives. The most straightforward way to map data to or from the device is using the \texttt{map} directive in conjunction with the \texttt{target} directive, which amounts to transferring the necessary data contextually to the kernel execution. The scope of the data is limited to the target region, meaning that the memory is deallocated and cannot be referenced anymore after the kernel terminates. Note that the \texttt{nvfortran} compiler automatically maps both variables and arrays that are declared locally in the subroutine. The same applies to scalar variables that are members of an object contained within a mapping directive. In contrast, array members still need to be mapped explicitly. The \texttt{map} directive requires the programmer to specify the clause that is used to transfer the data, which can be either \texttt{to} or \texttt{from}. The first corresponds to transferring data from the host to the device before the kernel is started, while the second performs the operation in the opposite direction once the kernel has terminated its execution. These two clauses can be merged into \texttt{tofrom} to specify that a data transfer should be carried out in both directions, before and after kernel execution. An example is the following.

\begin{verbatim}
integer,allocatable::x(:)
integer::i,size
allocate(x(size))
!$omp target map(from:x)
do i=1:size
    x = 10
enddo
!$omp end target
\end{verbatim}
Although this approach is the easiest and is almost always guaranteed to work, it comes with several limitations and is not efficient in most cases, as the data transfer overhead cannot be amortized. It is beneficial to decouple this operation from the kernel execution, so that memory latency can be hidden by running other instructions on the CPU. This can be achieved with a combination of \texttt{enter data, exit data} and \texttt{update} directives. This is the most unstructured way of mapping data to the device, since it does not require to specify the region of code where the data is effectively present on the device. Particular attention should be paid when mapping derived types to the GPU, since their internal structure can be rather complex. First, each array requires memory allocation on the device with the \texttt{alloc} clause. Then, the whole object instance is mapped with the \texttt{enter(to)} directive, which automatically maps all the scalar variables of the object. The GPU memory will be valid until explicitly de-allocated with a proper directive, or the program terminates. Then, the array data should actually be transferred to the device with the \texttt{update to} directive. After kernel execution, all arrays that need to be mapped back to the host should be inserted inside an \texttt{update from} directive. Ultimately, to ensure proper handling of device memory, all regions that have been allocated on the device should be freed with the \texttt{exit data map} together with the \texttt{delete} clause when they are no longer needed.  
An example is as follows.

\begin{verbatim}
module particle_mod
    implicit none
    type particle
        real*8,allocatable::partx(:)
        real*8,allocatable::partv(:)
        integer::size
        integer::species
    end type particle

    contains

    subroutine init_pos(this,size)
        implicit none
        type(particle)::this
        integer::i
        !$omp target teams distribute parallel do
        do i=1:this%size
            this%partx = 0.0
        enddo
        !$omp end target teams distribute parallel do
    end subroutine init_pos
end module particle_mod

program main
    use particle_mod
    implicit none
    type(particle)::p
    integer::size
    size=1000
    allocate(p%partx(size))
    p%size=size
    !$omp target enter data map(to:p)
    !$omp target enter data map(alloc:p%partx)
    call init_pos(p,size)
    !$omp target update from(p%partx)
    ! do other computations ...
    !$omp exit data map(delete:p%partx)
    !$omp exit data map(delete:p)
end program
\end{verbatim}

This approach guarantees more flexibility and proves to be more efficient, as it can minimize the number of memory movements when multiple kernels accessing the same data are called consecutively. The programmer can make sure that only necessary data are transferred every time the control is passed from the host to the device and vice versa. 
\chapter{GPU Offloading of TRIMEG} 
\label{gpu-offloading-of-trimeg}%
\graphicspath{Images/}

The main focus of GPU offloading is on the particle push procedure, where the particle markers are pushed along their trajectories in the grid, and several physical quantities are computed at their local positions. This portion of the code is suited for GPU offloading as all the particles can be treated and evolved independently, given that there is no interaction between them. This exposes a high degree of parallelism, which can be exploited because of the massive number of independent workers on the GPU. Data transfers between the host and the device are limited, and this portion of the code has a high arithmetic intensity, as shown by the profiling data in Section \ref{profiling-data}. For a high number of particle makers, e.g., in the order of $10^7$ or $10^8$, the particle pushing operation, including field interpolation calculations, can take up to 70 or 80\% of the total execution time of a simulation. The remainder of the execution time is devoted to the field solver, particle-to-grid and other grid-to-particle operations. The field solver is implemented with the PETSc library, which has very limited GPU support, and is usually not relevant from a computational point of view, reason why it has not been considered for GPU acceleration. \\
A general remark is that any speedup gained thanks to GPU offloading is only applied to the share of the execution time of the accelerated portion of code. Suppose an overall program execution time of 1000 seconds, 80\% of which is taken by the particle pushing and the g2p functions related to it. Assume that the GPU kernel achieves a speedup of 10 versus the CPU version. Let us compute the impact of the speedup on the total simulation time. The particle pushing takes 800 seconds, which is then reduced to only 80 seconds when offloaded to the GPU device. Nonetheless, the overall execution time is still 280 seconds, which translates to an overall speedup of $100/280\approx3.57$. This would be the best-case scenario, but if the percentage of time taken for the accelerated section is reduced to 70\%, the total speedup would be reduced to $1000/370\approx2.70$. For this reason, the choice of which portion should be offloaded to the GPU and the effort to realize the goal should be proportional to the potential performance gains. \\
Moreover, although OpenMP offloading seems straightforward at first, applying it to existing and complex codes is not a trivial task. While in principle the programmer does not need to write actual code, besides adding the necessary compiler directives, the offloading procedure might require extensive code restructuring and debugging. As previously stated, OpenMP is a high-level API, lacking low-level control that proves useful when the complexity of the code goes beyond the OpenMP offloading support offered by the used compiler. Moreover, it is quite hard to debug, as there is no direct control of the underlying CUDA or HIP kernels that are generated by the API. Finally, incomplete support of OpenMP offloading features by available Fortran compilers also contributed to increasing the difficulty of GPU porting. For the reasons mentioned, most of the work carried out in this thesis consisted of circumventing compiler limitations and finding the root cause of seemingly inexplicable problems.

\section{Particle Push Kernel} 
The particle pushing is the main focus of the GPU acceleration, as previously said. It is computed in a loop over the number of particle markers, where each marker evolves independently of the others by solving the gyrocenters' equations of motion. The structure of the loop body is quite complex, the stack of function calls is very deep and involves functions from external libraries, as well as a Fortran implementation of dynamically growing arrays that mimics the C++ vectors from the standard template library. The functions that are called from inside the loop body do not contain loops with a high number of iterations, which would be candidates for an additional layer of parallelism. 
The offloading strategy features preemptive device memory allocation for the data that will be accessed in the kernels, so that the latency at runtime is minimized. Moreover, after the initialization phase, all the data structures that are static throughout execution are transferred to the GPU once, to avoid unnecessary data movements. The entire particle loop is wrapped inside an \texttt{!\$omp declare target} directive, so that it constitutes a single large kernel. This brings the advantage of exposing the parallelism over all the particle markers, without any loop-carried dependency, which allows partitioning the iteration space into chunks and distributing them across OpenMP teams, where the iterations assigned to each team are processed in parallel by its threads. Moreover, teams and threads need to be spawned only once, thus reducing the overhead to a minimum. However, this comes at a great implementation cost in terms of difficulty, since every function inside the loop needs to be callable from the GPU. Furthermore, there is a constraint on the number of particle markers that need to be processed on a single node. If too many particles are assigned to each node, its GPU memory will not be sufficient to hold all the necessary data at once. \\
When it comes to the code, a necessary step prior to the particle push has already been described in section \ref{Particle positioning method} and consists of computing the intermediate grid, which is used to reduce the complexity of the triangle search algorithm by the means of bounding boxes. A two-dimensional dynamic array is used to encode all boxes, where each element is itself a vector storing all the triangular elements that overlap with the corresponding box. Each box is identified with its starting \texttt{x} and \texttt{y} coordinates, that is, the coordinates of its bottom left vertex. The array is populated at the beginning of the simulation inside the \texttt{container2d} class. The triangular mesh is usually much finer than the intermediate grid, so that a single box will superimpose on several triangles in the finite element mesh.
Now let us discuss the structure of the loop that positions the particle markers on the mesh, computes the equilibrium configuration of the magnetic field, interpolates the field quantities to the particle positions, solves the gyrocenters' equations of motion and computes the weights that determine the contribution of each marker of various physical quantities density, like density or current. First, the triangular element where each particle is located needs to be found. The \texttt{search\_tria\_aux} subroutine has the purpose of finding the mesh element that encapsulates the particle and calculating its local coordinates inside of it.
It first computes the global coordinates of the particle as a function of the container and element size. This results in the discretization of the particle position with respect to the intermediate grid, so that its coordinates coincide with those of a bounding box, i.e., its bottom left vertex. These values are used to index the array holding the box-triangle mapping and allow access to the correct element in O(1) time. This corresponds to a vector itself, which contains all the triangles that superimpose on it. A loop iterates over such elements to find the correct one by calling a utility function from an external math library, which computes if the particle is actually inside of the selected triangle. When the correct one is found, the local coordinates of the particle are computed, as described in section \ref{C1 finite elements in triangula meshes}. 
In order to obtain the dominant terms of the equations of motion for the reduced $\delta f$ model, first the equilibrium part is computed, then the perturbed contribution of the particle velocity along the toroidal direction, as well as its parallel acceleration. The equilibrium contribution is computed at the initial time step, i.e. $t=0$ and labeled with the corresponding subscript, as in Eqs.\eqref{eq:position_equation_of_motion} and \ref{velocity_equation_of_motion}. 
Then, the equilibrium configuration is computed as functions of $(R,Z)$ in the cylindrical coordinates, starting from the magnetic equilibrium constructed from the EQDSK output file. The $B_R,B_Z$ components of the equilibrium magnetic field are calculated inside the \texttt{equil\_cls\_calc\_BRZ} subroutine. The toroidal component of the magnetic field is calculated as $B_\phi=F/R$ by means of b-spline interpolation of the poloidal current function $F$. Then, it is possible to compute the derivatives of these components with respect to the $(R,Z)$ coordinates. Then, it is necessary to compute the perturbed part of the magnetic field, a fundamental contribution to the gyrocenters' equations of motion, which describes the motion of the particles inside the plasma. 
As previously stated, the code implements a mixed variable model for the pullback scheme, which has been defined in Eq.~\eqref{mixed variable}. If the input parameters are set to use the electromagnetic model instead of the purely electrostatic one, the magnetic vector potential is computed by interpolating the grid values to the particle positions.
The next contribution to be computed is the parallel velocity of the gyrocenter. Let us recall its definition.
$$u_{||}=v_{||} +\frac{q_s}{m_s}\langle\delta A_{||}^H\rangle$$
If the chosen model does not use the mixed variable scheme, then the vector potential consists only of the Hamiltonian contribution, since the $p_{||}$ formulation is used. 
The electrostatic scalar potential needs to always be computed, regardless of the input parameters. 
The perturbed part of the equation of motion requires the contribution $\delta G = \delta\Phi+ u_{||}\delta A_{||}$, where $\delta\Phi$ is the perturbed electrostatic scalar potential. 
If the rigorous gyrocenter's equation is solved, the high-order terms need to be calculated that depend on $\nabla\times {\bf b}$. This requires computing the terms involving the curl of the magnetic field in the $R,Z,\varphi$ directions, which is done with the \texttt{getcurlbR,getcurlbZ,getcurlbphi} subroutines. The gyrocenters' equations of motion are used to obtain the unperturbed components of the effective magnetic field in cylindrical coordinates, see Eqs.~\eqref{unperturbed_magnetic_field_R_equations_of_motion}, \eqref{unperturbed_magnetic_field_Z_equations_of_motion}, and \eqref{unperturbed_magnetic_field_phi_equations_of_motion}. Subsequently, the perturbed component is computed as described by Eqs.~\eqref{perturbed_magnetic_field_R_equations_of_motion}, \eqref{perturbed_magnetic_field_Z_equations_of_motion}, and \eqref{perturbed_magnetic_field_phi_equations_of_motion}. Finally, the gyrocenter position in cylindrical coordinates is computed following Eqs.~\eqref{R_component_equation_of_motion}, \eqref{Z_component_equation_of_motion}, and \eqref{phi_component_equation_of_motion} for the equilibrium part, 
\eqref{R_perturbed_equation_of_motion}, \eqref{Z_perturbed_equation_of_motion}, and \eqref{phi_perturbed_equation_of_motion} for the perturbed part, respectively. The same procedure is applied to compute the parallel velocity of the gyrocenter of each particle, by solving Eq.~\eqref{velocity_equation_of_motion} for the equilibrium contribution and Eq.~\eqref{velocity_perturbed_equation_of_motion} for the perturbed contribution. This concludes the calculation of the complete form of the equations of motion. Finally, the weights for the deposition of particle quantities on the grid are computed and the loop ends. In total, there are five output arrays, called \texttt{dRdt}, \texttt{dZdt}, \texttt{dphidt}, \texttt{dvpardt}, and \texttt{dwdt}, that need to be transferred from the device back to the host, each one of size equal to the number of particle markers. The values of these arrays are used to assess the correctness of the GPU implementation. 

\section{Other Kernels}
\subsection{Pullback Kernel}
Offloading the particle pusher was the most difficult part, as it was necessary to adapt a large portion of the code for GPU execution. After porting was achieved successfully, the next bottleneck was a set of subroutines that perform g2p-like operations in other sections of the code. Let us consider the subroutine that applies the pullback treatment to the parallel velocity and weight function, even if its contribution is minor. The pullback is a coordinate transformation applied to the perturbed distribution function, used to map the particle phase space to the gyrocenter phase space. If the selected model is electromagnetic and uses the mixed variable or the Hamiltonian model, it is necessary to compute a correction term for the magnetic potential $\delta A^h$ for each particle marker. This is done using a g2p function, which is quite computationally expensive and is the reason why this subroutine benefits from GPU porting, especially for a high number of particle markers, on the order of $10^7$ and beyond.

\subsection{Density Calculation Kernel}
To calculate the local charge and current densities represented by the coefficients of the finite element basis functions, a calculation analogous to the scattering/deposition is performed using the g2p subroutine. For this reason, it was natural to offload this small kernel to the GPU as well.

\subsection{g2p2g Kernel}
Once again, the same g2p subroutine is called to interpolate the distribution function at each particle position. The kernel has the same structure as those for pullback and density calculation.

\section{Code Compilation}
The starting point consisted of adapting the code to compile correctly on both target architectures. During the development phase, it was not trivial to preserve a unified code version, due to different compiler limitations and support for advanced programming features, highlighting the challenge of achieving portability to both NVIDIA and AMD GPU platforms. For reference, the versions that have been mainly used during development are 21.0.0 for the \texttt{amdflang} compiler, while 24.9 and 25.1 for the \texttt{nvfortran} compiler.

\subsection{Compilation on AMD}
The only compiler that supports OpenMP offloading to AMD GPUs is \texttt{amdflang}, developed by AMD itself. The Viper GPU cluster is the platform on which the AMD version was developed and tested. The AMD ROCm software stack is available as an environment module, and includes the \texttt{amdflang} compiler. To automatically link the code against the MPI library, the \texttt{mpifort} alias is available. It should be noted that the \texttt{amdflang} compiler is still in a rather early stage of development and not always suitable for production builds, especially when the programmer cannot exploit the advantage of USM. To be compiled successfully, the original code required some changes due to slightly different syntactic requirements of the compiler, but the overall structure was preserved. The \texttt{--offload-arch=gfx942:xnack+} option is necessary to enable offloading for the specific architecture of the MI300A APU with USM support, which is the hardware set of the Viper GPU cluster, while the environment variable \texttt{export HSA\_XNACK=1} needs to be set to enable USM at runtime. It will cause page faults to automatically trigger page table lookups, which are fundamental when using a unified memory space. OpenMP support is activated with the compiler flag \texttt{-fopenmp}. The code has been optimized with the \texttt{-O3} option, although the time required to apply link-time optimizations is extremely long for a project the size of TRIMEG, as it is on the order of 5 minutes. For this reason, the optimization level \texttt{O2} was used during the development phase, since it has a more reasonable - but still long - linking time of $\approx 2$ minutes. The code performance at runtime appears to be minimally affected by the higher optimization level, namely \texttt{O3} versus \texttt{O2}, but no quantitative analysis has been performed in this respect. The complete command used to compile the files for production simulations is \texttt{mpifort -O3 -fopenmp --offload-arch=gfx942:xnack+ -D\_SHARED\_MEM}, where the last string defines a preprocessor macro to activate the code portions to run the code with USM, by means of preprocessor directives. 

\subsection{Compilation on NVIDIA}
Compilation on NVIDIA-based clusters has been carried out with the MPI wrapper around the \texttt{nvfortran} compiler, together with the compiler options \texttt{-O3 -D\_NVIDIA\_ARCH \\-mp=gpu -Minfo=mp -gpu=cc80}. The NVIDIA-based clusters used to test the code are Raven of the MPCDF, and Pitagora of Cineca. The last string of the compilation command defines a preprocessor macro analogous to the one described in the previous section. The second compiler option enables openMP offloading, while the last specifies the compute capabilities for the A100 GPU, which define hardware features and supported instructions for a specific architecture. For the Pitagora cluster, equipped with H100 GPUs, \texttt{-gpu=cc90} was used instead, as the newer Hopper architecture corresponds to an updated compute capability. The \texttt{-Minfo=mp} option is particularly useful as it gives detailed feedback on how the compiler processes OpenMP directives, whether it performs any implicit data transfers, and which kernels and target functions are generated. As in the previous case, the code required minor changes, including the use of the IEEE library for a handful of arithmetic operations, but the overall code structure was left intact. Finally, these changes were integrated with those made to conform the code to the \texttt{amdflang} syntax requirements, to obtain a single code version.  

\section{Modus Operandi}
The procedure of offloading the code incurred several problems. For this reason, it was beneficial to create a testing repository in which to reproduce the structure of the code portions that generated the problems, so that it would be easier to pinpoint the root cause. In addition, debugging tools were necessary to understand runtime CUDA errors. Especially on the targeted NVIDIA architectures, which do not offer USM support, it was not trivial to assess how the compiler manages data transfers of complex and nested data structures, e.g., arrays of structures. The most immediate way of gathering runtime debug information is to use the \texttt{NV\_ACC\_NOTIFY} environment variable, a bit mask with the following values and the corresponding information: 
\begin{itemize}
    \item 1: kernel launches
    \item 2: data transfers
    \item 4: wait operations or synchronizations with the device
    \item 8: region entry/exit
    \item 16: data allocate/free
\end{itemize}
The most useful options have been the second and last, which correspond to \\ \texttt{NV\_ACC\_NOTIFY=18}. This prints detailed information about the size and address of the memory regions allocated on the device, to which variable each device pointer is associated, whether a variable is already present on the GPU device, and all the implicit transfers of scalar and array variables in the form of attachments. \\
However, this approach comes with big limitations, as it requires the programmer to manually inspect the standard error, which can be quite cumbersome. NVIDIA offers a professional debugging tool called \texttt{compute-sanitizer}, which is part of the CUDA toolkit. It is a suite for checking the functional correctness of the code and includes the following tools.
\begin{itemize}
    \item \textbf{memcheck} for memory access errors and leak detection, in the form of out-of-bounds and misaligned accesses
    \item \textbf{racecheck} to check for memory hazards, that is, race conditions. This tool is still experimental and might give false positives in some situations, especially if the kernel size is big, since tracking memory accesses to the same physical address is not an easy task
    \item \textbf{initcheck} for uninitialized global memory
    \item \textbf{synchcheck} to check for thread synchronization hazards
\end{itemize}

If the programmer wants to use the memcheck tool to debug an executable, the following command should be used. \\
\indent\texttt{compute-sanitizer --tool memcheck ./executable\_name} \\
If the first kernel is launched only after several seconds, the process attached to compute-sanitizer needs to be notified to avoid program exit. \\
\indent\texttt{compute-sanitizer --tool memcheck --launch-timeout 60 ./executable\_name}\\
To exploit the racecheck tool and target all active processes, use the following command. \\ 
\indent\texttt{compute-sanitizer --tool racecheck --target-processes all ./executable\_name}

\section{Implementation Challenges and Code Restructuring}\label{implementation-problems-code-restructuring}
This section is dedicated to the problems that have been encountered during the procedure of adapting the code for offloading on AMD and NVIDIA architectures. The errors refer to the corresponding compilers, \texttt{amdflang} and \texttt{nvfortran}, respectively. No reference documentation is available that clearly states which features are actually implemented by the compilers when it comes to OpenMP offloading. Some information may be found on specialized websites, but the overall lack of it makes OpenMP offloading with Fortran less intuitive than a programmer would think at first. For this reason, compiler limitations will be reported in this section, as well as workarounds to make the code compile and run successfully.\\

When it comes to the former compiler, most issues occurred during the linking phase. Several linker errors appeared stating an \texttt{undefined reference to function\_name}, referring to functions that make use of runtime host functions, even the most trivial and basic ones, such as integer allocation. The solution is to include the linker flag \texttt{-lflang\_rt.hostdevice}, which instructs the linker to link against both host and device Fortran runtime libraries. 
Also, the function designated to allocate device memory during the initialization phase did not compile correctly, giving the error \texttt{not yet implemented: derived type} for the memory allocation directives. These are not specifically necessary on AMD platforms when using USM, but in general they are. The solution is to use the \texttt{select type} construct on the passed-object dummy argument of the function. This construct is used to either safely downcast from a polymorphic type to a specific derived type, or branch based on the dynamic type of a polymorphic object. However, in the specific case of TRIMEG, the dummy argument is not polymorphic, so applying this construct to it should theoretically make no difference, but it was the only way to make the code compile successfully. \\

Now, let us discuss the limitations of the \texttt{nvfortran} compiler.
The OpenMP standard prescribes that a subroutine called from a target region should be declared as target-callable. One would expect a specific and structured compiler error when this prescription is not respected. In contrast, the compiler gives a generic error \texttt{undefined reference to 'subroutine\_name'}. This error could also be generated by the fact that a subroutine is declared as private in its module.\\
Another error has an output structure resembling this: \texttt{error: use of undefined value '@class\_name\_\_\_\_derived\_type\_name\_'}. 
This is caused by a stringent limitation of \texttt{nvfortran} due to the lack of dynamic typing support. \\
The declaration of the invoking object inside the subroutine, known as the passed-object dummy argument in Fortran jargon, cannot be \texttt{class(derived\_type\_name)} and needs to be changed to \texttt{type(derived\_type\_name)}. This equates to a non-polymorphic approach, since the type of the passed-object dummy argument, i.e., the object used to invoke the function, needs to be exactly the one specified in the function definition, and cannot be one that extends it. Moreover, the former declaration is fundamental to enable the use of type-bound procedures, a fundamental feature of extensible data types, in terms of inheritance and polymorphism. They allow overriding a base function in the extended derived type, customizing its behavior to the object specific needs.
Nevertheless, this feature is rarely used inside TRIMEG, since in most cases it is not necessary to make the derived type extensible. In order to solve the problem and circumvent this compiler limitation, it is sufficient to remove the extensibility feature from the subroutines to be accelerated. Nonetheless, this imposes a constraint on future developments of the code, so the decision to remove polymorphism should be carefully evaluated. Here is a comparison of the structure that yields the error:
\begin{verbatim}
module particle_mod
    implicit none
    
    type particle
        integer, allocatable :: partx(:),partv(:)
    contains
        procedure::compute
    end type particle
    
    contains
        subroutine compute(this)
            implicit none
            class(particle) :: this
            ...
end module particle_mod
\end{verbatim}
The structure needs to be modified as follows in order for \texttt{nvfortran} to successfully compile it:
\begin{verbatim}
module particle_mod
implicit none

type particle
integer, allocatable :: partx(:),partv(:)
end type particle

contains
subroutine compute(this)
implicit none
type(particle) :: this
...
end module particle_mod
\end{verbatim}
All calls to the \texttt{compute()} function also need to be modified, since it is no longer a type-bound procedure, e.g.
\begin{verbatim}
call this%dummy_type%compute()
\end{verbatim}
becomes
\begin{verbatim}
call compute(this%dummy_type)
\end{verbatim}

\begin{figure}
    \centering
    \includegraphics[width=0.3\linewidth]{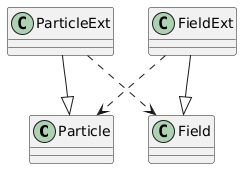}
    \caption{Circular dependency between particle and field classes}
    \label{circular-dependency}
\end{figure}

Still, there are cases where polymorphism is needed and cannot be removed. Specifically, the \texttt{particle class} contains all data related to the particle object, as well as several basic routines. The same is true for the \texttt{field class}. The issue is that the former needs to access the data and subroutines of the latter, and vice versa. This situation creates a circular dependency, which is not allowed and needs a workaround. In order to solve this, the data and basic functions are declared and implemented in the base classes, while the more complex functionalities are implemented in the derived classes, called \texttt{particle\_ext\_cls} and \texttt{field\_ext\_cls}, respectively. This allows each extended class to include the base class of the other type, thus granting access to its data and fundamental functions, without the issue of circular dependency. The dependency pattern is illustrated in Fig. \ref{circular-dependency}.

\noindent This does not actually leverage polymorphism, but uses the \texttt{extend} feature to artificially solve the previous problem. Nevertheless, future developments of TRIMEG involving the implementation of different particle species would actually require this feature. However, another issue arises from this trick. Some of the routines that are part of the base class are also called from the extended class. Specifically, they are called inside the kernel encapsulating the loop for particle pushing; consider the following example. The subroutine \texttt{particle\_cls\_gettemnorm} computes the normalized temperature, necessary to solve the weight equation, and needs to be called both from the base class and the extended particle class. The kernel needs to access this function through an extended particle object. 
Since type-bound procedures are not allowed inside a GPU kernel, the function call should be modified from \texttt{call this\%particle\_cls\_gettemnorm(...)}, where \texttt{this} is an extended particle object, to \texttt{call particle\_cls\_gettemnorm(this,...)}. However, the object on which the function is invoked is the extended particle object, while the subroutines can only be accessed when invoked on a base particle object, since it is not a type-bound procedure anymore. Moving the whole function to the extended class is not possible, since it needs to also be callable from base class subroutines. The solution would be duplicating the function in the extended class, since there needs to be a version that can be invoked on an object of extended particle type. However, duplicating code is known to be bad practice, since any modification to the function should be done twice, increasing the chance of bugs and the difficulty of maintaining the code. The final choice was to duplicate the definition in the extended class and encapsulate the function body inside an external file, then use the Fortran \texttt{INCLUDE} directive to read the statements from that file for both definitions. The Fortran \texttt{INCLUDE} should not be confused with the preprocessor include directive. It instructs the compiler to stop reading statements from the current file and instead read them from an included file---such as in this case---or from another module.
This stratagem enables the modification of the body of all function duplicates in a single time, while their definitions and signatures still need to be modified individually. \\

The lack of polymorphism support caused yet another problem. The code relies on an external library that implements several \texttt{bspline} classes to interpolate the solution computed on the poloidal cross section along the toroidal direction, and are based on different polynomial degrees depending on the necessary accuracy. All of these extend a base \texttt{bspline class} and contain a procedure pointer named \texttt{evaluate}, pointing to the function that evaluates a bspline interpolate, given the coordinates of the evaluation point and the derivative of the piecewise polynomial to be evaluated. A procedure pointer is intuitively a pointer to a procedure rather than a data object. It can point to different functions at runtime, allowing changing the function called without altering the code structure. It provides runtime polymorphism, while for type-bound procedures the call is resolved based on the type of the object at compile time. The structure is as follows.

\begin{verbatim}
type,extends(bspline_class),public :: bspline_1d
    ...
    contains 
    procedure,public :: evaluate => evaluate_1d
    ...
end type bspline_1d

type,extends(bspline_class),public :: bspline_2d
    ...
    contains
    procedure,public :: evaluate => evaluate_2d
    ...
end type bspline_2d

...
\end{verbatim}

\noindent \texttt{nvfortran} does not support procedure pointers inside OpenMP target regions. Suppose \texttt{F1d\_spline} is an object of type \texttt{bspline\_1d}; the function call \texttt{F1d\_spline\%evaluate(...)} is not allowed. It is required to call directly the function itself, whose name is different for each bspline class. In addition, the type declaration of the bspline object in the encapsulating derived type should be changed from \texttt{type(bspline\_[1-6]d)} to \texttt{class(bspline\_[1-6]d), allocatable}, and the object should be allocated in the subroutine dedicated to class initialization. \\
The compiler limitations are not only related to polymorphism. There exist syntax constructs that are not supported when producing the target version of a subroutine. One constraint is that the compiler struggles to understand that the size of a dynamic array needs to be consistent across function calls. This can cause an illegal memory access error during the execution of the CUDA kernel. If the size of an array is declared to be equal to a runtime variable inside a subroutine that is called from inside the kernel and is passed along a chain of function calls, the programmer must ensure that all declarations specify the actual size, while avoiding the more generic \texttt{dimension(:)}.
Another feature that is not supported, no matter how trivial it may seem, is the assignment of a default value in a variable declaration. For example, declaring a parameter as \texttt{integer::param=3} yields the generic compilation error \texttt{No device symbol for address reference} when the compiler tries to produce the target-compatible version of a function that contains such a syntax construct. \\

A major issue affected the structure that contains the mapping between the bounding boxes and the mesh triangles. It is a two-dimensional array of structures, where the two dimensions correspond to those of the intermediate grid. Each element is itself a dynamic vector containing all the mesh elements that superimpose with the corresponding box. Since no implementation of dynamically growing arrays is provided by vanilla Fortran, a simple library implementing them is included in TRIMEG. It mimics the behavior of C++ vectors of the standard library and defines a \texttt{VECTOR} object, which is composed of a one-dimensional array called \texttt{data} that stores the elements, and a scalar that stores its current size. As previously said, the mapping is constructed during the initial phase of the simulation. The vector structure is needed since the number of triangles superimposing with a single box is unknown \textit{a piori}. Once the mapping is generated, the vectors do not change, potentially allowing for a single memory transfer to the GPU. 
However, transferring such a structure to the GPU device is not trivial, since the array members of an object need to be explicitly mapped. In short, first it is necessary to allocate the whole two-dimensional array called \texttt{tria\_vec\_obj}, then map the \texttt{VECTOR} object corresponding to each element, so that the scalar variable is transferred, as well as allocating the \texttt{data} array and transferring its content. The result is the following.
\begin{verbatim}
!$omp target enter data map(alloc:tria_vec_obj)
do ftc=1,nx-1
    do fvc=1,ny-1
      !$omp target enter data map(to:tria_vec_obj(ftc,fvc))
      !$omp target enter data map(alloc:tria_vec_obj(ftc,fvc)%data)
      !$omp target update to(tria_vec_obj(ftc,fvc)%data)
    enddo
enddo
\end{verbatim}
where \texttt{nx} and \texttt{ny} are the dimensions of the intermediate grid. 
The ensemble of OpenMP directives to map the whole structure of arrays is cumbersome and causes runtime issues. In principle, it is sufficient to transfer the data structure only once, after the mapping has been computed during the initialization phase. However, the OpenMP runtime cannot preserve the device pointer to its memory region between different kernel calls, thus generating a CUDA illegal memory access error. The only workaround compatible with this array of structures consists of allocating the device memory before each kernel call, then transferring the data, executing the kernel, and finally deleting the data and deallocating the memory. While this solution works, it generates an overhead that decreases code performance. In the end, the adopted solution is to flatten the array of structures into a 1D array, an operation that can be done once, after each vector has been populated during the mapping construction. c
\begin{verbatim}
!$omp target enter data map(alloc:tria_vec_obj_flattened)
!$omp target update to(tria_vec_obj_flattened)
!$omp target enter data map(alloc:tria_vec_obj_index)
!$omp target update to(tria_vec_obj_index)
\end{verbatim}

\noindent These can be specified before the loop that computes the time evolution of the particles, avoiding the need to map them at every time step, as the compiler can correctly keep track of the device pointers.  

Once all of the subroutines and data structures were adapted to run on both architectures, significant numerical errors were present in the output arrays produced by the kernels when compared to the CPU results. The cause was a race condition inside the bspline library, which once again proved not to be suitable for GPU offloading. The \texttt{evaluate} internally makes use of temporary arrays for intermediate calculations. Though they are not declared locally inside the function, but rather as members of the bspline object used as a passed-object dummy argument; hence, it is shared among all threads that call the function in parallel on the GPU. It was sufficient to switch to a locally declared array, whose size is known at runtime, but depends on dummy arguments of the bspline object. Fortran allows automatic stack allocation under these conditions, so it was not necessary to declare it as an \texttt{allocatable} and explicitly allocate it on the heap.
If this were not the case, the programmer would have to allocate and transfer it inside the \texttt{evaluate} subroutine with \texttt{target enter data map(alloc:)} and \texttt{target update to()} directives. However, \texttt{nvfortran} does not support any kind of target directive inside a target-callable function, making it unfeasible to implement an easy solution to this issue.\\
The version compiled for NVIDIA platforms displayed another faulty behavior, that is, the simulation would stall during kernel execution, after a seemingly random number of calls. Tweaking the input parameters revealed a causality relation between the GPU grid size, that is, the number of GPU workers that process the kernel workload, and the number of kernel calls that occurred before the stall. Specifically, the larger the grid size, the greater the probability of stalling. This issue required a lot of debugging, which finally tracked down the root cause to yet another race condition in the bspline library. A shared variable was accessed inside a deeply nested subroutine by multiple threads at the same time, causing a general deadlock in the simulation. The code structure had to be changed to avoid this while preserving the semantics, and the problem was solved. It is still unclear why this did not affect the version compiled for the AMD MI300A APU, a factor that was misleading at first. The lesson to be learned from the multiple issues caused by the integration of the bspline library inside TRIMEG is the following. Integrating external code in an existing project implies a lack of full control of its runtime behavior. Moreover, the implementation design of external codes might not be suitable for GPU porting, especially if using a directive-based approach that does not require to write the kernels from scratch. The compromise between the increased simplicity of the approach and the decreased control over GPU kernels can often lead to unexpected behavior and time lost in debugging. \\

The resulting implementation features preemptive memory allocation on the device during the initialization phase of the code. Device memory allocation is a slow operation, compared to memory transfers, so it is crucial to do it only once for each data structure, as well as hiding its latency by performing other computations in the meantime.
The CPU implementation leverages shared memory across MPI processes for some data structures, so the memory footprint is reduced. Currently, there is no way to allocate shared memory on the GPU with this strategy, thus forcing the programmer to allocate private memory for each MPI rank. In principle, the data stored in each array requires being mapped only once, but the OpenMP runtime sometimes fails to keep track of it once the control is passed again to the CPU. For this reason, it is necessary to explicitly map some data structures before each kernel launch, namely the arrays stored inside the bspline objects. If this is not done, a CUDA error 700 for illegal memory access is thrown when the kernel is executed. \\

After completing the GPU porting of the particle push kernel and the g2p functions therein, it was necessary to assess the correctness of the output results. The kernel produces five output arrays that store the derivative of the space variables in cylindrical coordinates $(R,Z,\varphi)$, of the parallel velocity $v_{||}$, and of the weight coefficient $w$, used for the time integration of $(R,Z,\phi,v_\|,w)$ along $t$. To obtain an exact comparison between the CPU and GPU results, the same operations are executed twice in a row, and the CPU results are written in duplicates of the output arrays. After the execution of both the kernel and its CPU counterpart, the arrays are compared element-wise. Specifically, the absolute value of their difference is computed. The comparison revealed that the arrays often differ by a value greater than the epsilon machine, which for a double precision variable is $\approx 2.22\times10^{-16}$. This affects the code on both NVIDIA and AMD architectures. The numerical errors are in the order of $10^{-14}$ or $10^{-15}$, which is still very small, but the accumulation of these errors over time could lead to significant differences in the growth rate and structure of the physical modes. It was necessary to pinpoint the root cause of these numerical differences, although it was not an easy task due to the size of the kernel and the number of function calls inside it. After extensive debugging, several arithmetic operations were recognized as those that generate different results. They do not involve any special arithmetic operator such as square root or absolute value, which could have a slightly different GPU implementation. The first step was to remove all compiler optimization options, but it did not solve the issue. Then, more options were experimented, eventually leading to a solution for AMD platforms. 
When compiling with amdflang, the options \texttt{-O0 -fopenmp --offload-arch=gfx942:xnack+ -ffp-contract=off} were used, where \texttt{-ffp-contract=off} is again used to disable the generation of FMA instructions. Note that the default value of this option is \texttt{fast}, independently of the global optimization value being used. Explicitly setting it to \texttt{off} solves the issue on AMD, so that the kernel produces exactly the same results as the same set of operations executed on the CPU. After disabling this transformation, it was necessary to carry out an analysis to assess possible performance degradation, which revealed that this optimization does not influence the execution time of the kernels.

When targeting NVIDIA architectures, the nvfortran compiler options \texttt{-O0 -mp=gpu -Kieee -Minfo=all -Mnofma -Mr8 -gpu==cc80} were used, for the purpose of gathering more information on low-level optimizations and transformations, as well as disabling those that generate inexact arithmetic instructions. \texttt{-Minfo=all} generates verbose information on the low-level optimizations applied by the compiler, \texttt{-Kieee} enforces a strict rounding policy, in compliance with IEEE 754 standard, \texttt{-Mnofma} disables the generation of Fused-Multiply-Add instructions, and \texttt{-Mr8} promotes real variables and constants to double precision variables. This did not solve the issue on NVIDIA platforms.
Another possible cause is race conditions in deeply nested functions, which would be difficult to spot at first glance. The debugging tool \texttt{racecheck}, part of the compute-sanitizer suite, was used but did not reveal any such problems. Manual inspection yielded the same result; all the functions causing numerical differences have been identified and do not contain any race condition. The conclusion is that some arithmetic operations have a different hardware implementation on GPU architectures, and/or the OpenMP abstraction makes it extremely difficult to expose the low-level transformations that the compiler applies to the kernel code. The GPU version of TRIMEG is still capable of reproducing physically relevant results, but agreement with the CPU version can be hindered by several factors, e.g., low-$\beta$ and low-$n$ simulations, which are less numerically stable in general. The actual agreement of simulation results will be discussed in section \ref{chap:numerical_results}. 

\begin{figure}
    \centering
    \includegraphics[width=\linewidth]{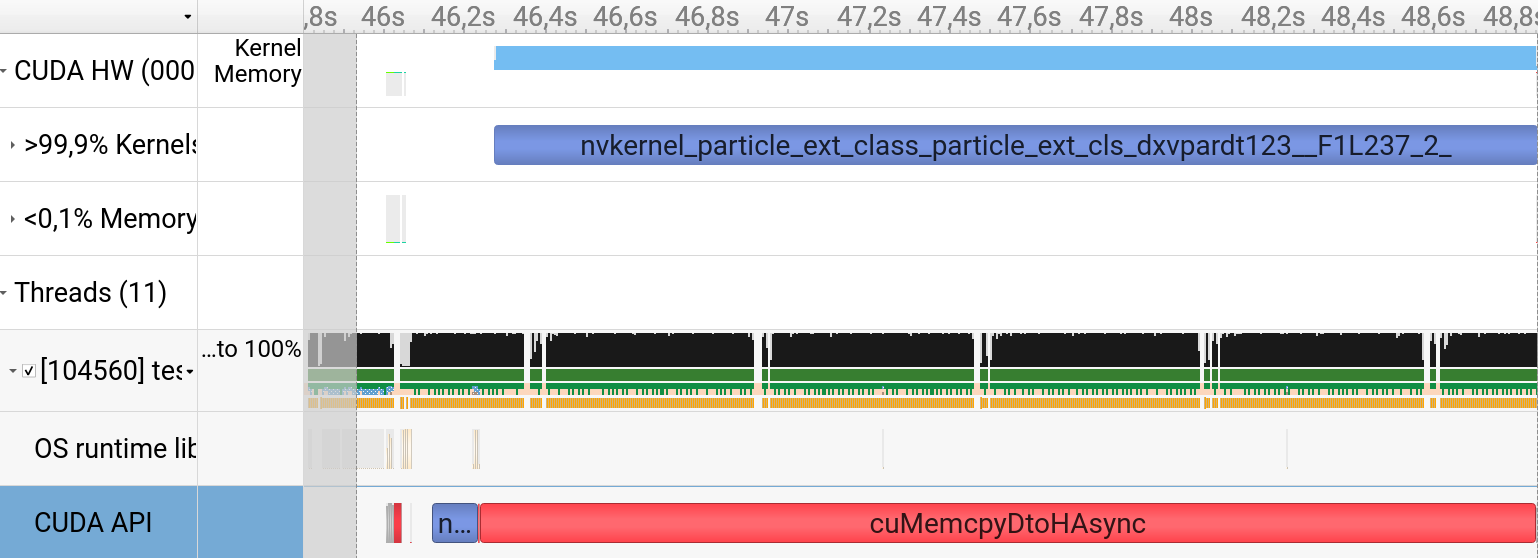}
    \caption{First profiling data of the particle pusher kernel.}
    \label{fig:nsys-faultly-profiling}
\end{figure}

\begin{figure}
    \centering
    \includegraphics[width=\linewidth]{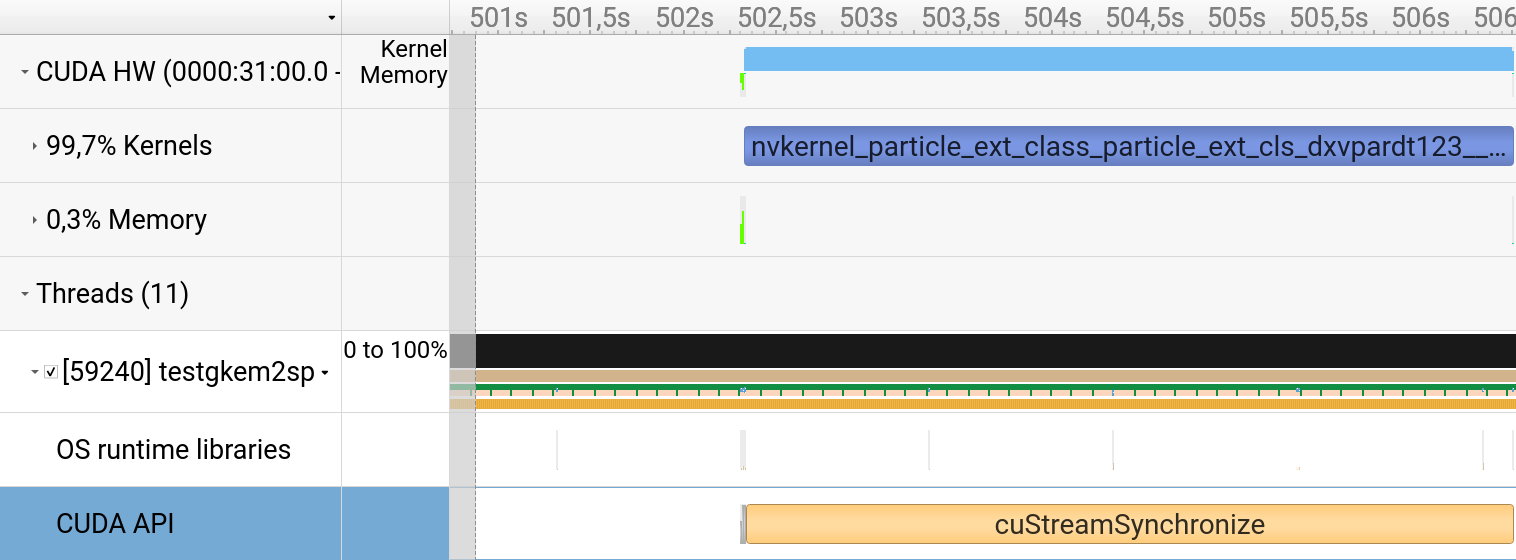}
    \caption{Second profiling data of the particle pusher kernel.}
    \label{fig:nsys-correct-profiling}
\end{figure}

The last major issue was the slowness of the NVIDIA GPU implementation. The particle pusher kernel was consistently slower than its original CPU version, making the GPU porting pointless. Pinpointing the issue required debugging, manual inlining of functions, and commenting chunks of the code to understand under which conditions the kernel was slowed down. The first possible cause was the cost of explicit memory transfers between the host and the device. There are multiple ways to understand if this is an actual issue. One is to comment out the whole kernel content but the memory transfers, and turn off compiler optimizations; otherwise if a data structure is not accessed inside the kernel, the compiler might avoid transferring it altogether. Another way is to profile the application with \texttt{nsys}, which gathers information about memory movements between host and device, kernel launches, CUDA events, and more. However, profiling information is not always to be trusted, as described below. Fig. \ref{fig:nsys-faultly-profiling} shows the results of the first profiling with nsys.  Apparently, the vast majority of the execution time on GPU is devoted to a huge memory transfer from the device back to the host, after the kernel has completed its task. However, the expectation is to transfer only five output arrays back to the CPU, with a relatively small size equal to the number of particle markers used in the simulation. Only $10^6$ particles were used for debugging purposes, which implies mapping five arrays of 1 million doubles each. Each one is $\approx 8$ MB using the SI convention, amounting for a total of $\approx 40$ MB to be transferred back to the host. The Raven cluster was used to carry out debugging, and it features a PCIe 4.0 link between the host and device on each node, providing a nominal transfer rate of about 32 GB/s in each direction. Supposing an actual bandwidth of $\approx 20$ GB/s, the movement of 40 MB back to the host would take $\approx 2$ ms. This is inconsistent with the profiling data generated by nsys, which shows a memory transfer back to the host that takes 2.618 seconds and dominates the total kernel execution time. This information was misleading and hinted at implicit memory transfers generated by the compiler. To rule this out, we first used the compiler option \texttt{-Minfo=mp}, which also shows whether the compiler inserts any implicit data mapping in the program. Indeed, no such transfer was generated.
Profiling again the application produced a significantly different outcome, as shown in Fig. \ref{fig:nsys-correct-profiling}. More than 99.7\% of the GPU execution time is taken up by the kernel itself, which is still extremely slow. Memory transfer events are $\approx 3$ orders of magnitude less expensive than kernel execution, which is why they cannot be seen on the dashboard without zooming in the CUDA API event timeline. The \texttt{cudaStreamSynchronize} corresponds to the kernel execution and accounts for the synchronization costs of GPU threads when they reach the implicit barrier at the end of the OpenMP target region. This clearly shows that the \texttt{nvfortran} compiler struggles in optimizing some features used internally by the kernel, which do not generate any slowdown when compiling with \texttt{amdflang} on AMD platforms with USM support. 

The next step was isolating the issue, which was not a trivial task since the kernel calls tens of subroutines, for a total of several thousands of lines. Finally, the problem was traced to the declaration of temporary arrays in the \texttt{evaluate\_[1-3]d} functions of the bspline library. These had been modified earlier to avoid race conditions, as previously described, and are allocated automatically inside the function. When one of these functions is called inside the GPU kernel, each thread allocates their own local instance of the temporary array, whose size is known only at runtime and depends on the order of the bspline functions provided in the input file. This generates a huge overhead, even though the reason why this happens only on NVIDIA platforms is unclear to date. The solution was switching to a compile-time size by taking the maximum possible size, which is 12 for the largest array. It is certainly not elegant, but is effective and does not incur performance degradation, as the size of the arrays is very small. To avoid reducing the flexibility of the code in terms of input parameters, this change is used only if compiling on NVIDIA architectures with GPU support, else the original declaration is used. The downside is code divergence, as portability proves once again to depend on the chosen compiler, rather than the GPU programming paradigm. 

\chapter{Performance Evaluation}
\label{chap:performance-evaluation}%
\graphicspath{Images/}

Evaluating the performance of the kernels is of paramount importance, given that the purpose of GPU porting is to accelerate the code. There exist several ways to compare the speed of the CPU and GPU versions, and most of them depend on a set of assumptions, as there is no universal strategy to carry them out. 
First, we analyzed each kernel individually, with an increasing number of MPI processes on a single node. A hybrid MPI-OpenMP parallelization approach is fundamental for running simulations with TRIMEG, which scales almost perfectly when the number of MPI tasks is increased. Note that the code does not implement OpenMP multithreading, so each core is usually mapped to an MPI task, to fully use the available hardware resources. We evaluated the performance degradation due to oversubscription of GPU resources as a function of the number of MPI processes, to understand if the hybrid parallelization can work efficiently for large-scale simulations. \\
Before testing the code in a realistic scenario on a GPU-equipped cluster, it is important to understand which is the best grid size configuration to maximize GPU hardware utilization while keeping register pressure under control. We chose a suitable problem size and performed GPU grid size exploration to tune the configuration parameters correctly. Finally, we outlined the overall performance of TRIMEG by conducting a strong scaling analysis, using the most efficient configuration parameters, and focusing the examination on the GPU-accelerated portion of the code. \\
For reference, the hardware capabilities of the clusters used for testing purposes are listed below. The GPU implementation targeting AMD architectures was executed on the Viper GPU cluster, where a single node is equipped with an AMD MI300A APU, with 24 cores, 128 GB of HBM3, 228 compute units, 14592 stream processors, and a theoretical performance of 61.3 TFLOPS for FP64 vector operations. \\
The GPU implementation targeting NVIDIA architectures is run on two different clusters, Raven and Pitagora, featuring different GPU architectures. A Raven GPU node is equipped with four NVIDIA A100 GPUs, each with 40 GB of HBM2. Unfortunately, this cluster is heavily oversubscribed, which is why no multi-node test has been performed on it. Both the CPU and GPU nodes are equipped with an Intel Xeon IceLake Platinum 8360Y with 72 cores and a theoretical peak performance of 5530 GFLOPS per node. The second NVIDIA-based cluster is Pitagora, which offers four H100 SXM 80GB HBM2e on each GPU node. This computational power is immense, but the supercomputer is still in the pre-production phase, so the actual performance might be lower than its theoretical potential. Also, each node of the booster partition is equipped with 2 Intel Emerald Rapids 6548Y 2.4 GHz with 32 cores each, for a total of 64.\\
The CPU-based cluster is TOK, where a single node consists of an AMD EPYC 9554, with 128 cores, and a maximum boost clock of 3.75 GHz. \\
In the various analyses, we compared individual kernels and overall code performance against the CPU counterpart, considering its execution on the same hardware, that is, the booster partitions, as well as on CPU-dedicated nodes. A fair comparison is not trivial to achieve, since on a GPU node, the GPU-accelerated version can harness both the available MPI tasks on the CPU, as well as the accelerator resources. The latter approach implies a comparison between two fundamentally different sets of hardware, where single-core performance can vary significantly. In addition, we chose to provide a comparison against several CPU versions compiled with different toolchains, highlighting the importance of choosing the right compiler and library version for a given architecture to achieve the best performance. For each, we used all optimization options that do not involve aggressive arithmetic approximations, which would hinder the high precision needed in the calculations.\\

\section{Individual Kernel Analysis} \label{individual-kernel-analysis}
First of all, we evaluated the performance of each kernel individually, without considering the rest of the code and the MPI layer of parallelism. This is useful to understand whether the workload is suitable for GPU acceleration and the speed gain when offloading is done in ideal conditions, i.e., each MPI process targets a single GPU and can fully utilize its resources. However, in realistic scenarios, TRIMEG needs to allocate as many MPI processes as cores on each node, as previously said. Constraints on cluster resources do not allow one to allocate a GPU node for each MPI process, as the total node count would be too high, even though this would be necessary to achieve the best performance with the GPU implementation of TRIMEG. For this reason, an assessment of kernel performance degradation was carried out when using an increasing number of MPI processes on a single GPU node. Each MPI process sets up a separate context on the GPU, and the GPU driver serializes their execution streams. Kernels can still be interleaved, but each context switch can generate a non-negligible overhead and reduce the overall performance. Moreover, memory transfers are also serialized on the PCI Express bus that connects the host to the device. Each kernel is smaller, as the total workload size is distributed across the MPI ranks, thus leading to underutilization of the GPU resources. All of these factors need to be considered when opting for GPU acceleration of a code that is only parallelized with MPI, where CPU cores cannot be mapped to OpenMP threads. \\
Table \ref{pusher-kernel-hybrid-parallelism-viper} shows the particle pusher kernel speedup on Viper GPU with USM with respect to two CPU versions executed on the same hardware. The workload size is $10^6$ particle markers, which is small for a full node, but is computationally expensive to process on a single core. This specific analysis aims to evaluate the advantage of using accelerators when the available CPU parallelism is the same. The execution times reported include the time that each MPI rank waits for the kernel to be launched on the GPU resources shared with all the other ranks, kernel interleaving, and the eventual performance degradation. We would have opted for a larger workload size for this analysis, which would result in greater parallelism exposure, but constraints on cluster resources and available time did not allow such a choice for this extensive analysis. A reduced performance study using a larger workload is presented subsequently. The kernel execution times are obtained as the average over a number of runs equal to $\rm{\#MPI\_procs}\cdot 20$, both for GPU and CPU runs. The first row corresponds to the kernel execution under ideal conditions, where a single MPI process can use a full GPU. Even though the size of the kernel is halved every time the number of MPI processes is doubled on the same GPU node, as the workload is distributed across MPI ranks, reduction in the execution time is not to be expected, as kernels are serialized on the accelerator. Nonetheless, oversubscription of the GPU does not bear significant performance degradation, even when the number of MPI tasks is increased to the maximum of 24 available on the MI300A APU. In contrast, there is a slight gain when increasing MPI parallelism, possibly due to the ability of interleaving kernels when stalls occur. The CPU versions exhibit perfect scaling when increasing MPI processes, progressively reducing the advantage of using hybrid MPI/OpenMP offloading acceleration. The CPU version built with the same toolchain as the GPU version, that is, amdflang, is faster than the one built with Intel, as expected. As the former compiler is developed by AMD, it offers better optimizations for their own hardware. \\
We conducted the same analysis by comparing the GPU model on the AMD platform against the CPU cluster TOK, and the results are shown in Table \ref{pusher-kernel-hybrid-parallelism-tok}. It describes a more realistic situation, where the GPU implementation is run on a GPU-dedicated supercomputer, while the CPU version is executed on a CPU-based cluster, taking into account the notable hardware differences on the CPU side. The last row corresponds to using the maximum number of cores as MPI processes on both a Viper GPU node and a TOK node, which are 24 and 128, respectively. The single core performance of the AMD EPYC 9554 is lower than that of the MI300A, but the higher core count of the former almost nullifies the advantage of GPU acceleration, as the CPU-based cluster offers significantly more MPI parallelism. However, the kernel size is very small, and realistic simulations require tens or hundreds of million particles. Such figures would allow better resource utilization on the GPU, making offloading more advantageous. An important note is the performance difference between the GCC-compiled and Intel-compiled versions executed on TOK, where the former outperforms the latter on the AMD Genoa architecture, up to a factor of $\approx 3.4$ on a full node compared to the Intel version. This is beyond reasonable, even if the support that Intel compilers offer for AMD hardware is limited and lacks optimizations, but we were unable to pinpoint the issue, which arises only when using this particular hardware and toolchain combination. The GCC toolchain cannot be used on the Viper GPU cluster, as there is no version of the PETSC library built with the \texttt{gfortran} compiler.\\
The same comparison was carried out for the kernels that contribute to the pullback, particle density calculation, and distribution function interpolation, and the results are shown in Tables \ref{calcN-pullback-kernel-hybrid-parallelism-viper} and \ref{calcN-pullback-kernel-hybrid-parallelism-tok}. They have been aggregated as the structure of the kernels is the same. Even though the workload is smaller, the magnitude of the speedup values is comparable to that achieved by the pusher kernel. It is reasonable to deduce that a large portion of the total kernel time is devoted to kernel launch overhead, driver scheduling, synchronization costs, and page faults triggered during execution, as the problem size is very small. Again, execution on a CPU-based cluster drastically reduces the speedup of the GPU-accelerated implementation for the given problem size, as can be observed in Table \ref{calcN-pullback-kernel-hybrid-parallelism-tok}. 

Then, we conducted a reduced analysis for the NVIDIA version executed on both Raven and Pitagora clusters.
Table \ref{pusher-kernel-hybrid-parallelism-raven} shows the results of the particle pusher kernel on a Raven GPU node against a TOK CPU node. The first row corresponds to the performance achieved with one A100 on a GPU node, and one MPI process for both CPU and GPU versions. The last row corresponds to testing the implementations on Raven with all 72 MPI processes, whereas 128 are used on TOK. All four A100 are allocated when using the full GPU node. Oversubscribing the GPU causes a noticeable performance decrease, and the kernel execution time on this hardware set is slower than that on the MI300A by a factor of $\approx 2.8$ of a full GPU node.
Analogously, the performance of the pusher kernel on a Pitagora GPU node versus a TOK CPU node is reported in Table \ref{pusher-kernel-hybrid-parallelism-pitagora}. The same rationale is used for the performance comparison. When allocating a full GPU node on Pitagora, 64 MPI processes and four H100 GPUs are used. The computational power of this chip is significantly higher than its predecessor, the A100, and the kernel achieves a speedup of $2.8$ on a Pitagora GPU node when compared with a Raven one, and is approximately as fast as the MI300A on Viper GPU.
As before, the speedup of the three smaller kernels is aggregated and the performance of the GPU model on both Raven and Pitagora is compared against the CPU version on TOK, and the results are shown in Table \ref{other-kernels-hybrid-parallelism-pitagora}. Oversubscription of GPU resources proves to be beneficial in this case, even though the advantage of GPU acceleration is marginal when compared to a full CPU node on Raven and TOK for a problem size of $10^6$ electrons.

However, the previous analysis was quite expensive, as it included workload processing by a single MPI task, and does not provide sufficient parallelism exposure for efficient GPU utilization. Therefore, we also tested a larger simulation of $32\cdot10^{6}$ electrons on a full node, a magnitude needed to achieve realistic particle-in-cell simulations with TRIMEG.
We analyzed the performance on both Viper GPU and Pitagora against TOK, and the results are reported in Table \ref{pusher-kernel-hybrid-parallelism-big-case}. Overall, the speedup achieved by the GPU implementation is significantly higher for a realistic workload size. The performance difference between one MI300A and four H100 is significant, even though the theoretical performance per-gpu is lower. However, multi-GPU support by OpenMP offloading depends on the chosen compiler, and we should not expect perfect workload distribution and parallelization across the available GPUs. Indeed, when reducing the GPUs per node to just one on Pitagora, the average kernel execution time is approximately the same, hinting at the fact that the \texttt{nvfortran} compiler support for multi-GPUs with OpenMP is in the early stage of development, or absent altogether. To confirm this, we inserted a call to the OpenMP routine \texttt{omp\_get\_num\_devices()} inside a target region, which should return the identifier of the GPU device on which the kernel is executing. However, such a call generates a \texttt{Fatal error: NOT IMPLEMENTED}, confirming that multi-GPU execution is not yet supported by nvfortran 24.9, which is the version installed on the Pitagora cluster.
The same analysis is carried out for the three smaller kernels. As shown in As shown in Table \ref{other-kernels-hybrid-parallelism-big-case}, a Pitagora GPU node outperforms a Viper-GPU node by a factor of $\approx 1.6$. As the workload of these kernels is smaller than the one for particle pushing, the speedup is also lower. Again, there is no performance difference when allocating one H100 instead of four. The absence of multi-GPU support by the used compilers is another disadvantage of the OpenMP offloading paradigm, as it drastically reduces the ability to fully use the available hardware.

\begin{table}[H]
\centering
\caption{Pusher kernel speedup on AMD-based Viper GPU with USM with respect to two CPU versions executed on the same hardware, for a problem size of $10^6$ electrons.}
\label{pusher-kernel-hybrid-parallelism-viper}
\begin{tabular}{|cccccc|}
\hline
\rowcolor{bluePoli!40}
\#MPI Procs &  Viper GPU (s) & CPU Intel (s) & Speedup & CPU amdflang (s) & Speedup \T\B \\
\hline \hline
1 & 0.163 & 42.5 & 261 & 33.6 & 207 \\
2 & 0.120 & 20.8 & 173 & 16.7 & 139 \\
4 & 0.110 & 10.3 & 93.6 & 8.39 & 76.3 \\
8 & 0.117 & 5.18 & 44.3 & 4.18 & 35.7 \\
16 & 0.105 & 2.71 & 25.8 & 2.09 & 19.9 \\
24 & 0.102 & 1.99 & 19.5 & 1.40 & 13.7 \T\B\\
\hline 
\end{tabular}
\end{table}

\begin{table}[H]
\centering
\caption{Pusher kernel speedup on AMD-based Viper GPU with USM with respect to two CPU versions executed on TOK, for a problem size of $10^6$ electrons.}
\label{pusher-kernel-hybrid-parallelism-tok}
\begin{tabular}{|cccccc|}
\hline 
\rowcolor{bluePoli!40}
\#MPI Procs & Viper GPU (s) & TOK Intel (s) & Speedup & TOK GCC (s) & Speedup \T\B \\
\hline \hline
1  & 0.163 & 55.2 & 338 & 38.6 & 237 \\
2  & 0.120 & 29.4 & 245 & 19.4 & 162 \\
4  & 0.110 & 17.9 & 163 & 9.76 & 88.7 \\
8  & 0.117 & 9.16 & 78.3  & 4.81 & 41.1 \\
16 & 0.105 & 4.86 & 46.3  & 2.35 & 22.4 \\
max* & 0.102 & 1.15 & 11.3  & 0.334 & 3.3 \T\B \\
\hline
\end{tabular}
\\[3pt]
*24 tasks on a GPU node, 128 on a CPU node
\end{table}

\begin{table}[H]
\centering
\caption{Speedup of the kernels for pullback, density calculation and distribution function interpolation on AMD-based Viper GPU with USM with respect to two CPU versions executed on the same hardware, for a problem size of $10^6$ electrons. Only the minimum and maximum number of MPI tasks on a single GPU node have been reported.}
\label{calcN-pullback-kernel-hybrid-parallelism-viper}
\begin{tabular}{|cccccc|}
\hline
\rowcolor{bluePoli!40}
\#MPI Procs & Viper GPU (s) & CPU Intel (s) & Speedup & CPU amdflang (s) & Speedup \\
\hline \hline
1  & 0.0334 & 9.83 & 294 & 6.14 & 184 \\
24 & 0.0362 & 0.582 & 16.1  & 0.254 & 7 \\
\hline
\end{tabular}
\end{table}

\begin{table}[H]
\centering
\caption{Speedup of the kernels for pullback, density calculation and distribution function interpolation on AMD-based Viper GPU with USM with respect to two CPU versions executed on TOK, for a problem size of $10^6$ electrons.}
\label{calcN-pullback-kernel-hybrid-parallelism-tok}
\begin{tabular}{|cccccc|}
\hline
\rowcolor{bluePoli!40}
\#MPI Procs & Viper GPU (s) & TOK Intel (s) & Speedup & TOK GCC (s) & Speedup \\
\hline \hline
1  & 0.0334 & 19.58 & 586 & 8.96 & 268 \\
max* & 0.0362 & 0.445 & 12.3  & 0.0737 & 2 \\
\hline
\end{tabular}
\\[3pt]
*24 tasks on a GPU node, 128 on a CPU node
\end{table}

\begin{table}[H]
\centering
\caption{Pusher kernel speedup on a Raven NVIDIA GPU node versus Intel-compiled CPU version executed on a Raven CPU node versus GCC-compiled CPU version on TOK. The problem size is $10^6$ electrons.}
\label{pusher-kernel-hybrid-parallelism-raven}
\begin{tabular}{|cccccc|}
\hline
\rowcolor{bluePoli!40}
\#MPI Procs & Raven GPU (s) & Raven CPU (s) & Speedup & TOK GCC (s) & Speedup \\
\hline \hline
1  & 0.182 & 44.3 & 243 & 38.6 & 212 \\
max* & 0.280 & 0.635 & 2.7 & 0.334 & 1.2 \\
\hline
\end{tabular}
\\[3pt]
*72 tasks on a Raven GPU node and a Raven CPU node, 128 on a TOK CPU node
\end{table}

\begin{table}[H]
\centering
\caption{Pusher kernel speedup on a Pitagora NVIDIA GPU node versus Intel-compiled CPU version executed on a Raven CPU node versus GCC-compiled CPU version on TOK. The problem size is $10^6$ electrons.}
\label{pusher-kernel-hybrid-parallelism-pitagora}
\begin{tabular}{|cccccc|}
\rowcolor{bluePoli!40}
\hline
\#MPI Procs & Pitagora GPU (s) & Raven CPU (s) & Speedup & TOK GCC (s) & Speedup \\
\hline \hline
1  & 0.111 & 44.3 & 399 & 38.6 & 348 \\
max* & 0.100 & 0.635 & 6.4 & 0.334 & 3.3 \\
\hline
\end{tabular}
\\[3pt]
*64 tasks on a GPU node, 72 on a Raven CPU node, 128 on a TOK CPU node
\end{table}

\begin{table}[H]
\centering
\caption{Speedup of the kernels for pullback, density calculation, and distribution function interpolation on a Pitagora NVIDIA GPU node versus Intel-compiled CPU version executed on a Raven CPU node versus GCC-compiled CPU version on TOK. The problem size is $10^6$ electrons.}
\label{other-kernels-hybrid-parallelism-pitagora}
\begin{tabular}{|cccccc|}
\hline
\rowcolor{bluePoli!40}
\#MPI Procs & Pitagora GPU (s) & Raven CPU (s) & Speedup & TOK GCC (s) & Speedup \\
\hline \hline
1  & 0.0897 & 10.23 & 114 & 8.96 & 100 \\
max* & 0.0393 & 0.081 & 2 & 0.0737 & 1.9 \\
\hline 
\end{tabular}
\\[3pt]
*64 tasks on a GPU node, 72 on a Raven CPU node, 128 on a TOK CPU node  
\end{table}

\begin{table}[H]
\centering
\caption{Pusher kernel speedup on AMD-based Viper GPU and NVIDIA-based Pitagora against GCC-compiled CPU version on TOK, which proved to have the fastest CPU node. The problem size has been increased to $32\cdot10^{6}$ electrons.}
\label{pusher-kernel-hybrid-parallelism-big-case}
\begin{tabular}{|ccccc|}
\hline
\rowcolor{bluePoli!40}
TOK CPU (s) & Viper GPU (s) & Speedup & Pitagora GPU (s) & Speedup \\
\hline \hline
6.22 & 0.42 & 14.8 & 0.21 & 29.6 \\
\hline
\end{tabular}
\end{table}

\begin{table}[H]
\centering
\caption{Speedup of the kernels for pullback, density calculation, and distribution function interpolation on AMD-based Viper GPU and NVIDIA-based Pitagora against GCC-compiled CPU version on TOK. The problem size has been increased to $32\cdot 10^{6}$ electrons.}
\label{other-kernels-hybrid-parallelism-big-case}
\begin{tabular}{|ccccc|}
\hline
\rowcolor{bluePoli!40}
TOK CPU (s) & Viper GPU (s) & Speedup & Pitagora GPU (s) & Speedup \\
\hline \hline
0.78 & 0.337 & 2.3 & 0.21 & 4.7 \\
\hline
\end{tabular}
\end{table}

\clearpage
\section{Parameter Exploration}
\begin{figure}
    \centering
    \includegraphics[width=0.8\linewidth]{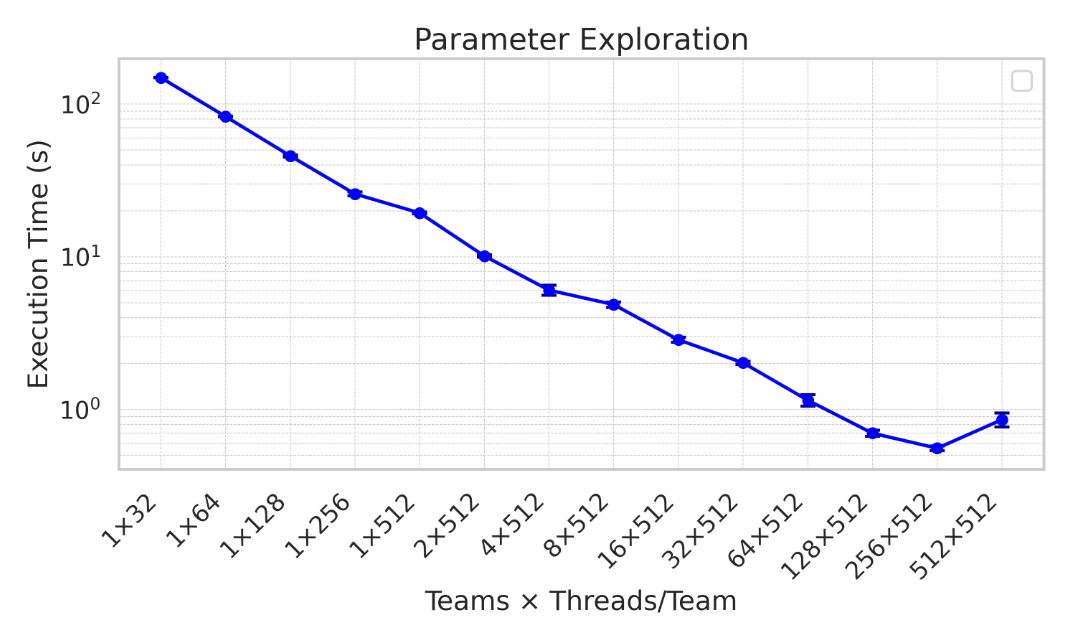}
    \caption{Variation of the kernel execution time as a function of the GPU configuration parameters on a Viper GPU node. The workload consists of $32\cdot 10^6$ electrons.}
    \label{fig:grid-size-exploration}
\end{figure}

Let us consider the most expensive kernel, which encloses the loop for particle pushing and several g2p subroutines. In realistic simulations, it takes up at least 50\% of the total execution time. It is fundamental to explore how the kernel performance varies for different block and grid configurations, given a fixed input size. The grid configuration consists of the number of GPU workers that are used to process the workload within the kernel. This procedure is useful to understand how well the workload maps to the GPU hardware and how to tweak the configuration parameters to achieve maximum efficiency in resource utilization. Note that the OpenMP abstraction does not provide fine-grained control over block size and grid size. Theoretically, each block is three-dimensional, and the same applies to the grid. Low-level GPU programming languages like CUDA let the programmer specify each dimension, allowing for the most efficient mapping of the workload to the available resources. The abstraction introduced by OpenMP does not provide such a level of control, and these parameters can be controlled as previously described in Section \ref{GPU_offloading_subsubsection}. \\
We considered a simulation with $32\cdot10^6$ electrons and performed parameter exploration for the particle push kernel on Viper GPU. A full node is allocated to reflect the resources that would be used in a large-scale simulation. Fig. \ref{fig:grid-size-exploration} shows the result of the grid and block size exploration for the pusher kernel. It does not include all the configurations that we experimented, as it would have been too chaotic. Each data point is obtained as the average of $\rm{\#MPI\_procs}\cdot10 = 240$ kernel executions. In addition, an error bar is plotted at each point to represent the standard deviation of the execution time. The scale of the y-axis is logarithmic for better readability. The best configuration for the given workload consists of 256 teams and 512 threads per team.

\newpage
\section{Strong Scalability}
\begin{figure}
    \centering
    \includegraphics[width=0.6\linewidth]{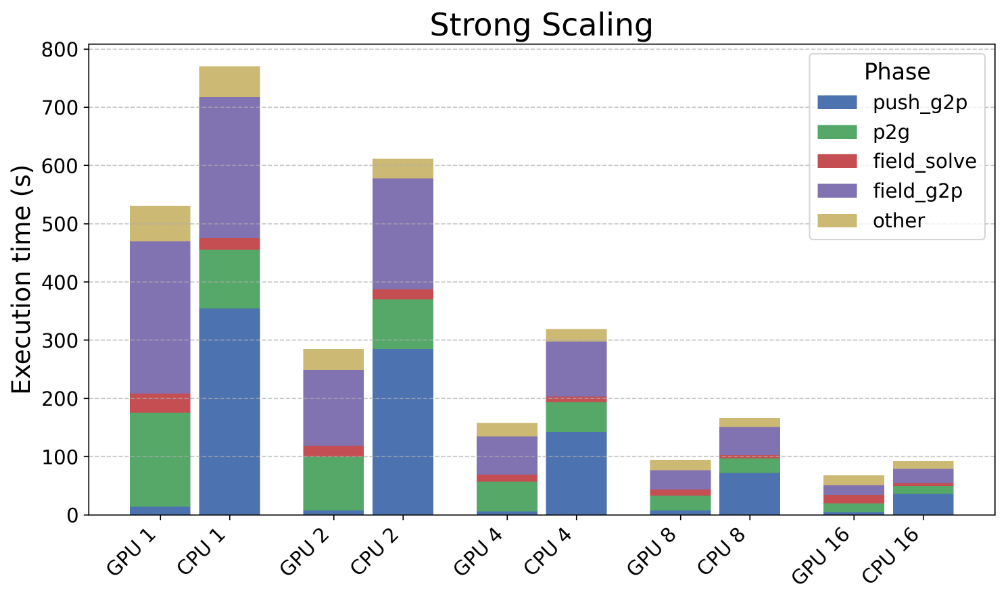}
    \caption{Strong scaling of the GPU implementation and amdflang-compiled CPU version on AMD-based Viper GPU.}
    \label{fig:strong-scaling-viper}
\end{figure}

\begin{figure}
    \centering
    \includegraphics[width=0.6\linewidth]{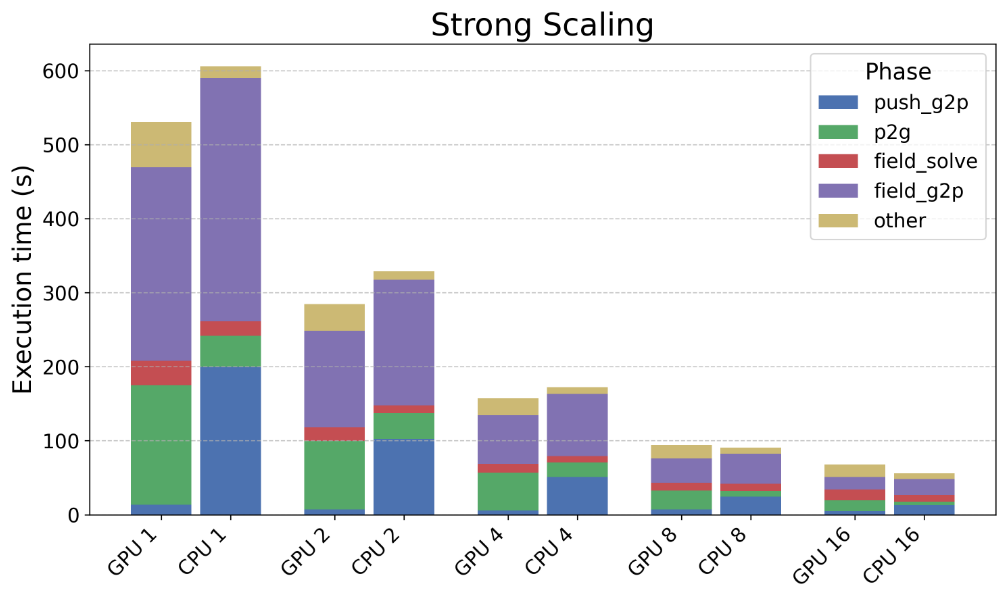}
    \caption{Strong scaling of the GPU implementation on AMD-based Viper GPU compared with the GCC-compiled CPU version.}
    \label{fig:strong-scaling-tok}
\end{figure}

After fine-tuning the kernel parameters for the number of particle markers to be used in a large-scale simulation, it is interesting to see how the workload scales across different nodes. Strong scalability aims to evaluate how much a fixed workload can be accelerated when more processing power is added. This is primarily a benchmark for the MPI parallelization, and might seem out of scope for evaluating the GPU acceleration. However, this analysis should confirm that GPU offloading can efficiently work in harmony with MPI on multiple nodes and might give insights into new bottlenecks of the code. We evaluated the scalability of a simulation with $32\cdot 10^6$ electrons and $10^6$ ions for an increasing number of nodes, up to 16. The simulation was run for only 10 time steps, which is why the particle pushing procedure is not overly dominant, as the initialization phase takes up a large portion of the total time. A realistic simulation needs a number of time steps in the order of thousands or tens of thousands to produce physically relevant results, so that the execution time proportions of the different phases would be in favor of that which is GPU-accelerated. We limited the number of time steps to avoid wasting computational resources and saving time. Note that the analyzed version did not include the smaller g2p kernels, as it was preliminary to understand which code sections could benefit from GPU offloading, besides the pusher. Again, we carried out the comparison by considering two different hardware sets, that is, Viper GPU and TOK. Both CPU versions are compiled with the best performing toolchains for the respective architectures, i.e., amdflang and GCC, respectively. A strong scaling analysis of the GPU implementation versus the CPU version, both on Viper GPU nodes, is shown in Fig. \ref{fig:strong-scaling-viper}. The different execution phases are shown in the form of a stacked bar graph. The execution time of each phase has been gathered thanks to the timer system already implemented in TRIMEG. The \texttt{push\_g2p} portion, colored blue, corresponds to the GPU-accelerated kernel. Note that the \texttt{field\_g2p} portion, colored purple, also partially accounts for subroutines called inside the kernel, as the time measurements cannot be easily decoupled. For this reason, the \texttt{field\_g2p} portion also slightly benefits from GPU acceleration, and is the main bottleneck of the code. However, g2p functions are scattered across the code, making their GPU acceleration challenging and resource-demanding. Some of them were offloaded following this analysis, amounting to three new kernels, whose performance is described in Section \ref{individual-kernel-analysis}. The \texttt{p2g} portion, colored green, is the next biggest performance concern, but its structure is not suitable for GPU offloading.\\
The same is done by comparing the GPU implementation on Viper GPU with the CPU version executed on TOK, and the result is shown in Fig. \ref{fig:strong-scaling-tok}. The performance gain of the accelerated code is marginal, primarily due to the fact that the \texttt{p2g} phase strongly benefits from the 128 MPI tasks per node that the AMD 9554 chip offers. For this reason, the advantage fades out for an increasing number of nodes, where the overall performance of the GPU implementation is comparable to that of the CPU version, even though there are notable differences in time consumption by the different phases. Anyway, the result confirms the performance gain of the particle pushing procedure thanks to GPU acceleration, and the kernel performance does not significantly degrade for an increasing number of nodes.

\clearpage
\section{Profiling Data} \label{profiling-data}
\begin{figure}
    \centering
    \includegraphics[width=\linewidth]{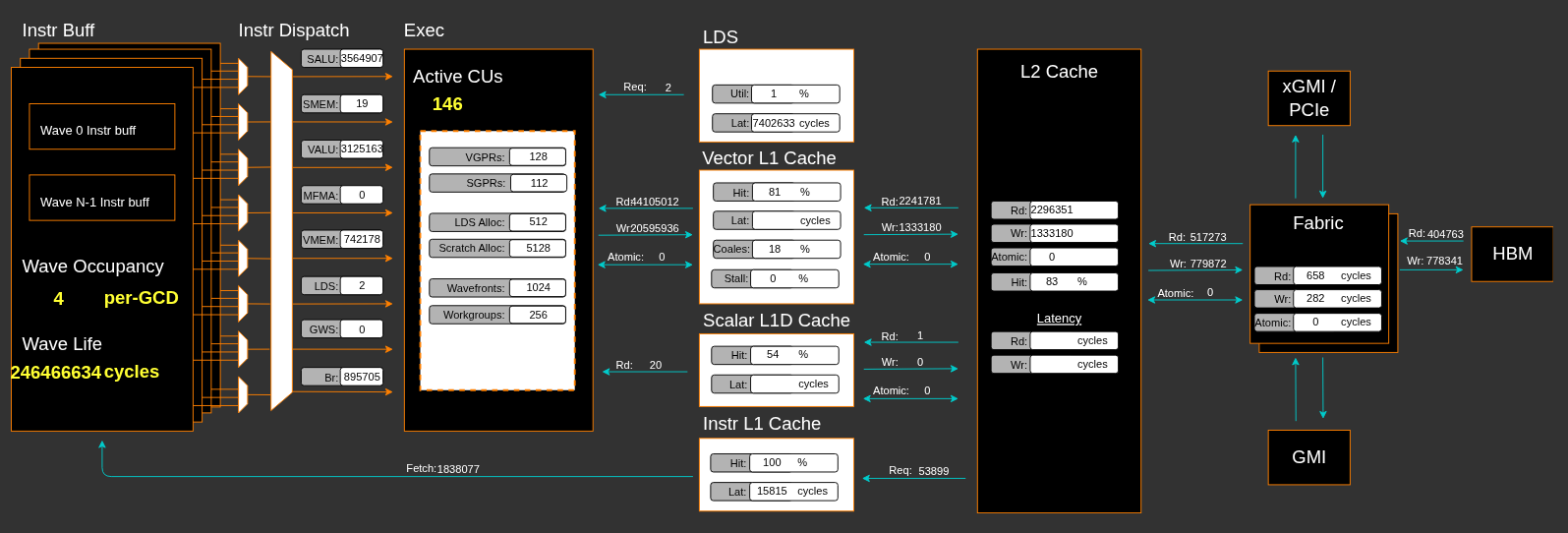}
    \caption{Overview of the GPU architecture, hardware utilization, including CUs, wavefronts and workgroups, dispatched instruction types, and amount of data transfers between the different memory levels for the pusher GPU implementation.}
    \label{fig:overall-dxvpardt-kernel-profiling}
\end{figure}

\begin{figure}
    \centering
    \includegraphics[width=\linewidth]{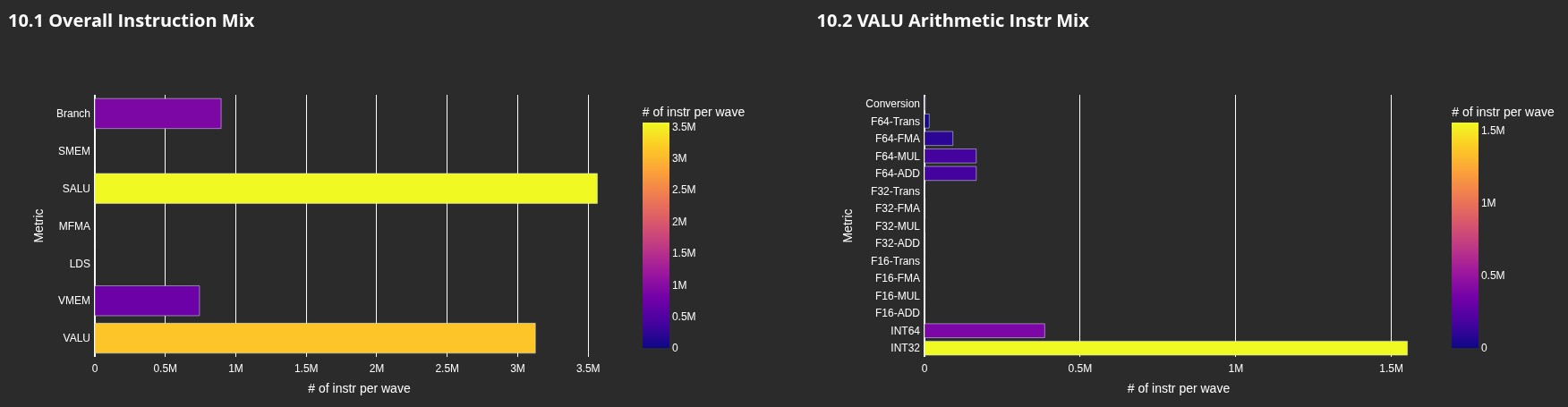}
    \caption{Overall and VALU instruction mix per wavefront for the pusher GPU implementation.}
    \label{fig:overll-instruction-mix}
\end{figure}

Detailed information on kernel performance can be generated with profiling tools specifically designed by hardware vendors. Several metrics can be collected, e.g., bytes read and written to the global memory, L1 and L2 hit rates, number of vector and scalar instructions executed, and occupancy of the hardware resources. \texttt{rocprof-compute} is the tool developed to profile kernel behavior on AMD GPU hardware. We profiled a simulation of $4\cdot10^6$ electrons and $10^6$ ions evolved for a single time step, which is sufficient to run all kernels at least once. It is important to keep the computational cost of the simulation low, since harvesting all the performance and utilization metrics requires running the program several times. For the sake of brevity and to avoid redundancy, we report only profiling data of the particle push kernel, the most computationally expensive one. Fig. \ref{fig:overall-dxvpardt-kernel-profiling} gives an overview of GPU resource utilization and data transfers of the particle pusher kernel. The memory hierarchy is clearly represented, as well as the flow of data. The number of active CUs is only 146 out of the 228 that are available on the MI300A, probably due to register pressure, as each thread needs a large context to store all the variables that are actively accessed in the particle push kernel. Note that the Local Data Share (LDS), analogous to shared memory on NVIDIA GPUs, is not used. However, not much can be done in this respect, as the OpenMP API does not allow to explicitly make use of it. This represents a potentially important limitation, since LDS is a low-latency, high-performance storage type.
Fig. \ref{fig:overll-instruction-mix} shows the overall and Vector Arithmetic Logic Unit (VALU) instruction mix for the particle push kernel. Note that all threads in a wavefront execute the same instruction at each clock cycle in a SIMD manner, a fundamental feature to achieve good performance and power efficiency. The dominant instruction types are Scalar ALU (SALU) and VALU instructions, which amount to $\approx80\%$, whereas branches and Virtual Memory Instructions (VMEM) are less common. This is important to achieve high arithmetic intensity, a fundamental metric to evaluate the computational requirements of an algorithm. VALU instructions are primarily done on integer variables, while FP64 instructions, corresponding to the Fortran type \texttt{real*8}, are much lower.\\
Memory operations incur much higher latency than arithmetic instructions, hence optimizing them is of paramount importance. The L1 vector data cache hit rate averages 80.51\%, and the L2 hit rate averages 83.12\%. This implies that only $(100-80.51)\cdot(100-83.12)\approx3.3\%$ of the read and write operations go to global memory, a great indicator. Only 18\% of the accesses is coalesced, but the memory access patterns in the kernel cannot be easily changed to increase this metric. The scalar data cache is almost unused, as shown in Fig. \ref{fig:overall-dxvpardt-kernel-profiling}. Overall, the memory access patterns prove to be quite efficient thanks to high locality and data reuse.

\chapter{Numerical Results}
\label{chap:numerical_results}

\section{Study of the Cyclone Case}
\begin{figure}
    \centering
    \includegraphics[width=0.7\linewidth]{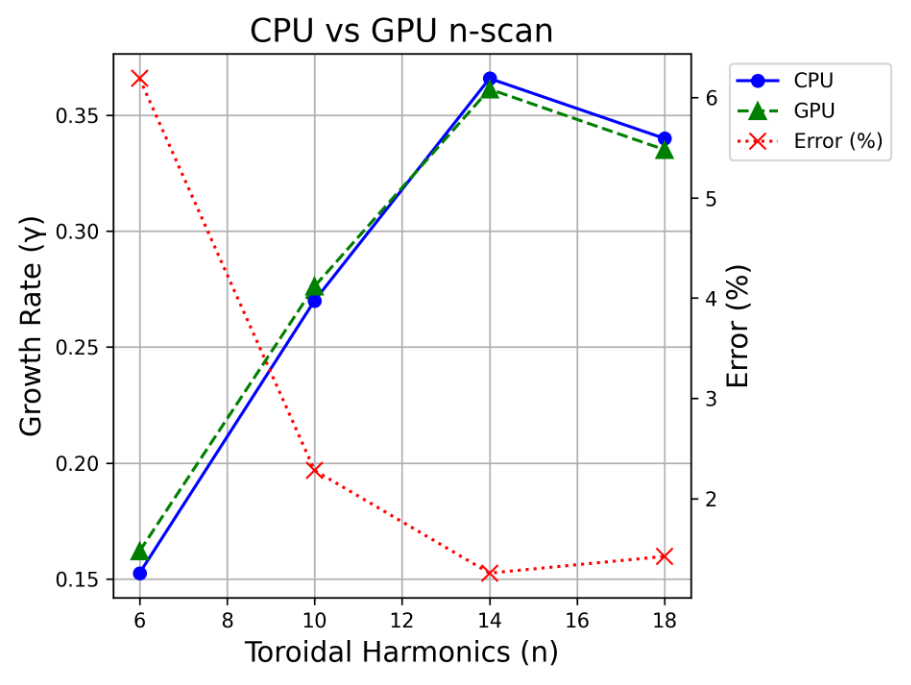}
    \caption{Energy growth rate comparison between CPU and GPU versions.}
    \label{fig:growth-rate-comparison}
\end{figure}

\begin{figure}
    \centering
    \includegraphics[width=0.5\linewidth]{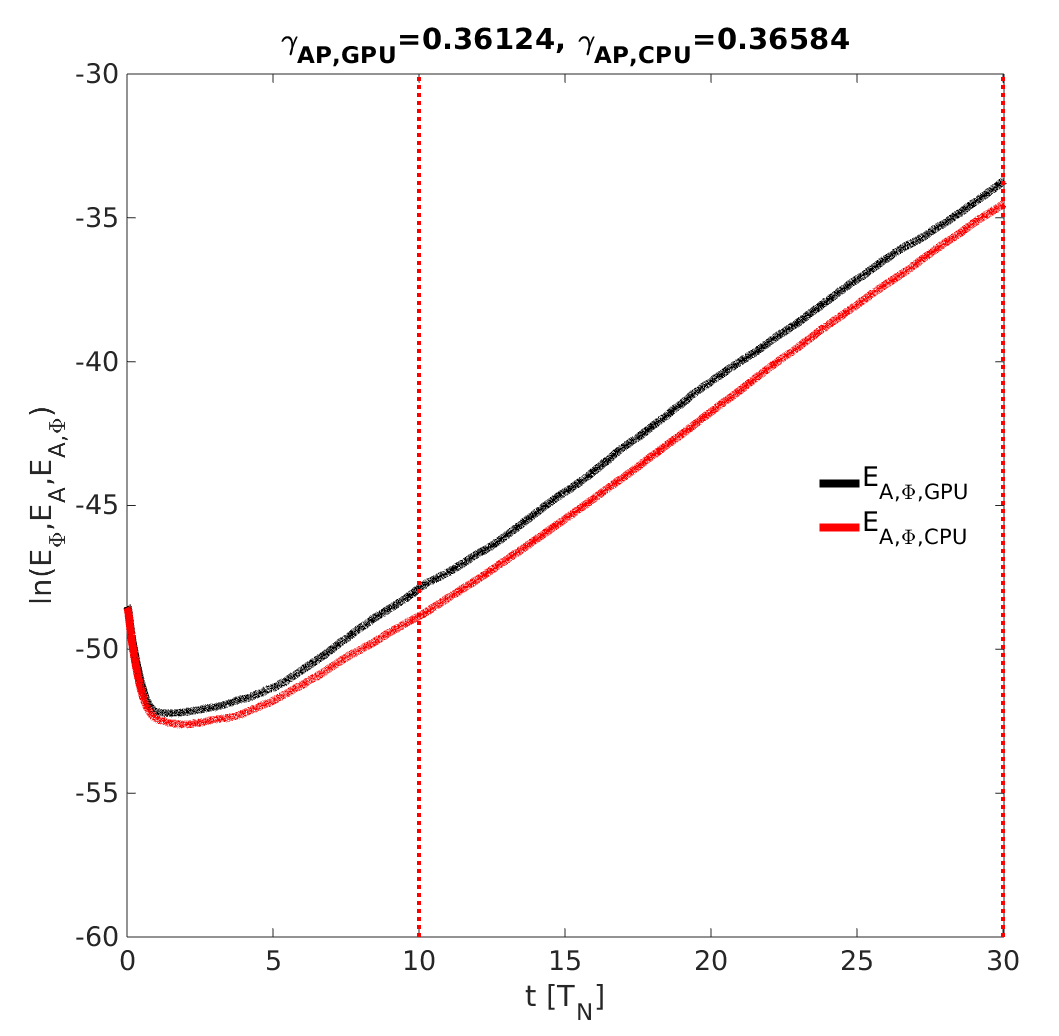}
    \caption{Comparison of the CPU and GPU time evolution of the total energy stored in the electromagnetic field for $n=14$.}
    \label{fig:total-energy-evolution}
\end{figure}

\begin{figure}[h!]
    \centering
    \begin{minipage}[t]{0.48\linewidth}
        \centering
        \includegraphics[width=\linewidth]{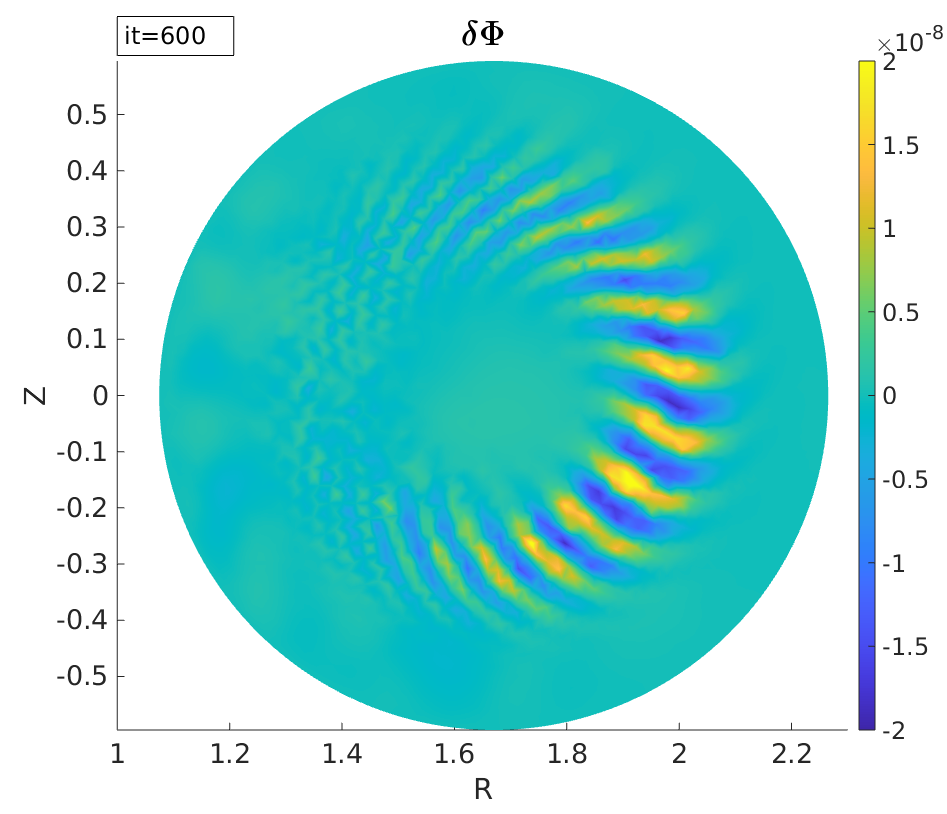}
        \caption{GPU version}
        \label{fig:gpu}
    \end{minipage}
    \hfill
    \begin{minipage}[t]{0.48\linewidth}
        \centering
        \includegraphics[width=\linewidth]{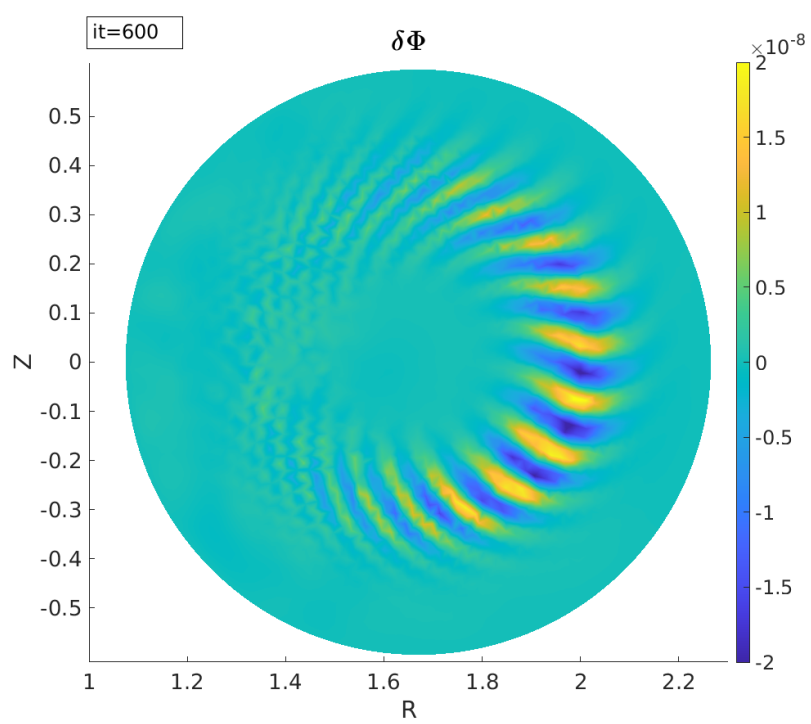}
        \caption{CPU version}
        \label{fig:cpu}
    \end{minipage}
    \caption{Comparison of the two-dimensional mode structure of the electric field on the poloidal cross section for $n=14$.}
    \label{2d-mode-structure-comparison}
\end{figure}

To compare the simulation results with and without GPU offloading, the instabilities in tokamak plasmas are simulated. 
In the previous work of TRIMEG-C1 \cite{lu2024gyrokinetic}, the Cyclone case that models the plasma core is used to simulate the ion temperature gradient (ITG) mode in the low $\beta$ limit \cite{gorler2016intercode}. 
The simulation parameters have been tweaked to reduce the computational cost by choosing $\rho^*=1/60$, $m_{\rm i}/m_{\rm e}=100$, where $m_{\rm i}$ is the mass of the ion and $m_{\rm e}$ is that of the electron. The gyro average of electrons and ions is switched off. A low $\beta_N=0.004$ is chosen for the plasma to simulate the ITG mode. The simulation uses $8\cdot10^6$ electron markers and $2.5\cdot 10^5$ ion markers to model the plasma behavior.  The radial grid number is $N_r=16$ for $n=2$ and $N_r=32$ otherwise. We carried out a scan over the number of toroidal harmonics, specifically for the values $n=6,10,14,18$. The $n$-scan for the Cyclone case is used to show that the overall implementation of the GPU-accelerated code is correct, including the structural changes that were carried out during the course of this work. The GPU version was run on the Viper GPU, while the CPU version was run on TOK. Fig. \ref{fig:growth-rate-comparison} shows the agreement between the growth rate $\gamma$ produced by the CPU and GPU versions.  The blue line corresponds to the values of $\gamma$ produced by the CPU version, while the green line identifies those generated by the GPU model. The red dotted line corresponds to the percentage difference between each pair of data points. Note that consecutive CPU runs do not produce exactly the same results, so there is a physiological deviation in the $\gamma$ values. This is due to the different initialization of the internal random number generators by the MPI ranks. For this reason, the red line should not be intended as the GPU model failing to reproduce the correct results, but rather as a metric of intrinsic randomness in the simulations. Specifically, the error is inversely proportional to $\gamma$, as a higher value of $\gamma$ implies that the simulation is more numerically stable. Anyway, it has been assessed that the GPU kernels produce the exact same results as if the computations were executed on the CPU, for the same set of inputs, assuring there is no race condition within the kernels.\\
The time evolution of the total field energy is demonstrated in Fig. \ref{fig:total-energy-evolution}. The data corresponds to the case with 14 toroidal harmonics. Early stages of the evolution show significant differences, but the trend eventually becomes similar once the exponential phase is reached. Note that the scale of the y-axis is logarithmic, hence why the trend looks linear. The growth rate of the two versions in the selected time interval, identified by the vertical dotted lines, is reported at the top of the figure. \\
A comparison of the 2D mode structure is shown in Fig. \ref{2d-mode-structure-comparison}. The tokamak geometry used for this case is simplified, and corresponds to an ideal torus. Again, the data corresponds to the case with 14 toroidal harmonics. The overall structure is preserved, even though the GPU version produces a result which is less smooth. However, the culprit of the coarser solution is not the GPU implementation itself, but rather the amdflang compiler. Compiling the original CPU code version with it and running it on the MI300A without GPU acceleration produces results that appear less smooth in general.

\clearpage
\section{Study of the TCV-X21 Case}
\begin{figure}
    \centering
    \includegraphics[width=0.7\linewidth]{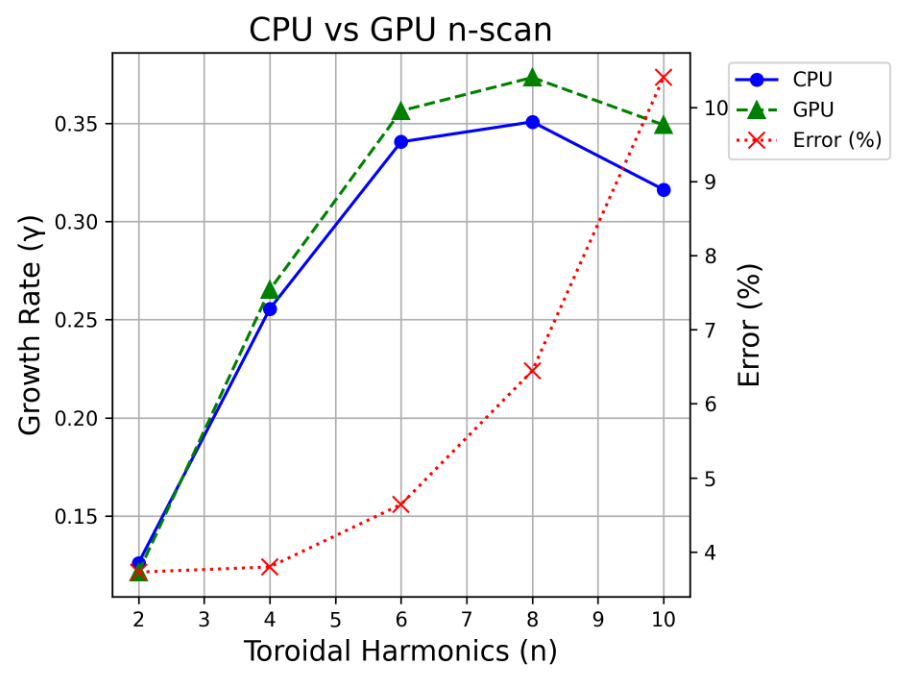}
    \caption{Growth rate comparison between CPU and GPU versions.}
    \label{fig:tcv-growth-rate}
\end{figure}

\begin{figure}
    \centering
    \includegraphics[width=0.55\linewidth]{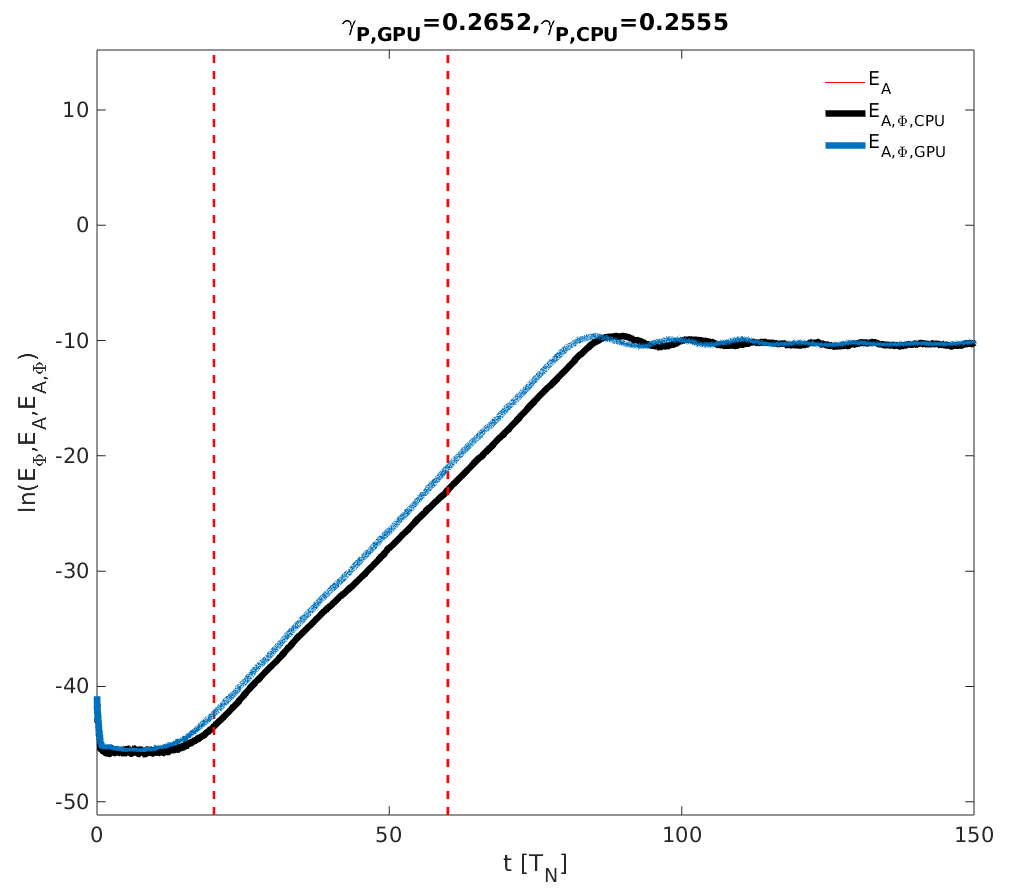}
    \caption{Comparison of the CPU and GPU time evolution of the total energy stored in the electric field for $n=4$.}
    \label{fig:tcv-energy-evolution}
\end{figure}

\begin{figure}[h!]
    \centering
    \begin{minipage}[t]{0.38\linewidth}
        \centering
        \includegraphics[width=\linewidth]{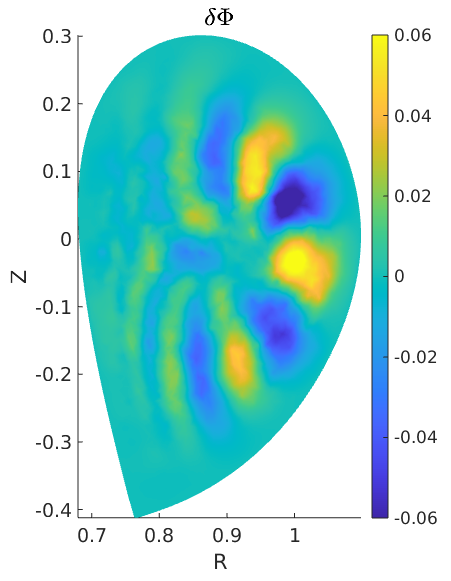}
        \caption{GPU version}
        \label{fig:gpu-tcv}
    \end{minipage}
    \hfill
    \begin{minipage}[t]{0.38\linewidth}
        \centering
        \includegraphics[width=\linewidth]{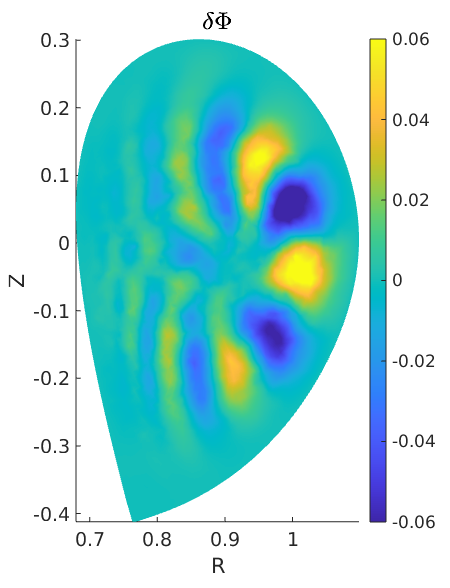}
        \caption{CPU version}
        \label{fig:cpu-tcv}
    \end{minipage}
    \caption{Comparison of the two-dimensional mode structure of the electric field on the poloidal cross section for $n=4$.}
    \label{tcv-mode-structure}
\end{figure}

In addition to the Cyclone case, the TCV-X21 case is also studied in this work. The TCV-X21 case has been studied experimentally and numerically for Tokamak à configuration variable (TCV) \cite{oliveira2022validation,body2022development,ulbl2023influence,lu2025trimeg}. This case is used by various codes such as GRILLIX and GENE-X for the studies of the transport and the profile generation with the consideration of the separatrix. In this section, the plasma simulation is performed to demonstrate the features of nonlinear ITG instability. The electrostatic model is adopted for single-$n$ simulations. The reference Larmor radius is $\rho_{\rm ref}=0.01$ cm instead of the nominal value $\rho_{\rm ref}=0.3539$ cm, to save computational resources. The reference magnetic field is $B_{\rm{ref}}=0.90727$ T. \\
Again, the aim of the $n$-scan is to reproduce the results achieved by the original CPU version with the GPU-accelerated implementation. The former was run on TOK, while the latter was run on Viper GPU.
The number of toroidal harmonics used for the $n$-scan is $2,4,6,8,10$. To keep the simulations cheap, the number of electrons is limited to $10^6$, and the number of ions to $2.5\cdot10^5$. This compromise was necessary to complete the $n$-scan in time, but the accuracy is reduced, and the agreement of the results could be improved by increasing these parameters.
The agreement of the growth rate produced by the GPU and CPU versions is shown in Fig. \ref{fig:tcv-growth-rate}. As before, the blue line corresponds to the values of $\gamma$ produced by the CPU version, while the green line identifies the values generated by the GPU model. The red dotted line plots the percentage difference between each pair of data points. The agreement deteriorates for higher values of $n$, and could be improved by increasing the number of particle markers. Nonetheless, the overall trend in the growth rate is correctly reproduced by the GPU implementation.\\
The time evolution of the electric field from the exponentially growing stage to the nonlinear saturation stage is shown in Fig. \ref{fig:tcv-energy-evolution}. The data corresponds to the case with 4 toroidal harmonics. The linear growth rate shows good agreement, and the saturation level of the nonlinear stage is the same for both simulations, indicating eventual convergence to the same value even if the previous stages are not identical. The two vertical dotted lines identify the time interval used to compute the linear growth rate, reported at the top of the plot.\\
Finally, a comparison of the 2D mode structure in the poloidal cross section is shown in Fig. \ref{tcv-mode-structure}. The tokamak geometry is more realistic than the one used for the Cyclone case, and includes the separatrix. Again, the data corresponds to the case with 4 toroidal harmonics. The GPU version reproduces the overall mode structure correctly, even though there are minor differences.

\ifpdf
    \graphicspath{{Chapter4/Figures/PNG/}{Chapter3/Figures/PDF/}{Chapter4/Figures/}}
\else
    \graphicspath{{Chapter4/Figures/EPS/}{Chapter3/Figures/}}
\fi

\chapter{Summary and Outlook}
\label{chap:summary}%
\graphicspath{Images/}

In this work, we successfully ported the most computationally expensive portions of the TRIMEG code to GPU architectures with OpenMP. The particle pusher and subroutines that perform grid-to-particle operations have proven to be suitable for GPU offloading. However, the concept of portability comes with several limitations. One should not expect to seamlessly port an OpenMP-offloaded code to different architectures, even if the vendor is the same. This also applies to clusters having the same GPU hardware, where correct functioning of the code can be hindered by a different compiler version, which is the actual variable on which portability depends. 
The TRIMEG GPU implementation works correctly on both NVIDIA and AMD architectures, and the two versions have been merged into a single solution. Both exhibited several constraints on the supported features, so we explained the workarounds to successfully compile and link Fortran programs that make use of object-oriented features and polymorphism. We reported incorrect runtime behaviors, the modus operandi we used to systematically debug them, the root causes, and the proposed solutions. The result is a surrogate of the missing compiler-specific documentation, namely for nvfortran and amdflang, whose absence slows down any attempt of porting complex Fortran codes to GPUs using the OpenMP API. All of the above information should be taken into consideration when choosing between different GPU porting alternatives, to avoid transforming this process into a time-consuming and excruciating compiler exploration, with no guarantee of eventually obtaining a working, and most importantly, high-performance solution. 
This is the reason why we were unable to deliver a functioning version on AMD platforms without USM support, because the amdflang compiler is still in the early stages of development, and showed once again the limitations of portability.\\
Performance analysis is a crucial step in developing a GPU implementation, as it highlights bottlenecks and provides insightful information to fine-tune the kernel configuration parameters and resource allocation for a given workload size. 
In light of the above, we carried out GPU parameter exploration to find the fastest configuration for a given problem size, and analyzed kernel performance and memory usage metrics with available profiling tools.
Various performance studies have been conducted, in terms of single-node performance and scalability tests. We compared the GPU implementation against several CPU versions, compiled with different toolchains and executed on different hardware sets.
We isolated the performance of the kernels from the rest of the code, and showed that the workload can be accelerated significantly. Conversely, multi-node scalability tests do not specifically benchmark the OpenMP kernels, but still provide a comprehensive view of the strengths and weaknesses of running large-scale simulations on GPU-capable clusters. Specifically, booster partitions tend to have a smaller core count per node, hence why programs that scale well with MPI parallelization might not benefit from GPU acceleration under the constraint of having less parallelism across CPU cores. For this reason, we assessed performance degradation due to oversubscription of GPU resources by multiple MPI processes, which is necessary for TRIMEG, since the code does not implement OpenMP multithreading.
The GPU-accelerated sections of the code also scale well on multiple GPU nodes, proving that a hybrid MPI-OpenMP offloading parallelization approach can still work efficiently and deliver performance improvements under certain conditions. \\
Finally, a crucial step was to assess the correctness of the structural changes that were necessary for the GPU implementation to work. To achieve this, we performed well-established benchmarks in the gyrokinetic field, namely the Cyclone case for simulating the ITG mode in simplified tokamak geometry, and the TCV-X21 case for simulating nonlinear electrostatic ITG instabilities in realistic tokamak geometry, up to the separatrix.
In both cases, we proved that our implementation can reproduce the results with high fidelity. The differences in the energy growth rate and two-dimensional mode structure can be attributed to the different compiler toolchains and library versions used for the CPU and GPU versions, which generate code that behaves in a slightly different way, and to different initialization of the random number generators. 

\newpage
\bibliography{References/references}

\appendix
\chapter{Appendix}
\section{Mixed Variable and Pullback Scheme}\label{mixed-variable-and-pullback-scheme}
In gyrokinetics, there exist two approaches to represent the particle distribution function, as mentioned in Chapter \ref{subsec:structure_code}. 

\begin{definition}
    The distribution function $f$ of a plasma is the probability of finding a particle at a given position, with a given velocity at a given time step. It is a statistical representation of the plasma particles.
\end{definition} 

The full $f$ method describes the whole distribution function, while the $\delta$f approach decomposes it into an equilibrium component and a perturbed component, that is, $f(z,t) = f_0(z,t) + \delta f(z,t)$ \cite{lu2023full_f_delta_f}. In a collisionless plasma, the full $f$ distribution is a constant along the particle trajectory according to Liouville's theorem \cite{chen2022evolution}. As a result, using the full $f$ scheme, no additional equation is needed for $f$, while using the $\delta f$ scheme, the equation of $\delta f$ must be solved over time. 
Due to the noise level, the first approach is more computationally expensive, especially when dealing with small-magnitude perturbations and small-scale phenomena. The $\delta$f approach focuses primarily on turbulent transport and is more efficient, while retaining physical accuracy if the perturbations do not affect a significant portion of the plasma particles. To apply the $\delta$f approach, the gyro center's equations of motion are decomposed into an equilibrium and perturbed part as shown in Eqs. \ref{eq:position_equation_of_motion} and \ref{eq:velocity_equation_of_motion_deltaf}.
Given that the equilibrium contribution is time independent, $f_0$ corresponds to the steady-state solution and its time derivative is equal to zero. \\
As mentioned previously, the pure $p_\|$ gyrokinetic model suffers from the cancellation problem, which affects the calculation of the parallel component of Ampère's law. 
There exist two approaches to solving the gyrokinetic equation that have been described in the literature. The first is the $p_{||}$-formulation, or Hamiltonian model, where the parallel momentum, defined as $p_{||} = v_{||} + \frac{q_s}{m_s}\delta A_{||}$ is used as a phase space variable. Here, both the electrostatic potential $\phi$ and the magnetic field perturbation $\langle A_{||}\rangle$ contribute to the Hamiltonian, and the perturbation is calculated in the gyrokinetic equation. This numerical issue caused by the cancellation problem becomes less severe as one gets closer to the electrostatic limit, that is, for $A_{||}\rightarrow0$. The advantage of the $p_{||}$-formulation is that Poisson's equation retains its standard unperturbed form. \\
The second approach is the $v_{||}$-formulation, or simplectic model, which uses the parallel velocity as a phase space variable. The cancellation problem does not appear in the parallel Amp\`ere's law, but perturbation terms appear in Poisson's equation and the gyro center's equations of motion. Specifically, it leads to the presence of terms proportional to $\partial\langle A_{||}\rangle/\partial t$ in the equation of motion, which is a source of numerical instabilities, especially when explicit schemes are used to solve the ODEs of particle motion. 

The mixed variable for the pullback scheme is a more general form that recovers these two formulations.
The parallel component of the scalar potential $\delta A_{||}$ is decomposed into the symplectic part and the Hamiltonian part. \cite{lu2024gyrokinetic}
\begin{equation}\label{mixed variable}
\delta A_{||} = \delta A_{||}^s+\delta A_{||}^h\;\;,
\end{equation}
where the simplectic part is usually chosen so as to satisfy the ideal Ohm's law involving the electrostatic scalar potential $\delta\phi$ as follows. \cite{lu2024gyrokinetic}
$$\partial_t\delta A_{||} + \partial_{||}\delta\phi = 0\;\;,$$
where the parallel derivative is defined as $\partial_{||}=b_0 \cdot \nabla$ \cite{mishchenko2017mitigation}. The parallel velocity coordinate of the gyro center of motion is defined as follows.
$$u_{||}=v_{||}+\frac{q_s}{m_s}\langle \delta A_{||}^h\rangle\;\;,$$
where $q_s$ and $m_s$ are the charge and mass of a particle of species $s$, and the operator $\langle\dots\rangle$ represents the gyro average. \\

The general form of the nonlinear pullback scheme is the following. \cite{hatzky2019reduction,lu2023full_f_delta_f}
$$f_{s,v}(v_{||})=f_{s,u}\left(v_{||}+\frac{q_s}{m_s}\langle A_{||}^h\rangle\right)\;\;.$$

Ampère's law in $v_{||}$ space is given by \cite{lu2023full_f_delta_f}
$$-\nabla_\perp^2\delta A_{||}=\mu_0\delta j_{||,v}\;\;,$$
where 
$$\delta j_{||,v}(\textbf{x}) = \sum\limits_{s} q_s\int d^6 z\delta f_{v,s}\delta(\textbf{R}+\rho-\textbf{x})v_{||} \;\;.$$ 

The cancellation problem still appears, as in the case of the original $p_{||}$-formulation, but the terms to be canceled are now proportional to $\delta A_{||}^{\rm h}$. The cancellation problem is weaker in regimes with $|\delta A_{||}^{\rm h}| \ll |\delta A_{||}|$, meaning that the contribution on the right-hand side should be a good guess of the actual magnetic vector potential. 
In the pullback scheme, the equations for $\delta f$ are the following. \cite{mishchenko2014pullback,hatzky2019reduction,lu2023full_f_delta_f}

$$
\delta A_{||,new}^s = \delta A_{||,old}^s + \delta A_{||,old}^h\;\;,
$$

$$
u_{||,new} = u_{||,old} - \frac{q_s}{m_s}\langle \delta A_{||,old}^h\rangle\;\;,
$$

$$
\delta f_{new} = \delta f_{old} + \frac{q_s \langle \delta A_{||,old}^h\rangle}{m_s}\frac{\partial f_{0s}}{\partial v_{||}}\;\xrightarrow[f_{0s}=f_M]{\text{Maxwellian}} \;\delta f_{old}-\frac{2v_{||}}{v_{ts}^2}\frac{q_s\langle\delta A_{||,old}^h\rangle}{m_s}f_{0s}\;\;,
$$
where the subscripts \textit{old} and \textit{new} refer to the variables before and after the pullback transformation. The last equation is the linearized version for the calculation of the $\delta f$ pullback.

The full $f$ scheme requires only the first two equations, as the perturbed contribution is neglected. The normalized equation for the pullback treatment is formulated as follows. \cite{lu2023full_f_delta_f}

$$
\delta\bar A^{\rm{s}}_{\|,\rm{new}} = \delta\bar A^{\rm{s}}_{\|,\rm{old}} + \delta\bar A^{\rm{h}}_{\|,\rm{old}}\;\;,
$$

$$
u_{\|,\rm{new}} = \bar u_{\|,\rm{old}} - \frac{\bar{q}}{\bar{m}_s} \left\langle\delta\bar A^{\rm{h}}_{\|,\rm{old}} \right\rangle\;\;,
$$

$$
\delta f_{\rm{new}} = \delta f_{\rm{old}} + \frac{\bar{q}_s\left\langle\delta\bar A^{\rm{h}}_{\|,\rm{old}} \right\rangle}{\bar{m}_s}\frac{\partial f_{0s}}{\partial \bar{v}_\|}\;\;,
$$

$$
\xrightarrow[f_{0s}=f_{\rm{M}}]{\text{Maxwellian}} \delta f_{\rm{old}} - \frac{2 \bar{q}_s}{\bar{T}_s} \bar{v}_\| \left\langle\delta\bar A^{\rm{h}}_{\|,\rm{old}} \right\rangle f_{0s}\;\;,
$$
where the factor $2$ in the equation for the perturbed variable is consequential to the normalization of $T$ to $T_{\rm{N}} = m_{\rm{N}}v_{\rm{N}}^2/2$.

\section{Magnetic Flux and Field-aligned Coordinates}\label{field-aligned-coordinates}
Let us introduce the following concepts to better characterize the magnetic flux coordinate system.
\begin{definition}
    The \textbf{magnetic flux} through a surface, denoted as $\Phi$, is the integral over the surface of the normal component of the magnetic field.
\end{definition}
The $(R,\varphi,Z)$ coordinate system is used to solve the field equations and, additionally, to describe the gyrocenters' equations of motion \cite{lu2023full_f_delta_f}.

\begin{definition}
    In the context of magnetic confinement fusion, the surface identified by the magnetic field lines is called \textbf{flux surface}.
\end{definition}

\begin{figure}
    \centering
    \begin{subfigure}{0.4\linewidth}
        \centering
        \includegraphics[width=\linewidth]{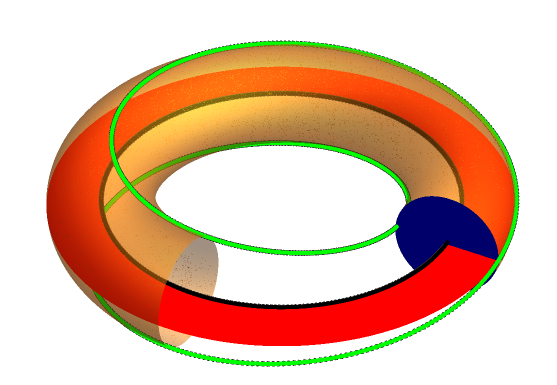}
        \caption{Flux surface \cite{wikipediaFluxSurface}}
        \label{fig:flux-surface}
    \end{subfigure}
    \hfill
    \begin{subfigure}{0.4\linewidth}
        \centering
        \includegraphics[width=\linewidth]{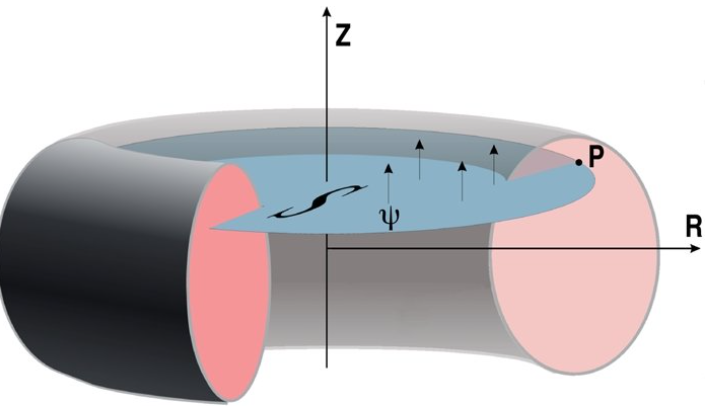}
        \caption{Poloidal flux \cite{article}}
        \label{fig:poloidal-flux}
    \end{subfigure}
    \label{fig:comparison}
\end{figure}

In a tokamak device, flux surfaces are nested and can be described by either the poloidal or toroidal magnetic flux. The poloidal flux is the magnetic flux passing through a surface encircling the center axis of a torus. The toroidal flux is the magnetic flux passing through a circle that encloses the magnetic axis of a torus. They are essentially integrals over the poloidal and toroidal components of the magnetic field, respectively. 
Periodic magnetic field lines identify rational surfaces; the magnetic field lines are continuous and do not intersect the device material. This is true for the plasma core, up to the separatrix line. Beyond the separatrix boundary, the surface are not closed, since that magnetic field lines connect to the internal surface of the device.
In Fig.~(\ref{fig:flux-surface}), the portion of a flux surface is depicted in yellow, while the magnetic axis is shown in black. The magnetic flux passing through the red and blue surfaces is called poloidal and toroidal magnetic flux, respectively.
Another visual representation of the poloidal flux function is given in Fig.~(\ref{fig:poloidal-flux}). Given a point $p$ in the poloidal section, the poloidal flux $\Psi_p$ computed at that point is defined as the total flux passing through the surface $S$, bounded by the toroidal circle tangent to the chosen point.

Let us now describe the meaning of the magnetic flux coordinates. The coordinate $\psi$ represents the \textbf{poloidal flux function} enclosed in a given magnetic surface, as previously stated. Such magnetic surfaces are nested in a tokamak, thus $\psi$ is often used as a radial-like coordinate, with $\psi=0$ at the magnetic flux axis and increasing outward as one gets closer to the plasma boundary \cite{walker2006emerging}. The coordinate $\varphi$ identifies the \textbf{toroidal angle} around the symmetry axis of the tokamak itself. Due to the tokamak geometry, this coordinate is periodic and usually ranges from $0$ to $2\pi$. The coordinate $\theta$ is a \textbf{poloidal angle} used to parameterize positions on a magnetic surface. Due to the lack of symmetry along this direction, such coordinate is usually a label for position along a field line or a flux surface. Since magnetic field lines lie on flux surfaces (where $\psi = \text{const}$), a second coordinate is needed to label different points on a given surface. The coordinate $\theta$ can be chosen to be either geometric, i.e. the usual poloidal angle in toroidal coordinates, or such that field lines appear straight in a given coordinate system. 

In the right-handed coordinates $(R,\varphi,Z)$ and $(\psi,\varphi,\theta)$, using the EFIT (Equilibrium Fitting code) convention from \cite{lao1985reconstruction_efit}, the magnetic field is described as follows.
$$\textbf{B} = \nabla\psi \times\nabla\varphi+F\nabla\varphi\;\;,$$
where $F$ is the poloidal current function \cite{lu2019development}. In the $(\psi,\varphi,\theta)$ coordinates, the safety factor is defined as $q = \textbf{B} \cdot\nabla\varphi / (\textbf{B}\cdot\nabla\theta) = JF/R^2$. \\

\begin{definition}
    The \textbf{safety factor q} described the stability of plasma confinement. Given that magnetic field lines wind helically around the tokamak, it is the ratio at which an individual magnetic file line travels around the toroidal direction for each revolution along the poloidal direction.
\end{definition}

In the $(R,\varphi,Z)$ coordinates, the equilibrium magnetic field can be expressed as:
$$B_R = -\frac{1}{R}\frac{\partial\psi}{\partial Z}\;,\;B_Z=\frac{1}{R}\frac{\partial\psi}{\partial R}\;,\;B_\varphi=\frac{F}{R}\;.$$
\begin{definition}
    The \textbf{toroidal mode number} describes the symmetry and periodicity of instabilities in the toroidal direction of a plasma inside a tokamak. This means that a mode with toroidal mode number $n$ repeats $n$ times along the toroidal direction of the device.
\end{definition}

Starting from the poloidal-toroidal coordinates, one can compute the Clebsch coordinate numerically: $\eta = \eta(r,\phi,\theta)$, where $r = \sqrt{(\Phi - \Phi_{axis})/(\Phi_{edge}-\Phi_{axis})}$. Therefore, the equations can be represented in the Clebsch coordinates $(r,\eta,\phi)$. These are tailored to represent stream functions of magnetic fields. In the context of plasma physics, a stream function is any function that is constant along a field line. A magnetic field line can be identified with a curve lying at the intersection of two surfaces belonging to two different stream functions. More specifically, the Clebsch coordinates allow to write the magnetic field as the cross product of gradients of the stream functions: $\textbf{B} = \nabla \alpha \times\nabla\beta$ \cite{d1991clebsch}. Since $\alpha$ and $\beta$ are constant along magnetic field lines, they can be effectively used as field line labels. They are useful because they do not depend on the existence of flux surfaces, allowing the description of more general magnetic field configurations. Furthermore, such coordinates facilitate the formulation of the magnetic field in terms of scalar potentials, which allows a simplification of the mathematical treatment of problems in plasma physics, especially those involving turbulence. Strong deformations of the poloidal grids are avoided. Moreover, the periodicity along $\theta$ and $\phi$ is satisfied.
This is exploited in TRIMEG by a recently developed finite element method in which the basis functions are aligned to the magnetic field lines.

\section{Transformation in Clebsch Coordinates}
Suppose we need to solve the electrostatic gyrokinetic equation, where it is necessary to compute the derivative of the electrostatic scalar potential. In cylindrical coordinates, the parallel component of the derivative is defined as follows.

\begin{equation}
  \partial_{||}\delta\phi=\left(b_R\partial_R + b_Z\partial_Z+\frac{b_\phi}{R}\partial_\phi\right)\delta\phi  \;\;,
  \label{electrostatic_potential_toroidal}
\end{equation}
where $\delta\phi$ is the perturbed component of the electrostatic scalar potential, $b_R = B_R/B,\;b_Z=B_Z/B $ and $b_\phi=B_\phi/B$ are the normalized magnetic field components in cylindrical coordinates.

The calculation of the parallel derivatives of the electrostatic potential can also be computed with respect to the field-aligned coordinates. Along the magnetic field line \textbf{B}, the Clebsch coordinates $(\chi,\xi,l)$ are computed as follows. \cite{lu2019development}
\begin{align*}
    \left. \frac{dR(\chi,\xi,l)}{d\phi} \right|_{\chi,\xi} &= R\frac{B_R}{B_\phi}\;\;, \\
    \left. \frac{dZ(\chi,\xi,l)}{d\phi} \right|_{\chi,\xi} &= R\frac{B_Z}{B_\phi}\;\;, \\
    \left. \frac{dl}{d\phi} \right|_{\chi,\xi} &= R\frac{B}{B_\phi}\;\;,
\end{align*}
where the $(\chi,\xi)$ coordinates identify a magnetic field line and can alternatively be converted to $(R,Z)$ at $\varphi=0$ for convenience, while $l$ is the coordinate along the magnetic field \textbf{B}. The derivative is then simplified to the following expression.
\begin{equation}
  \partial_{||}\delta\phi=\left. \frac{\partial}{\partial l}\right|_{\chi,\xi} \delta\phi\;\;.  
  \label{electrostatic_potential_clebsch}
\end{equation}

The accuracy of the two equations \eqref{electrostatic_potential_toroidal} and \eqref{electrostatic_potential_clebsch} can be comparable for high toroidal mode numbers when the high-order finite element method is adopted with sufficient grid numbers. However, using field-aligned coordinates and Eq.~\eqref{electrostatic_potential_clebsch}, the parallel grid resolution can be greatly reduced without losing the high accuracy of the parallel derivative. 

\section{Normalized Gyro Centers' Equations of Motion}
First, it is convenient to introduce a suitable normalization for the equations at play, so that the variables in the equilibrium input file can be used without the need for further normalization. The equilibrium input file is mainly determined by the EQDSK output which provides the macroscopic variables such as the on-axis magnetic field, the minor and major radii, and the dimension of the simulation domain. Since the length unit in EQDSK output is $1$ meter, TRIMEG adopts $1$ m as the length unit, which is also motivated by the intention to study macroscopic modes in tokamak plasmas, such as low $n$ Alfv\'enic modes. 

The particle markers are sampled from a Maxwellian distribution, which can be used to describe a system with a large number of identical physical particles in thermodynamic equilibrium. In the context of plasma physics, it is the most widely used velocity distribution function for a homogeneous and stationary medium. It represents the distribution function of the particle velocities inside an ideal gas in thermodynamic equilibrium. It can also be generalized to cases involving collisions, as well as  plasmas with long-range electrostatic interactions between charged particles \cite{belmont2019introduction}. Generally, it provides an excellent approximation of the velocity distribution for ideal plasmas, whose density is sufficiently low. The velocity of each particle is initialized from a random value by sampling the Maxwellian distribution. The Maxwellian distribution used in TRIMEG is defined as follows.

$$
f_M = \frac{\overline{n_0}}{\overline{v}_t^3\pi^{3/2}}\text{exp}\left({-\frac{mv_{||}^2+2\mu B}{2T}}\right) = \frac{\overline{n_0}}{\overline{v}_t^3\pi^{3/2}}\text{exp}\left(-\frac{\overline{m}\overline{v}_{||}^2}{\overline{T}} - 2\frac{\overline{m}\overline{\mu}_{||}\overline{B}}{\overline{T}} \right)\;\;,
$$
and the corresponding time derivative of the logarithm is

$$
\frac{d}{d\overline{t}} \text{ln}f_M = \delta \dot{\overline{\textbf{R}}} \cdot \left[\vec{\kappa}_n + \left(\frac{\overline{m}\overline{v}_{||}^2}{\overline{T}} + 2\frac{\overline{m}\overline{\mu}_{||}\overline{B}}{\overline{T}} - \frac{3}{2} \right)\vec{\kappa}_T - 2\frac{\overline{m}\overline{\mu}\overline{ B}}{\overline{T}} \vec{\kappa}_B\right]- \frac{2\overline{m}\overline{v}_{||}}{\overline{T}}\delta\dot{\overline{u}}_{||} \;\;,
$$
where $\kappa$ is the magnetic curvature, defined as follows.
\begin{definition}
    \textbf{Magnetic curvature} is a vector along the magnetic field, and is defined by $\vec{\kappa} = \vec{b}\cdot\vec{\nabla}\vec{b} $. $\kappa$ points towards the local center of curvature of B, and its magnitude is equal to the inverse radius of curvature. A plasma is said to be stable against curvature-driven instabilities, e.g. ballooning modes, when $\vec{\kappa}\cdot \vec{\nabla}p < 0 $, where $p$ is the pressure.
\end{definition}

Along the parallel velocity ($v_\|$) direction, a normal distribution is used to generate random numbers, where the mean is $x_0 = 0$ and the standard deviation is $\sigma = \sqrt{\overline{T}/2\overline{m}}$\;\;. \\

Now let us introduce the normalized version of the gyro centers' equations of motion. Their purpose is to calculate the time evolution of the gyro center velocity components in cylindrical coordinates, as well as its parallel acceleration, i.e. the acceleration component that is parallel to the magnetic field lines, to enable the analysis of particle behavior and plasma evolution over time.
The equations are the following. \cite{mishchenko2014pullback,hatzky2019reduction,lu2024gyrokinetic}
$$
 \dot{\boldsymbol R}_0 
  = u_\parallel {\boldsymbol b}^*_0 + \frac{m\mu}{qB^*_\parallel} {\boldsymbol b}\times\nabla B \;\;, 
$$

$$
  \dot u_{\parallel,0}
  = -\mu {\boldsymbol b}^*_0\cdot \nabla B \;\;,
$$

$$
  \delta\dot{\boldsymbol R}
  = \frac{{\boldsymbol b}}{B^*_\parallel}\times \nabla \langle \delta\Phi -u_\parallel \delta A_\parallel\rangle 
  -\frac{q_s}{m_s}\langle\delta A^{\rm h}_\parallel\rangle {\boldsymbol b}^*\;\;, 
$$

$$
  \delta \dot u_\parallel
  =  -\frac{q_s}{m_s} \left({\boldsymbol b}^*\cdot\nabla\langle\delta\Phi-u_\parallel\delta A^{\rm{h}}_\parallel\rangle +\partial_t\langle\delta A_\parallel^{\rm{s}}\rangle \right) 
  -\frac{\mu}{B^*_\parallel}{\boldsymbol b}\times\nabla B\cdot\nabla\langle\delta A_\parallel^{\rm{s}}\rangle \;\;,  
$$
where ${\boldsymbol b}^*={\boldsymbol b}_0^*+\nabla\langle\delta A_\parallel^{\rm s}\rangle\times{\boldsymbol b}/B_\parallel^*$, ${\boldsymbol b}^*_0={\boldsymbol b}+(m_s/q_s)u_\parallel\nabla\times{\boldsymbol b}/B_\parallel^*$, ${\boldsymbol b}={\boldsymbol B}/B$ is the unit vector in the direction of the equilibrium magnetic field. The effective magnetic field $B^*$ differs from the physical magnetic field in that it incorporates the effects of the particle's parallel velocity. This allows us to get the effective magnetic field $B_{||}^*$ along the field lines, which is defined as $B_\parallel^*=B+(m_s/q_s)u_\parallel{\boldsymbol b}\cdot(\nabla\times{\boldsymbol b})$. 
The complete form of the equations in cylindrical coordinates $(R,\phi,Z)$ is described below. Note that all the variables are normalized, e.g. $m_s$. The reference component $\textbf{b}_0^*$ in $(R,\phi,Z)$ coordinates is defined as follows. \cite{lu2024gyrokinetic}

\begin{align}
    b_{0R}^* &= b_R-\frac{B_0}{B_\parallel^*}\rho_0 v_\parallel\partial_Zb_\phi \;\;, 
        \label{unperturbed_magnetic_field_R_equations_of_motion}\\ 
    b_{0Z}^* &= b_Z+\frac{B_0}{B_\parallel^*}\rho_0
    v_\parallel\left(
    \partial_Rb_\phi+\frac{b_\phi}{R} \right) \;\;, 
        \label{unperturbed_magnetic_field_Z_equations_of_motion}\\
    b_{0\phi}^* &= b_\phi+\frac{B_0}{B_\parallel^*}\rho_0 v_\parallel\left(
    \partial_Zb_R-\partial_Rb_Z \right) \;\;,
    \label{unperturbed_magnetic_field_phi_equations_of_motion}
\end{align}

\noindent The complete term is the following.

\begin{align}
    b_R^* &= b_{0R}^*
    +\frac{B_0}{B_\parallel^*}\rho_0 \left[\frac{b_Z}{R}\partial_\phi-b_\phi\partial_Z\right]\langle\delta A_\parallel^{\rm s}\rangle\;\;,
    \label{perturbed_magnetic_field_R_equations_of_motion}\\
    b_Z^* &= b_{0Z}^*
    +\frac{B_0}{B_\parallel^*}\rho_0 
    \left[b_\phi\partial_R-\frac{b_R}{R}\partial_\phi\right]\langle\delta A_\parallel^{\rm s}\rangle\;\;,
    \label{perturbed_magnetic_field_Z_equations_of_motion}\\
    b_\phi^* &= b_{0\phi}^*
    +\frac{B_0}{B_\parallel^*}\rho_0 
    \left[
    b_R\partial_Z-b_Z\partial_R\right]\langle\delta A_\parallel^{\rm s}\rangle\;\;.
    \label{perturbed_magnetic_field_phi_equations_of_motion}
\end{align}

The equations can be decomposed in the equilibrium part and the perturbed part, given that we are working with the $\delta$f formulation. The equilibrium part is defined as follows.

\begin{align}
\dot{R_0}&=b_{0R}^*u_{||} + C_d B_\phi\partial_ZB \;\;,
\label{R_component_equation_of_motion}\\
\dot{Z_0}&=b_{0Z}^*u_{||} + C_d B_\phi\partial_RB \;\;,
\label{Z_component_equation_of_motion}\\
\dot{\phi_0}&=\frac{b_{0\phi}^*}{R}u_{||} + \frac{C_d}{R}(B_Z\partial_RB- B_R\partial_ZB) \;\;,
\label{phi_component_equation_of_motion}\\
\dot{u}_{||,0} &= -\mu(b_{0R}^*\partial_RB+b_{0Z}^*\partial_ZB) \;\;,
\label{velocity_equation_of_motion}
\end{align}
where the coefficient $C_d$ is defined as $C_{\rm d}=(m_s/q_s)\rho_0 \mu B B_0/(B^2B^*_\parallel)\;\;.$

\noindent The perturbed contribution of the gyro centers' equations of motion is defined as

\begin{align}
    \delta\dot{R} &= C_{\rm E}\left(
        b_\phi\partial_Z\delta G-\frac{1}{R}b_Z\partial_\phi\delta G\right)
        -\frac{\bar q_s}{\bar m_s}\langle\delta\bar A_\parallel^{\rm h}\rangle {b}^*_R \;\;,
    \label{R_perturbed_equation_of_motion}\\
    \delta\dot{Z} &= C_{\rm E} \left(
        -b_\phi\partial_R\delta G+\frac{1}{R}b_R\partial_\phi\delta G\right)
        -\frac{\bar q_s}{\bar m_s}\langle\delta\bar A_\parallel^{\rm h}\rangle {b}^*_Z \;\;,
    \label{Z_perturbed_equation_of_motion}\\
    \delta\dot{\phi} &= \frac{C_{\rm E}}{R}\left(
        b_Z\partial_R\delta G-b_R\partial_Z\delta G\right) 
        -\frac{1}{R}\frac{\bar q_s}{\bar m_s}\langle\delta\bar A_\parallel^{\rm h}\rangle {b}^*_\phi\;\;,
    \label{phi_perturbed_equation_of_motion}
\end{align}

\begin{equation}
    \begin{aligned}
        \delta\dot{u}_\parallel &= -\frac{q_s}{m_s}\Big[ 
      b_R^*(\partial_R\delta\Phi-u_\parallel\partial_R\delta A^{\rm h}_\parallel) 
        + b_Z^*(\partial_Z\delta\Phi-u_\parallel\partial_Z\delta A^{\rm h}_\parallel) 
        +\frac{b_\phi^*}{R}(\partial_\phi\delta\Phi-u_\parallel\partial_\phi\delta A^{\rm h}) \\
        &+\partial_t\delta A_\parallel^{\rm s} 
        + \frac{m_s}{q_s} (\dot{R}_{\mu}\partial_R+\dot{Z}_{\mu}\partial_Z+\dot{\phi}_{\mu}\partial_\phi)\langle\delta A_\parallel^{\rm s}\rangle 
    \Big] \;\;,
    \end{aligned}
\end{equation}
where $\delta G = \delta\Phi - u_{||}\delta A_{||}$, $C_{\rm E}=\rho_0 B_0/B^*_\parallel$, $C_\mu=(m_s/q_s)\rho_0 \mu BB_0/(B^2B^*_\parallel)$, $\dot{R}_{\mu}=C_\mu B_\phi\partial_ZB$, $\dot{Z}_{\mu} =-C_\mu B_\phi\partial_RB$, $\dot{\phi}_{\mu} =\frac{C_\mu}{R}( B_Z\partial_RB-B_R\partial_ZB)$. 
By applying the ideal Ohm's law for the simplectic part of the magnetic potential we obtain

\begin{equation}
\begin{aligned}
    \delta\dot u_{\parallel} &= -\frac{q_s}{m_s} \Big[ 
        - b_Ru_\parallel\partial_R\delta A^{\rm h}_\parallel 
        - b_Zu_\parallel\partial_Z\delta A^{\rm h}_\parallel
        - ({b_\phi}/{R}) u_\parallel\partial_\phi\delta A^{\rm h}_\parallel 
        +\Delta b_R^*\partial_R\delta\Phi  \\
        &+\Delta b_Z^*\partial_Z\delta\Phi
        +\Delta b_\phi^*\partial_\phi\delta\Phi \Big] 
        -(\dot{R}_{\mu}\partial_R+\dot{Z}_{\mu}\partial_Z+\dot{\phi}_{\mu}\partial_\phi)\langle\delta A_\parallel^{\rm s}\rangle\;\;,
\end{aligned}
\label{velocity_perturbed_equation_of_motion}
\end{equation}
where $\Delta{\boldsymbol b}^*={\boldsymbol b}^*-{\boldsymbol b}$.

An approximation can be introduced by considering only the dominant terms of the gyro centers' equations of motion, which can be obtained by replacing \textbf{$b^*$} with \textbf{b}, the normalized magnetic field.
The equilibrium contribution of the gyro center motion in the $(R,\phi,Z)$ coordinates is defined as follows.

\begin{align}
\dot R_0 &= b_R u_\parallel + C_{\rm d} B_\phi\partial_Z B \;\;,
\label{R_reduced_equations_of_motion}\\
\dot Z_0 &= b_Z u_\parallel - C_{\rm d} B_\phi\partial_R B \;\;,
\label{Z_reduced_equations_of_motion}\\
\dot\phi_0 &= \frac{b_\phi}{R} u_\parallel + \frac{C_{\rm d}}{R}(B_Z\partial_R B - B_R\partial_Z B) \;\;,
\label{phi_reduced_equations_of_motion}\\
\dot u_{\parallel,0} &= -\mu (b_{0R}^*\partial_R B + b_{0Z}^*\partial_Z B) \;\;,
\label{velocity_reduced_equations_of_motion}\\
\end{align}
where $C_{\rm d} = (m_s/q_s){\rho_0}(v_\parallel^2 + \mu B)B_0/B^3$.

The perturbed component of the equations of motion is defined as follows.

\begin{align}
\delta \dot R &= C_{\rm E}\left(b_\phi\partial_Z\delta G - \frac{1}{R}b_Z\partial_\phi\delta G\right) - \frac{\bar q_s}{\bar m_s}\langle\delta\bar A_\parallel^{\rm h}\rangle {b}_R\;\;,
\label{R_perturbed_reduced_equations_of_motion}\\
\delta \dot Z &= C_{\rm E} \left(-b_\phi\partial_R\delta G + \frac{1}{R}b_R\partial_\phi\delta G\right) - \frac{\bar q_s}{\bar m_s}\langle\delta\bar A_\parallel^{\rm h}\rangle {b}_Z\;\;,
\label{Z_perturbed_reduced_equations_of_motion}\\
\delta \dot\phi &= \frac{C_{\rm E}}{R}\left(b_Z\partial_R\delta G - b_R\partial_Z\delta G\right) - \frac{1}{R}\frac{\bar q_s}{\bar m_s}\langle\delta\bar A_\parallel^{\rm h}\rangle {b}_\phi\;\;,
\label{phi_perturbed_reduced_equations_of_motion}\\
\end{align}

\begin{equation}
    \begin{aligned}
        \delta \dot u_{\parallel} &= -\frac{q_s}{m_s}\Big[ 
          b_R(\partial_R\delta\Phi - v_\parallel\partial_R\delta A^{\rm h}_\parallel) 
        + b_Z(\partial_Z\delta\Phi - v_\parallel\partial_Z\delta A^{\rm h}_\parallel) 
        + \frac{b_\phi}{R}(\partial_\phi\delta\Phi \\
        &- v_\parallel\partial_\phi\delta A^{\rm h}_\parallel) 
        + \partial_t\delta A_\parallel^{\rm s} 
        + \frac{m_s}{q_s}(\dot{R}_{\mu}\partial_R + \dot{Z}_{\mu}\partial_Z + \dot{\phi}_{\mu}\partial_\phi)\langle\delta A_\parallel^{\rm s}\rangle 
        \Big]\;\;,
    \end{aligned}
\end{equation}
where $\delta G = \delta\Phi - u_\parallel\delta A_\parallel$,$C_{\rm E} = {\rho_0}B_0/B$, $C_\mu = (m_s/q_s){\rho_0}\mu BB_0/B^3$, $\dot{R}_{\mu} = C_\mu B_\phi\partial_Z B$, $\dot{Z}_{\mu} = -C_\mu B_\phi\partial_R B$, $\dot{\phi}_{\mu} = \frac{C_\mu}{R}(B_Z\partial_R B - B_R\partial_Z B)$. If the ideal Ohm's law is used, we obtain the following.

\begin{equation}
{ \delta \dot u_{\parallel} } = +\frac{q_s}{m_s} \left( 
      b_R v_\parallel\partial_R\delta A^{\rm h}_\parallel 
    + b_Z v_\parallel\partial_Z\delta A^{\rm h}_\parallel
    + \frac{b_\phi}{R} v_\parallel\partial_\phi\delta A^{\rm h}_\parallel
    \right) - (\dot{R}_{\mu}\partial_R + \dot{Z}_{\mu}\partial_Z + \dot{\phi}_{\mu}\partial_\phi)\langle\delta A_\parallel^{\rm s}\rangle \;\;.
\label{velocity_perturbed_reduced_equations_of_motion}
\end{equation}

\listoffigures

\listoftables

\chapter*{Acknowledgements}
I would like to express my gratitude to my supervisors, Prof. Gianluca Palermo and Dr. Zhixin Lu, for their continuous support, useful insights, patience and encouragement during the period of my master's thesis. Their expertise in gyrokinetics and deep insight into the code played an important role in shaping the direction of this work.
I would also like to thank Dr. Matthias Hoelzl for their valuable suggestions and constructive conversations, which improved the quality of this work.
I further wish to acknowledge the JOREK group members, who offered technical and moral support, which was fundamental to overcome the hurdles along the way.
Finally, I am grateful for the ever-present love of my family, no matter the spatial distance between us. My parents Roberta and Ivano, and my brother Gregorio are - and will always be - the cornerstone of my strength and determination. 

\cleardoublepage

\end{document}